\documentclass[a4paper,fleqn]{cas-dc}

\usepackage{natbib}

\usepackage{graphicx}
\usepackage{xcolor}
\usepackage[utf8]{inputenc}
\usepackage[T1]{fontenc}
\usepackage{lineno}
\usepackage{placeins}
\usepackage{caption}

\def\tsc#1{\csdef{#1}{\textsc{\lowercase{#1}}\xspace}}
\tsc{WGM}
\tsc{QE}

\newcommand{\sfive}{S5~1044+71}
\newcommand{\fermi}{{\it Fermi}-LAT}
\newcommand{\gray}{$\gamma$-ray}
\newcommand{\grays}{$\gamma$-rays}

\begin{document}
\let\WriteBookmarks\relax
\def\floatpagepagefraction{1}
\def\textpagefraction{.001}

\shorttitle{A precessing jet from a supermassive black hole}    

\shortauthors{M. Cerruti et al.}  

\title [mode = title]{A precessing jet from a supermassive black hole: multi-wavelength observations of \sfive}  



%




\author[1]{M. Cerruti}[orcid=0000-0001-7891-699X]

\cormark[1]

\ead{cerruti@apc.in2p3.fr}


\affiliation[1]{organization={Université Paris Cité, CNRS, Astroparticule et Cosmologie},
            city={Paris},
            postcode={F-75013}, 
            country={France}}

\author[1]{P. A. Duverne}[orcid=0000-0002-3906-0997]

\cormark[1]

\ead{duverne@apc.in2p3.fr}

\author[2]{G. Ganesaratnam}[orcid=]

\cormark[1]

\ead{ganushan.ganesaratnam@ias.u-psud.fr}


\affiliation[2]{organization={Institut d’Astrophysique Spatiale},
            city={Orsay},
            postcode={F-91405}, 
            country={France}}

\author[3,1]{P. Goswami}[orcid=0000-0001-5430-4374]

\cormark[1]

\ead{pgoswami@lsw.uni-heidelberg.de}


\affiliation[3]{organization={ZAH Landessternwarte, Universität Heidelberg},
            addressline={Königstuhl 12},
            city={Heidelberg},
            postcode={D-69029}, 
            country={Germany}}

\author[4]{H. X. Ren}[orcid=0000-0003-0221-2560]

\cormark[1]

\ead{helena.ren@mpi-hd.mpg.de}


\affiliation[4]{organization={Max-Planck-Institut für Kernphysik},
            addressline={PO Box 103980},
            city={Heidelberg},
            postcode={D-69029}, 
            country={Germany}}

\author[5]{N. Sahakyan}[orcid=0000-0003-2011-2731]

\cormark[1]

\ead{narek.sahakyan@icranet.org}


\affiliation[5]{organization={ICRANet-Armenia},
            addressline={Marshall Baghramian Avenue 24a},
            city={Yerevan},
            postcode={AM-0019}, 
            country={Armenia}}

\cortext[1]{Corresponding author}


\begin{abstract}
 The bright \gray\ blazar \sfive\ has been identified as showing very significant quasi-periodic oscillations in the \fermi\ data  in recent studies, with a periodicity of about three years.
   With the completion of a new three-year \gray\ cycle, we aim to revisit the periodicity in \fermi\ data, and analyze all available multi-wavelength (MWL) data to search for possible correlations and time-lags. These observations will be used to test for the compatibility of the observed periodicity with a precessing jet from the supermassive black hole powering the blazar.
   We analyze data from \fermi\ (\grays), \textit{NuSTAR}, \textit{Swift}-XRT, and \textit{AstroSat}-SXT (X-rays), \textit{Swift}-UVOT, \textit{AstroSat}-UVIT, ASAS-SN, ZTF, Pan-STARRS (ultraviolet and optical), and NEOWISE (infrared). In addition we present an analysis from historical observations from the Palomar and Pulkovo observations. Single-band spectral variability, MWL correlations, and cross-correlations are computed. We then model the \fermi\ light curve with a precessing jet model, providing constraints on the geometry of the system and providing the evolution of the Doppler factor with time. The latter is used as a geometrical constraint in the modeling of the MWL spectral energy distributions, to test whether they are compatible with the jet parameters inferred from the light curve study.
   We confirm previous claims on the existence of a periodic \gray\ signal in \sfive. We detect significant spectral variability in \gray, X-rays, and optical/UV data. We detect significant correlation between low-energy (infrared/optical/ultraviolet) data and \grays, with a correlation index of about 1; the correlation between X-rays and \grays\ is milder, with a correlation index of about 0.3. We do not detect any significant time-lag between bands. The \fermi\ light curve is successfully fit by a precessing jet model. The fit to the spectral energy distributions indicate that \sfive\ is a typical blazar, in which the \gray\ emission is located beyond the broad-line region.
   The existence of a periodic \gray\ signal \sfive\ is confirmed. All the MWL observations we present in this work are consistent with the existence of a precessing relativistic jet from the supermassive black hole.
\end{abstract}





\maketitle


\section{Introduction }

The extragalactic \gray\ sky is dominated by Active Galactic Nuclei (AGNs), accreting supermassive black holes (SMBHs) with masses ranging from $10^6$ to $10^{10} M_\odot$, located  at cosmological distances. The large majority of \gray\ AGNs belong to the class of blazars \citep{4LAC+}, which are characterized by rapid variability, a spectral energy distribution (SED) showing a broad non-thermal continuum from radio to TeV energies, and a high degree of polarization in radio, optical, and X-rays \citep[see e.g.][for a recent review]{Blandford}. High angular resolution radio observations of these objects show a core-dominated morphology with a single jet-like structure \citep[e.g.][]{Lister09}. These observational characteristics are well explained within the AGN unified model assuming that blazars are radio-loud AGNs (i.e. objects in which the accretion of matter onto the SMBH is accompanied by the launching of two jets of plasma along the polar axis of the black hole) whose jet is closely aligned to the line of sight \citep{1995PASP..107..803U}. The small angle to the observer implies a large value of the relativistic Doppler factor $\delta$ that takes a value typically $\delta \geq10$. The emission from the jet is thus Doppler-boosted in the reference frame of the observer, and dominates blazar SEDs. From an observational point of view, blazars are classified into two subclasses, flat-spectrum-radio-quasars (FSRQs) and BL Lacertae objects, according to the presence (in the former) or absence (in the latter) of emission lines in the optical/UV spectrum, with the threshold between the two classes set at an equivalent width of the emission lines equal to $5\mathring{\textrm{A}}$. In blazar leptonic radiative models, the \gray\ emission is produced via inverse Compton scattering of low-energy photons by electrons/positrons in the jet. In FSRQs, the photons upscattered to \grays\ are likely external to the jet, and are either from the emission lines produced in the broad-line-region (BLR), or from the thermal black-body radiation of the dusty torus \citep[for further details see e.g.][for a recent review on radiative modeling in blazars]{Cerruti20}.\\

The Large Area Telescope \citep[LAT,][]{LAT} on board the \textit{Fermi} satellite has significantly improved our knowledge of \gray\ blazars, in particular thanks to its survey capabilities designed to monitor the whole sky about every three hours. Uninterrupted \gray\ light curves on thousands of blazars are thus available since the launch of the \textit{Fermi} satellite in June 2008, and represent a unique dataset to study blazar variability. \\

One of the unexpected discoveries in LAT AGN observations is the periodicity in the light curves of some of these \gray\ blazars, with periods ranging from months to years. The first hint (3$\sigma$) for periodic modulation was seen in PG~1553+113 by \citet{Ackermann15}. Following this first tentative detection, many authors have extensively studied periodic or quasi-periodic oscillations (QPOs) in LAT blazars \citep[see][and references therein]{Covino20, Penil20, Yang21}. Although many of these results are of low significance and become compatible with noise once the large trial factors are included \citep{Ren23QPO, Penil25}, some periodicities seem solid: besides PG 1553+113, among the significant QPO detections, it might be worth mentioning the one in PKS 2247-131 \citep{Zhou18}. The work presented in \citet{Ren23QPO} investigates QPOs in a sample made of the 35 brightest LAT AGNs, and identifies the source \sfive\ (an FSRQ at redshift $z=1.15$) among the most significant candidates. The object shows a QPO behavior with a period of about 3 years ($P = 1100 \pm 200$ days). The original work shows three full cycles of the periodic signal. The overall significance was lowered by the fact that the first supposed maximum was strongly suppressed compared to the next ones, likely due to the intrinsic blazar variability. The same periodicity in \sfive\ is identified by \citet{Wang22}.
A multi-wavelength (MWL) data analysis of the object was presented in \citet{Kun22}, covering the same data as in \citet{Ren23QPO}, i.e. four full cycles assuming the first one was suppressed due to blazar variability. VLBI radio observations are shown in \citet{Kun22}, showing that the kpc and pc jets are misaligned by about 90$^\circ$, suggesting jet precession.  A more recent follow-up study in \gray\ and radio by the same authors is presented in \citet{Kun24}, covering part of the fifth maximum (including data up to June 2023).\\

Several models have been developed to explain \gray\ QPOs in blazars. They can be split into two families: geometric models, in which the periodic modulations are due to changes in the viewing angle of the emitting region, and thus to changes in the Doppler factor $\delta$; and accretion-ejection coupling models, in which the periodicities are produced at the level of the accretion flow, and propagate into the jet and thus in the \gray\ light curve \citep[as discussed in][for an application to OJ~287]{Valtonen09}. Concerning geometric models, two main options arise: either the whole jet change its angle to the observer \citep[see][]{Begelman80, Sobacchi17}, or the \gray\ emitting region is a sub-structure in the jet, and the QPO track its motion in the jet, for example in the form of a helical path giving access to jet structure \citep{Rieger04}. In the first case, the fact that the jet is precessing, or that the SMBH itself is moving on a circular orbit, could imply the existence of a companion massive enough to produce this effect, and hence open the window to detect and study SMBH binaries in the Universe. \\

In this work, we present a new MWL analysis of \sfive\  that incorporates updated \fermi\ observations covering five full cycles, together with X-ray data \textit{NuSTAR}, \textit{Swift}-XRT and \textit{AstroSat}-SXT; optical/UV measurements from \textit{Swift}-UVOT and \textit{AstroSat}-UVIT, and additional archival optical and IR data from surveys including ASAS-SN, ZTF, Pan-STARRS1 and NEOWISE. Furthermore, we explored historical optical observations from Palomar and Pulkovo, obtained via the NAROO project \citep{Robert}. Following the results by \citet{Kun22}, that suggests that the periodicity is related to jet precession, we focus on this radiative scenario. The present work extends previous studies in three key directions. 
First, the detection of a new \gray\ maximum at the epoch predicted by \citet{Ren23QPO} provides an a-posteriori confirmation of the $\sim$3-year timescale of the QPO. Second, we perform a comprehensive MWL variability and correlation analysis over more than a decade of observations, allowing us to compare flux–flux relations and time-lag properties across multiple activity cycles. Finally, we combine the variability analysis with a time-dependent geometrical model and multi-epoch SED modeling, enabling us to constrain the jet precession scenario in a self-consistent framework. \\

This paper is structured as follows: in section \ref{sec:anal} we present the data analysis from all instruments; in section \ref{sec:var} we analyze the variability properties of the dataset, including spectral variability within a band, MWL correlations, search for time lags, and revisited QPO search; in section \ref{sec:four} we present a modeling of the LAT light curve; in section \ref{sec:five} we present a MWL modeling of the SED; in section \ref{sec:disc} we provide discussion and conclusions.\\

\section{Multi-wavelength observations}\label{sec:anal}

\begin{figure*}
    \centering
    \includegraphics[width=1\linewidth]{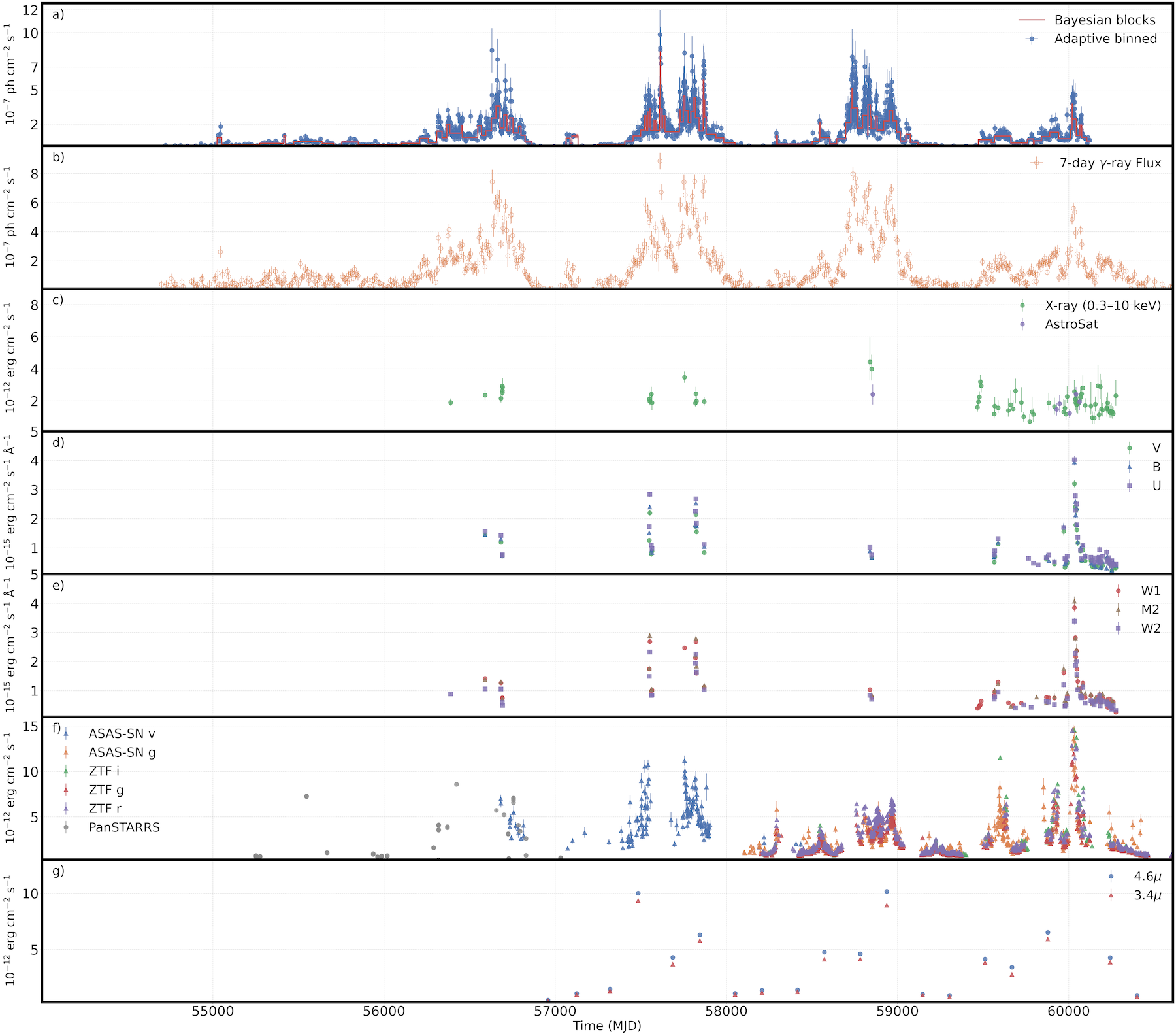}
    \caption{MWL light curve of \sfive. From top to bottom: \fermi\ integral photon flux (100 MeV to 100 GeV) using adaptive binning, and with a fixed 7-day time bin; \textit{Swift}-XRT and \textit{AstroSat} X-ray integral energy flux (0.3 keV to 10 keV); \textit{Swift}-UVOT flux density per filter (split in W1, M2, and W2, and V,B, and U, for visibility); optical spectral energy distribution from ASAS-SN, ZTF, and Pan-STARRS; NEOWISE spectral energy distribution at 3.4$\mu m$ and 4.6$\mu m$.}
    \label{fig:lc}
\end{figure*}

\subsection{\textit{Fermi}-LAT}

    The LAT \citep{LAT} instrument onboard the \textit{Fermi} satellite is a pair-conversion detector, sensitive to photons with energy ranging from 20~MeV to 300~GeV and it carries out an all-sky survey every 3 hours. 
    The \gray\ data used in this work were obtained from the Markarian Multiwavelength Data Center \citep[MMDC, \url{mmdc.am}, see][]{2024AJ....168..289S}. Specifically, we used the adaptively binned \gray\ light curve, and the results of spectral analyses performed over different periods.
    The data reduction is performed using \texttt{fermitools} (version 2.0.8) with the \texttt{P8R3\_SOURCE\_V3} instrument response functions. Events in the 100~MeV–300~GeV energy range are selected within a $12^\circ$ region centered on the \gray\ position of \sfive\ (RA$=162.11$, Dec$=71.73$). The dataset we use covers a time period starting from August 4, 2008, to August 4, 2025 (MET 239667417–775958405) and includes only events classified as {\it evclass=128} and {\it evtype=3}. To minimize contamination from Earth-limb photons, events with zenith angles larger than $90^\circ$ are excluded. Then, we apply a likelihood analysis that includes all sources within a $17^\circ$ radius of the target from the incremental version of the \fermi\ fourth source catalog (\texttt{4FGL-DR3}; \citealt{4LAC+}). The spectral parameters of sources within $12^\circ$ of the target are left as free parameters, while those outside this region are fixed to their catalog values. The model also includes the standard Galactic diffuse emission template \texttt{gll\_iem\_v07} and the isotropic background model \texttt{iso\_P8R3\_SOURCE\_V3\_v1} \footnote{\url{https://fermi.gsfc.nasa.gov/ssc/data/access/lat/BackgroundModels.html}}. The parameter optimization is performed through a maximum-likelihood fit, where the significance of the \gray\ signal is quantified using the test statistic, denoted TS:
        \begin{equation}
            TS = 2(\ln L_1 - \ln L_0),
        \end{equation}
     where $L_1$ and $L_0$ are the likelihood values with and without the source, respectively \citep{1996ApJ...461..396M}.

    The \gray\ emission of \sfive\ is investigated using the adaptive binning method \citep{2012A&A...544A...6L}. This approach allows to track the temporal variability of the \gray\ flux with high sensitivity, enabling the detection of even modest flux changes and facilitating the identification of distinct emission states of the source \citep[see, e.g.,][]{2013A&A...557A..71R, 2016ApJ...830..162B, 2017MNRAS.470.2861S, 2017A&A...608A..37Z, 2017ApJ...848..111B, 2018ApJ...863..114G, 2018A&A...614A...6S, 2021MNRAS.504.5074S, 2022MNRAS.517.2757S, 2022MNRAS.513.4645S}. Unlike traditional fixed-binning techniques, the adaptive method determines the time intervals of the light curve such that each bin achieves a predefined flux uncertainty -- here set to 20\% -- above the optimal energy threshold -- $E_{\rm opt}=185.6$ MeV in this work. For each interval, an unbinned likelihood analysis is then performed using the previous criteria.

    The \gray\ light curve of \sfive\ (above $185.6$ MeV) computed with the adaptive binning method is shown in panel a) of Figure \ref{fig:lc}. The source clearly shows strong variability with phases alternating between quiescent states and phases of enhanced activity during the period covered by the dataset. The average flux in the \gray\ band is $2.15 \times 10^{-7}\mathrm{ph\:cm^{-2}\:s^{-1}}$, while during bright states it increases up to a maximum \gray\ flux of $\sim9.84 \times 10^{-7}\mathrm{ph\:cm^{-2}\:s^{-1}}$, observed on MJD 57612.69. The light curve also reveals several distinct activity cycles. The source initially remains in a low activity state, followed by a pronounced active phase (e.g., between MJD $\sim 56100-56800$ and MJD $\sim 57300-58000$, each characterized by multiple flares. After these episodes, the source returns to a quiescent state before entering another active period with similarly strong variability. In total, at least four such cycles can be distinguished, each lasting for extended durations when flaring episodes are observed.

    In addition to the \gray\ light curve computed with the adaptive binning method, the 7-day binned light curve is also shown in panel b) of Figure \ref{fig:lc}. These light curves are extracted from the \fermi\ Light Curve Repository\footnote{\url{https://fermi.gsfc.nasa.gov/ssc/data/access/lat/LightCurveRepository/}}, where we keep only measurements with $TS > 4$ -- corresponding to a $\sim2\sigma$ detection. Although the flux values in this light curve are averaged over 7-day intervals, the overall variability pattern is consistent with that seen in the adaptively binned light curve. In particular, the source exhibits recurrent activity cycles, within which distinct flaring episodes are observed. Together, these two binning approaches provide complementary views of the temporal variability, with adaptive binning highlighting rapid changes and fixed bins confirming the long-term modulation.

    \subsection{\textit{NuSTAR}}
    
    The \textit{NuSTAR} satellite \citep{nustar} observed \sfive\ in the 3–79 keV energy range on 2023 April 3, with a total exposure of 23.1 ks. The \textit{NuSTAR} observations of \sfive\ were retrieved from the \texttt{MMDC} \citep{2024AJ....168..289S}. The data are analyzed using the \texttt{NuSTAR\_Spectra} tool \citep{2022MNRAS.514.3179M} which accesses the \textit{NuSTAR} data archive, retrieves the observations, and performs standard data reduction. In particular, the science-ready products are obtained with the \textit{nuproducts} pipeline. Source counts are extracted from a circular region with a radius of $40^{\prime\prime}$, while background counts are obtained from an annular region centered on the source position, with inner and outer radii of $200^{\prime\prime}$ and $300^{\prime\prime}$, respectively. Spectral fitting is performed assuming a power-law model modified by Galactic absorption. Galactic $N_H$ is fixed to $3.92\times10^{20}$ cm$^{-2}$ \citep{HI4PI}. The best-fit parameters are derived using the \texttt{XSPEC} package \citep{1996ASPC..101...17A} and the Cash statistic \citep{1979ApJ...228..939C}. 
    The analysis shows that the flux in the 3–10 keV band is $(1.45\pm0.07)\times10^{-12}\,\mathrm{erg\,cm^{-2}\,s^{-1}}$, while it is $(2.31\pm0.17)\times10^{-12}\,\mathrm{erg\,cm^{-2}\,s^{-1}}$ in the 10–30 keV band. The photon index measured over the \textit{NuSTAR} energy range is $\Gamma = 1.52\pm0.09$, indicating a hard X-ray spectrum. This spectral slope is consistent with the rising part of the inverse-Compton component, as commonly observed in FSRQs. The corresponding \textit{NuSTAR} spectrum is shown in the lower panel of Figure~\ref{fig:sed}.

    \subsection{\textit{Swift}-XRT}
    
    \sfive\ has been observed several times by the \textit{Swift} satellite \citep{Swift}, often in response to bright flaring states seen by \fermi. In addition to these pointed observations, the source is in the field of view of \textit{Swift} observation of the star KELT-24B, targeted to study its exoplanet \citep{Pillitteri}. Given that the observations of KELT-24B have very short individual exposures, we stacked the observations taken within a month.
    
    The \textit{Swift}-XRT data \citep{XRT} are processed using the standard \texttt{HEASOFT} (v.6.34) tools and the most recent \texttt{CALDB} files \footnote{\url{https://heasarc.gsfc.nasa.gov/FTP/caldb}}. For each observation, the event files are reprocessed with the task \texttt{xrtpipeline}, applying the standard screening criteria and generating cleaned event lists and exposure maps. Source and background spectra are extracted from the cleaned event files using circular regions. For each observation, the corresponding ancillary response files (ARF) are generated with \texttt{xrtmkarf}, making use of the exposure maps produced with \texttt{xrtexpomap}. The appropriate response matrix files (RMF) are taken from CALDB. 
    Given the generally limited photon statistics of individual observations, the spectra are grouped to a minimum number of 5 counts per bin and analyzed using Cash statistics. Spectral fitting is performed in \texttt{XSPEC} assuming an absorbed power-law model (\texttt{TBabs*powerlaw}), with the hydrogen column density fixed to the Galactic value along the line of sight $N_{H} = 3.92\times10^{20}$ cm$^{-2}$. For observations with sufficient statistics, both the photon index and normalization are left as free parameters during the fit. For the faintest observations, where the two free parameters could not be both meaningfully constrained, the photon index is fixed to $1.5$, and only the normalization is fitted. In all cases, the 0.3–10 keV fluxes are derived from the best-fit models, using the flux command in \texttt{XSPEC}.\\

    \subsection{\textit{AstroSat}-SXT}
    
    The Soft X-ray Telescope (SXT) onboard the \textit{AstroSat} satellite is a CCD imaging instrument with a 2~m focal length, an angular resolution of $\sim$2 arcmin, and operating in the 0.3–8 keV energy range \citep{2017JApA...38...29S}. \textit{AstroSat}-SXT observed \sfive\ in photon counting mode as a target-of-opportunity (ToO) in 2020. Following a successful proposal by the authors to a regular announcement of opportunity (AO) call, \textit{AstroSat} observed \sfive\ at several epochs in 2022–2023.
    We processed the data orbit-by-orbit starting from Level-1 products, retrieved from the ISRO Science Data Archive\footnote{\url{https://astrobrowse.issdc.gov.in/astro_archive/archive/Home.jsp.}}
    Each orbit is reduced with {\tt sxtpipeline} available in the SXT software (\texttt{AS1SXTLevel2}, version 1.4b), and the resulting Level-2 event lists are merged into a single cleaned event file using the SXT event-merger utility ({\tt SXTEVTMERGER }/ {\tt SXTMerger.jl}) provided by the SXT Payload Operation Center \footnote{\href{www.tifr.res.in/~astrosat_sxt}{www.tifr.res.in/~astrosat$\_$sxt}}. We used {\tt XSELECT} (version 2.4d) package built-in to \texttt{HEAsoft} (version 6.31.1) to extract image, light curves and spectra from the merged event files. A 16 arcmin circular source aperture is set from the SXT {\tt sxtEEFmake} tool that encloses $>95\%$ of the source's pixels. For timing analysis, the $0.3$--$10.0$ keV (to be consistent with the \textit{Swift}-XRT flux measurements) light curves are extracted using a 1-day bin size and are shown in Figure \ref{fig:lc}, panel c). For spectrum extraction we use the standard deep blank-sky background ({\tt "SkyBkg$\_$comb$\_$EL3p5$\_$Cl$\_$Rd16p0$\_$v01.pha"}) recommended for SXT to mitigate the broad point spread function. We generate source-specific ARFs with the {\tt sxtmkarf} tool, and we applied the standard RMF ({\tt "sxt\_pc\_mat\_g0to12.rmf"}). Each spectrum is grouped to a minimum of 40 counts per bin prior to fitting. The spectra are fitted in {\tt XSPEC} with the \texttt{TBabs*powerlaw} model, fixing the Galactic $N_H$ ($3.92\times10^{20}$ cm$^{-2}$) along the line of sight. The spectra from ToO and AO observations are shown in Figure \ref{fig:sed}.  

        \begin{table*}[ht!]
        \centering
        \caption{Photometric measurements from archival photographic plates. The telescope names provided in the table are the one provided in the image's header.}
        \label{tab:archival_photometry}
        \begin{tabular}{lccccccc}
        \hline
        Image & Mag & Err & Detection & Band & Time (UTC) & Telescope \\
        \hline
        038-7169-B & 19.7    & 0.5 & visible & g  & 1996-12-13 11:03:37 & Palomar Schmidt \\
        038-7811-R & 18.5    & 0.5 & visible & i  & 1999-04-11 04:22:14 & Palomar Schmidt \\
        062-7353-B & 20.7    & 0.5 & visible & B  & 1997-04-08 05:39:21 & Palomar Schmidt \\
        062-7750-R & 16.0    & 0.5 & visible & I  & 1998-12-18 11:55:36 & Palomar Schmidt \\
        063-7251-B & 18.9    & 0.5 & visible & g  & 1997-03-03 07:52:45 & Palomar Schmidt \\
        063-7278-R & 18.3    & 0.5 & visible & r  & 1997-03-09 07:54:09 & Palomar Schmidt \\
        05078      & $>$18.5 & --  & no      & B  & 1955-03-19 20:19:18 & PAO Normal astrograph \\
        05522      & $>$18.5 & 0.5 & no      & B  & 1957-03-22 20:17:26 & PAO Normal astrograph \\
        10265      & --      & --  & no      & -- & 1974-02-12 23:38:10 & PAO Normal astrograph \\
        13734      & $>$17.5 & --  & no      & B  & 1983-04-09 19:46:44 & PAO Normal astrograph \\
        D0121      & 18.8    & 0.5 & visible & B  & 1949-04-25 20:08:40 & PAO Normal astrograph \\
        D0906      & $>$18.5 & --  & no      & B  & 1953-04-05 20:46:17 & PAO Normal astrograph \\
        E0685      & 18.5    & 0.5 & visible & R  & 1953-03-07 06:33:00 & Palomar Schmidt 1.2-m \\
        E0714      & 19.0    & 0.5 & visible & r  & 1953-04-12 06:40:00 & Palomar Schmidt 1.2-m \\
        O0685      & 18.8    & 0.5 & visible & B  & 1953-03-07 07:20:00 & Palomar Schmidt 1.2-m \\
        O0714      & 18.7    & 0.5 & visible & V  & 1953-04-12 06:20:00 & Palomar Schmidt 1.2-m \\
        \hline
        \end{tabular}
        \end{table*}

    \subsection{\textit{Swift}-UVOT}
    
    The \textit{Swift}/UVOT data \citep{UVOT} are analyzed using the tool \texttt{uvotmaghist}. For each observation and available filter (V, B, U, UVW1, UVM2, and UVW2), source and background regions are defined and kept fixed across all epochs to ensure consistency. The \texttt{uvotmaghist} task is run on all available images to extract photometric measurements, producing time-resolved flux densities and associated statistical uncertainties for each filter. The analysis is performed following the standard UVOT prescriptions, and no additional filtering beyond the default quality criteria is applied. The correction for Galactic extinction is performed assuming $E_{B-V} = 0.056$ \citep{Schlafly11}. The resulting flux densities are extracted from the output products and converted into physical units, which are then used to build the optical and ultraviolet light curves in Figure~\ref{fig:lc}, and to complement the MWL SED.
    
    \subsection{\textit{AstroSat}-UVIT}
    
    The UltraViolet Imaging Telescope (UVIT) onboard the \textit{AstroSat} satellite observed the source between December 2022 and April 2023 with the F1 filter (F148W, effective wavelength $\sim$148 nm) of the FUV channel. The Level-1 data ae processed with the standard {\tt UVIT L2 pipeline} (version 6.3) to produce Level-2, astrometrically calibrated images (including drift correction, flat-fielding, and geometric distortion) \citep{2021JApA...42...29G}. Visual inspection of the Level-2 image shows no significant signal above the background at the source position. Consequently, aperture photometry is not performed and the data are not used in this work. As the source is not detected in the FUV F1 band we obtain an upper limit which is consistent with the UVIT sensitivity for the given exposure and filter.
    
    \subsection{Archival optical and IR data}
    In addition to the optical/UV data from \textit{Swift}-UVOT observations of \sfive, complementary optical measurements are available from several wide-field, all-sky survey facilities, including the All-Sky Automated Survey for Supernovae \citep[ASAS-SN;][]{2014ApJ...788...48S, 2017PASP..129j4502K}, the Zwicky Transient Facility \citep[ZTF;][]{2019PASP..131a8002B}, and the Panoramic Survey Telescope and Rapid Response System \citep[Pan-STARRS1;][]{2016arXiv161205560C}. These data are accessible through the \texttt{MMDC}, and the procedures used to retrieve the observations and convert magnitudes into fluxes are described in detail by \citet{2024AJ....168..289S}.

    The long-term optical light curve of \sfive\ is shown in panel~f) of Figure~\ref{fig:lc} which combines data from ASAS-SN (V and g filters), ZTF (i, g, and r filters), and Pan-STARRS (g, r, i, z, y; due to the lower number of measurements compared to the other telescopes, the Pan-STARRS filters have been merged in the light curve). This long-term optical monitoring reveals pronounced variability, characterized by multiple episodes of enhanced activity. During quiescent states, the optical flux remains at a baseline level of $\sim(2-3)\times10^{-12}\,\mathrm{erg\,cm^{-2}\,s^{-1}}$\, while during the highest activity states it increases to values exceeding 10$^{-11}\,\mathrm{erg\,cm^{-2}\,s^{-1}}$. The maximum optical flux of $(2.47\pm0.07)\times10^{-11}\,\mathrm{erg\,cm^{-2}\,s^{-1}}$ is observed in the ASAS-SN $g$ band on MJD~60028.5, while a peak flux of $(1.65\pm0.02)\times10^{-11}\,\mathrm{erg\,cm^{-2}\,s^{-1}}$ is recorded in the ZTF $r$ band on MJD~59895.4. 

    The comparison of the optical light curve (Figure~\ref{fig:lc}, panel~f) with the \gray\ light curves (Figure~\ref{fig:lc}, panels~a and~b) shows that several major optical brightening episodes occur contemporaneously with enhanced \gray\ activity. In particular, the optical brightening observed between $\mathrm{MJD}\:\approx 57000$ and $58000$ coincides with \gray\ flaring episode, while additional optical active states are observed between $\mathrm{MJD}\:\approx 59500$ and $60000$, coincident with \gray\ activity. However, this behavior is not uniform across all activity phases: despite the \gray\ flux reaching a high state during $\mathrm{MJD}\:\approx 58500$ to $59000$ the optical flux increases only about half of the other active periods. In contrast, during the last \gray\ activity period, which is characterized by a lower \gray\ flux, the optical flux increases much more significantly. 

    In addition to the optical data, panel~g) of Figure~\ref{fig:lc} presents the infrared (IR) light curve of \sfive\ obtained from observations with the Near-Earth Object Wide-field Infrared Survey Explorer \citep[NEOWISE;][]{2011ApJ...731...53M}. The data in the 3.4~$\mu m$ ~$\mu m$ and 4.6~$\mu m$ bands are retrieved from \texttt{MMDC} \citep{2024AJ....168..289S}. As NEOWISE observes a given sky position for only a few days approximately every six months, the individual observations separated by less than 10 days are stacked in order to improve the signal-to-noise ratio. Although the IR light curve does not provide continuous monitoring comparable to the \gray\ and optical light curves, it nevertheless shows that the IR flux increases during the \gray\ flaring periods. In the baseline state, the IR flux is at a level of $\sim 2\times10^{-12}\,\mathrm{erg\,cm^{-2}\,s^{-1}}$, while during the \gray\ brightening periods it increases up to $\sim 10^{-11}\,\mathrm{erg\,cm^{-2}\,s^{-1}}$.

    \subsection{Historical optical observations: Palomar and Pulkovo}

    We analyse archival photographic material from the Palomar and Pulkovo observatories covering the period 1949--1999 provided by the Naroo digitization program \citep{Robert}. The dataset consists of scanned photographic plates in which the target \sfive\ is present within the field of view. A total of 25 images are available; however, only 15 frames were deemed suitable for photometric analysis. We used  Astrometry.net \citep{2010AJ....139.1782L} to estimate the astrometric solution of the frames. Photometric measurements are obtained using the \texttt{STDweb}\footnote{\href{http://stdweb.favor2.info}{http://stdweb.favor2.info/}} tool \citep{2025AcPol..65...50K}, which estimates source magnitudes by comparing the measured fluxes of objects in the field with reference sources from photometric catalogues. We estimate the photometric calibration against the Pan-STARRS objects \citep{2016arXiv161205560C} and used the conversion relations implemented in the \texttt{STDweb}\footnote{\href{github.com/karpov-sv/stdpipe/blob/master/stdpipe/catalogs.py}{github.com/karpov-sv/stdpipe/blob/master/stdpipe/catalogs.py}} code to go from the Pan-STARRS photometric system (where the photometric bands are denoted with $g$, $r$, $i$ and $z$) to the Johnson-Cousins one (where the bands are denoted $B$, $V$, $R$, and $I$). To estimate the most appropriate band for the calibration of the images, we used the color-term diagnostic and took the band that had the minimum color-term. The final photometric measurements are reported in Table~\ref{tab:archival_photometry}. For the image \emph{10265} in Table~\ref{tab:archival_photometry}, we do no find a proper photometric calibration, preventing to have an estimation of \sfive\ magnitude, however, a visual inspection of the frame at the AGN's position reveal no visible source. The same visual inspection reveals no significant excess in the images \emph{05078}, \emph{05522}, \emph{13734} and \emph{D0906}, but a photometric calibration is found, allowing for an estimation of upper-limits for \sfive's luminosity.  The latter are estimated by using the background noise inside the aperture multiplied by five times the signal to noise ratio.

\section{Variability studies}

    \subsection{Spectral variability}
    \label{sec:var}
    
\begin{figure}[ht!]
    \centering
   \includegraphics[width=0.495\textwidth]{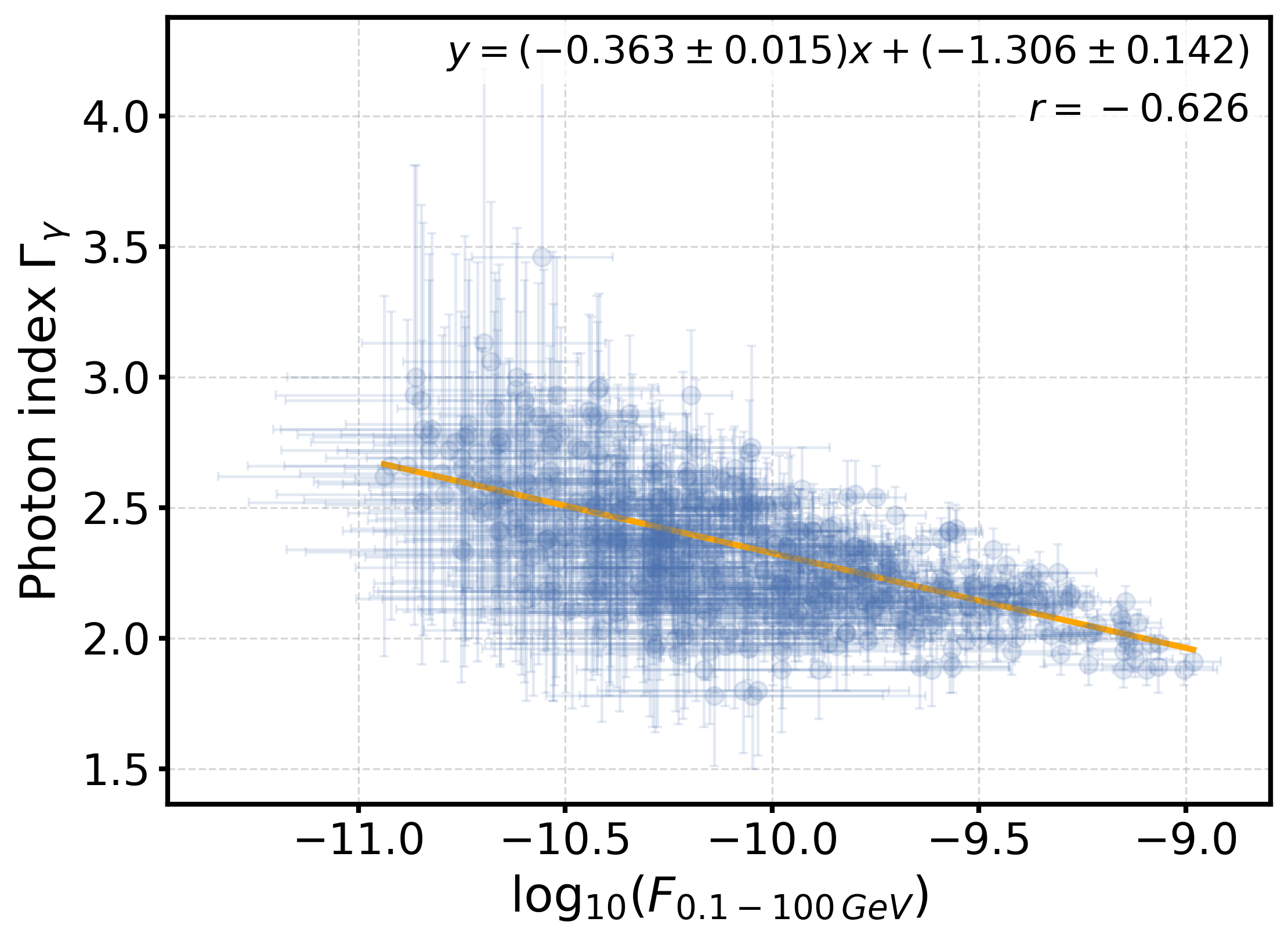}
    \includegraphics[width=0.5\textwidth]{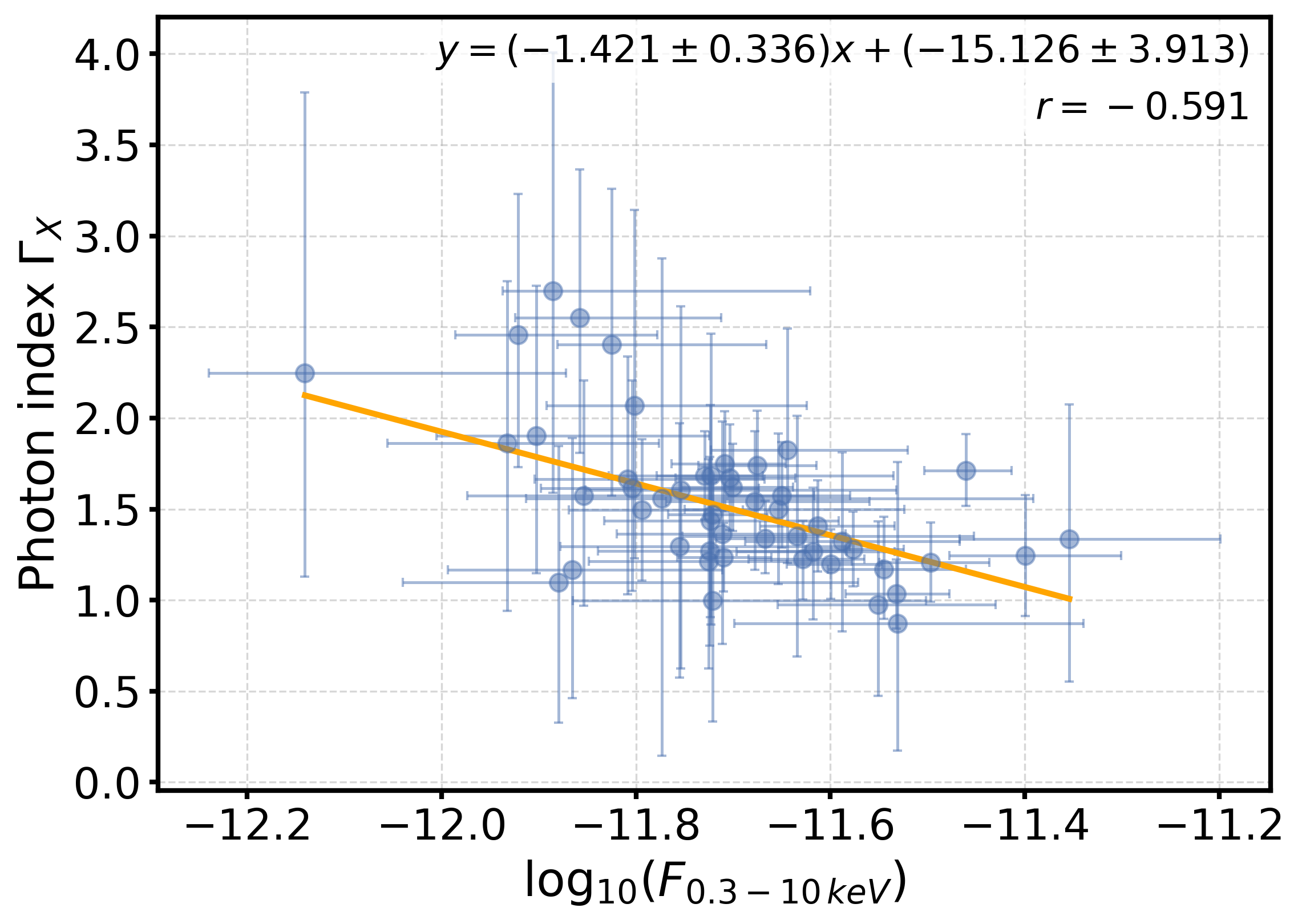}
    \includegraphics[width=0.5\textwidth]{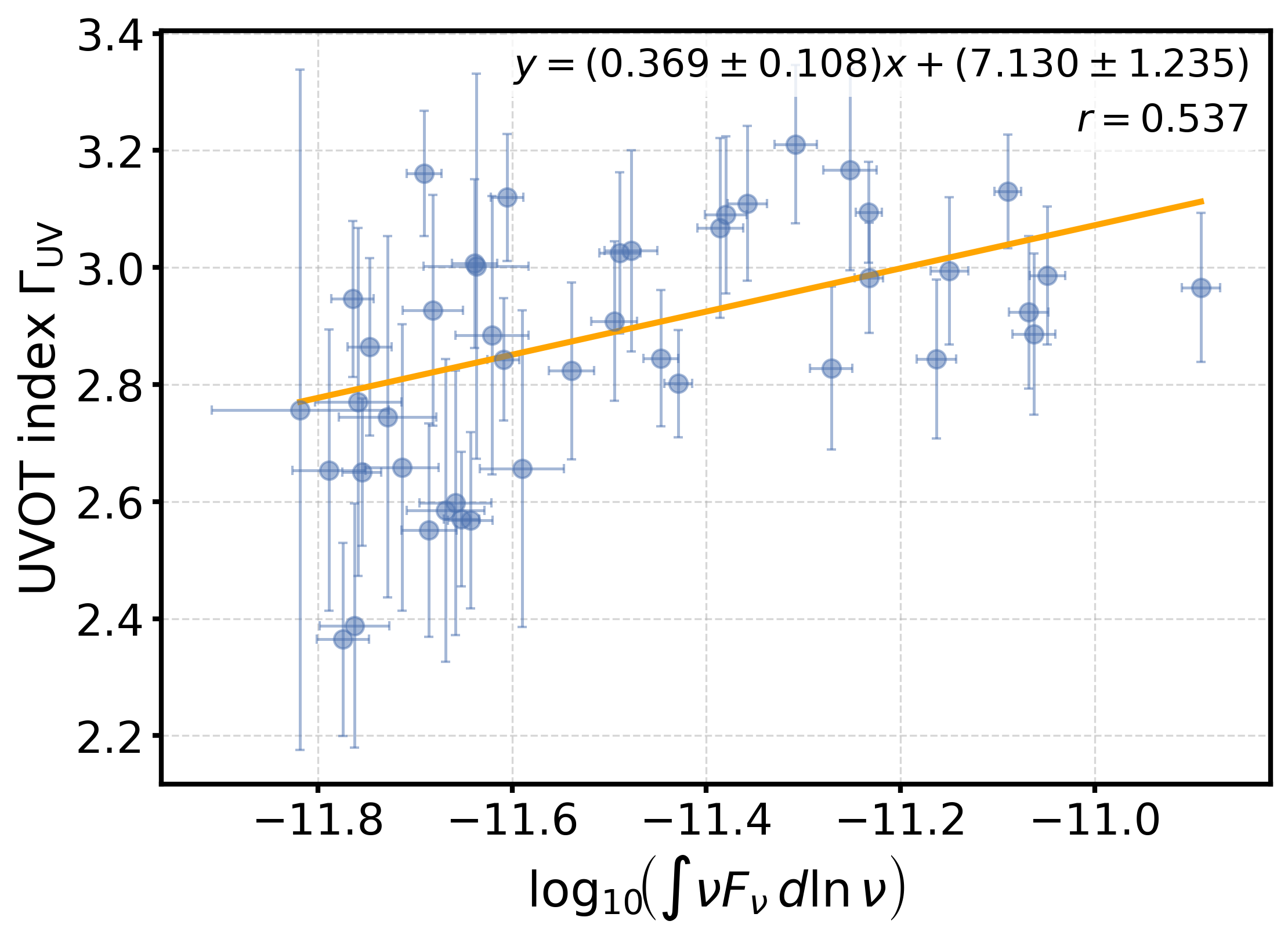}

    \caption{Spectral variability in \sfive. From top to bottom: $\gamma$-ray photon index vs the logarithm of the integral photon flux (100 MeV to 100 GeV); X-ray photon index vs the logarithm of the integral photon flux (0.3 keV to 10 keV); optical/UV photon index vs the logarithm of the integral photon flux (from V to W2 UVOT filter).    \label{fig:LAT_spectral_var}}
\end{figure}

    In addition to flux variability, the long-term monitoring of S5~1044+71 enables us to investigate spectral changes within individual bands, and to test whether the source exhibits systematic \textit{harder-when-brighter} or \textit{softer-when-brighter} trends. For the \textit{Fermi}-LAT and the X-ray band, we extract directly the best-fit photon indexes ($\Gamma_{\gamma}$ and $\Gamma_{\rm X}$), and the integral fluxes ($F_{\gamma}$ and $F_{\rm X}$). For UV/optical we perform a fit of the UVOT flux points only for the observations that contain all available filters, obtaining $\Gamma_{\rm{UVOT}}$, and the integral of the power-law over the UVOT spectral band, $F_{\rm{UV}}$. 
    For each band, we quantify spectral variability by fitting the photon index $\Gamma$ as a function of the logarithm of the flux $F$:
    \begin{equation}
    \Gamma = \beta\,\log_{10}(F) + \varepsilon.
    \end{equation}
     When computing the fit, we only use measurements with positive flux and well-defined logarithmic errors, requiring in particular $(F-\sigma_F)>0$. Pearson correlation coefficients are also reported in Figure~\ref{fig:LAT_spectral_var}.
    
    The LAT band shows a clear spectral trend with flux: $\Gamma_{\gamma}$ decreases as the $\gamma$-ray flux increases, as visible in the top panel of Figure~\ref{fig:LAT_spectral_var}, which corresponds to a harder spectrum during bright states. Fitting $\Gamma_{\gamma}$ as a function of $\log_{10}(F_\gamma)$ yields a negative slope with $\beta_{\gamma} = -0.363 \pm 0.015$, with Pearson $r=-0.626$. In the X-ray band, we also find evidence for spectral variability correlated with flux as $\Gamma_{\rm X}$ decreases for increasing X-ray flux, with $\beta_{X} = -1.42 \pm 0.34$, with Pearson $r=-0.591$. This also indicates a \textit{harder-when--brighter} trend, although the scatter is larger than in the LAT band.
    
    In contrast, the UV/optical proxy displays the opposite tendency as visible in the bottom panel of Figure~\ref{fig:LAT_spectral_var}. The UV/optical index increases with increasing UV/optical flux, corresponding to a softer spectrum in brighter optical/UV states: $\beta_{\rm{UVOT}} = 0.37 \pm 0.11$ with Pearson $r=0.537$. This behavior is naturally expected if the observed UV flux includes multiple components with distinct variability properties, for instance a relatively stable thermal contribution from the accretion disk and a variable synchrotron component from the jet.\\

    \subsection{Multi-wavelength correlations}

To characterize the MWL behavior of the source, we build correlation diagrams between the \fermi\ light curve and each MWL band. For the optical filters shared between multiple instruments (i.e. g and V), we merged them into a single light curve.  We first worked with integral fluxes in X-ray ($F_{X}[0.3-10~\rm{keV}]$) and \gray\ ($F_{\gamma}[0.1-100~\rm{GeV}]$). For optical/infrared data, that are measured from a narrow filter, we use their $\nu F_{\nu, band}$ value. For each MWL band, we associated every \fermi\ bin with the closest quasi-simultaneous measurement within a maximum time lag of 3.5 days (i.e. half of the width of a \gray\ bin). Given that the same \fermi\ bin might then be associated with multiple MWL flux points, we explicitly prevented this double counting by creating in these cases an average MWL flux value.
    
\begin{figure*}[ht!]
    \centering
    \includegraphics[width=0.32\textwidth]{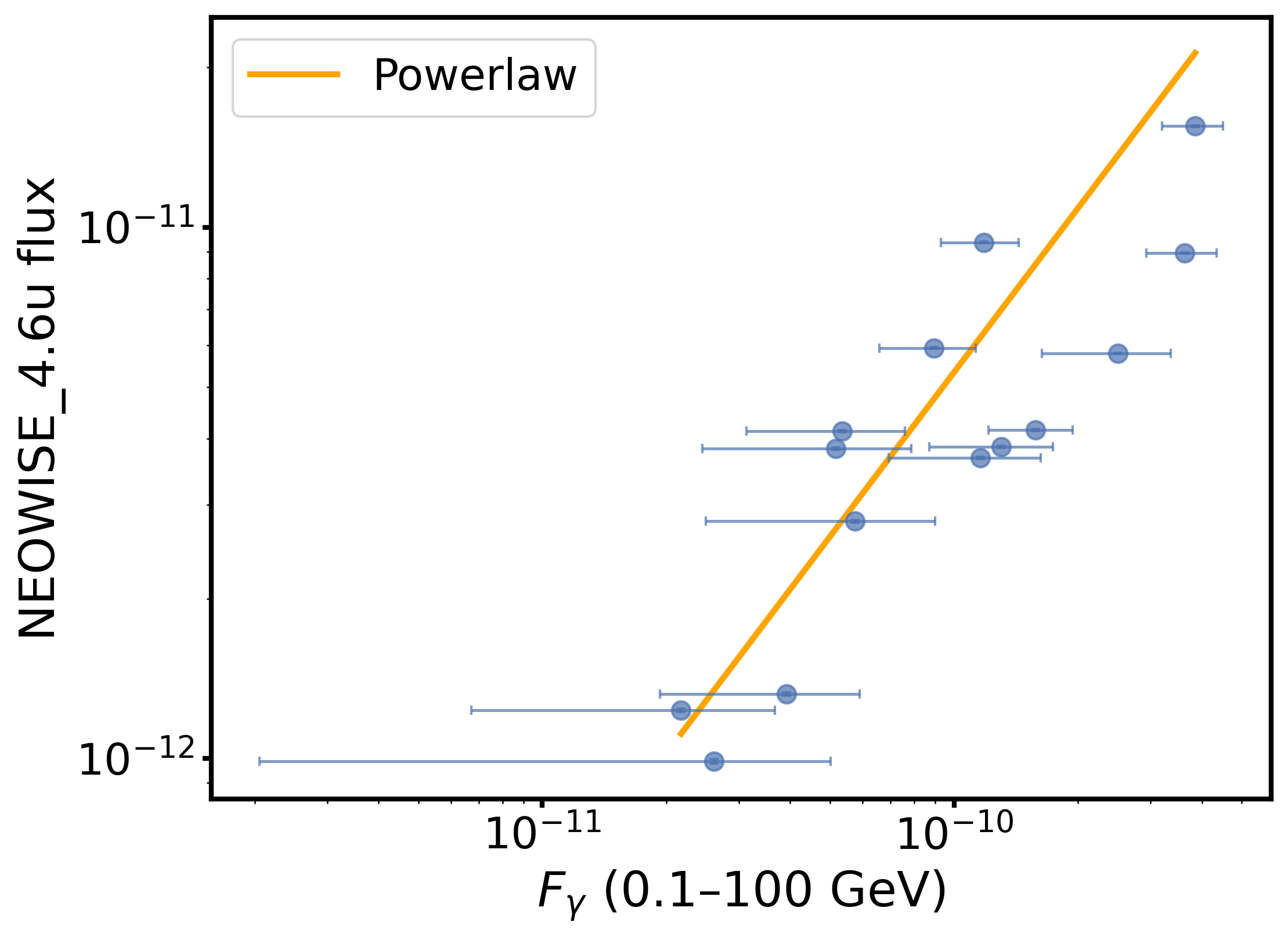}
    \includegraphics[width=0.32\textwidth]{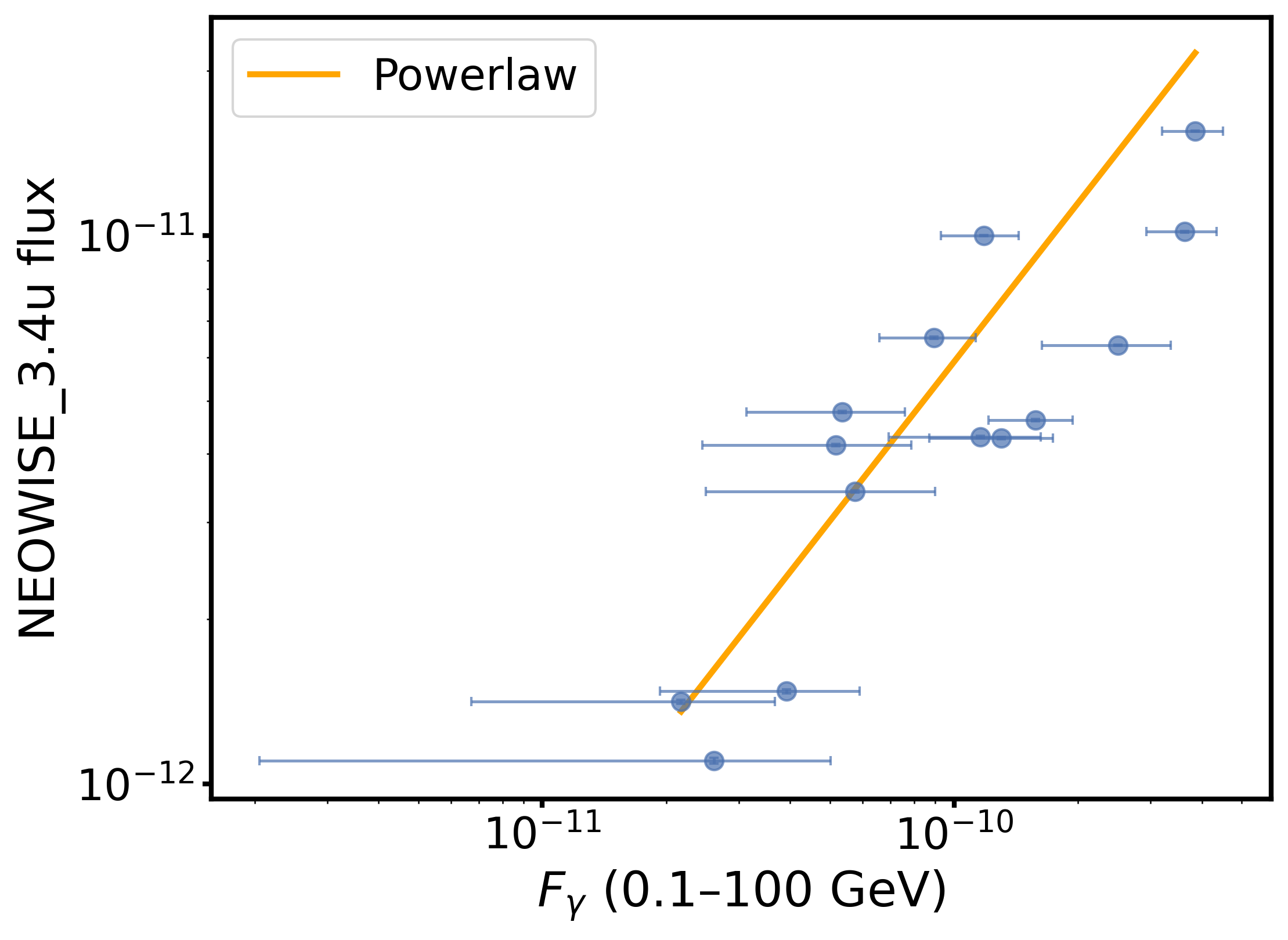}
    \includegraphics[width=0.32\textwidth]{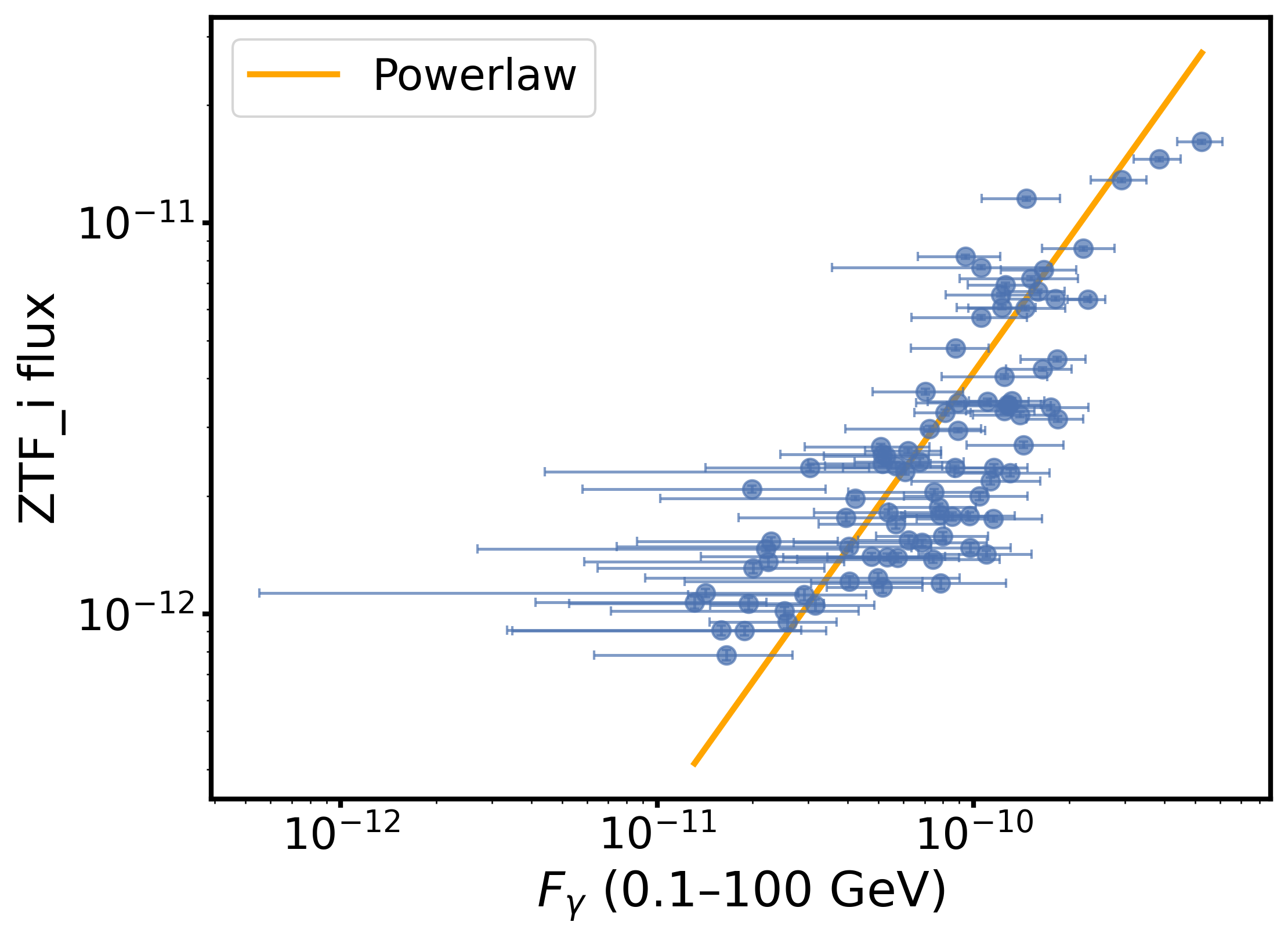}
    \includegraphics[width=0.32\textwidth]{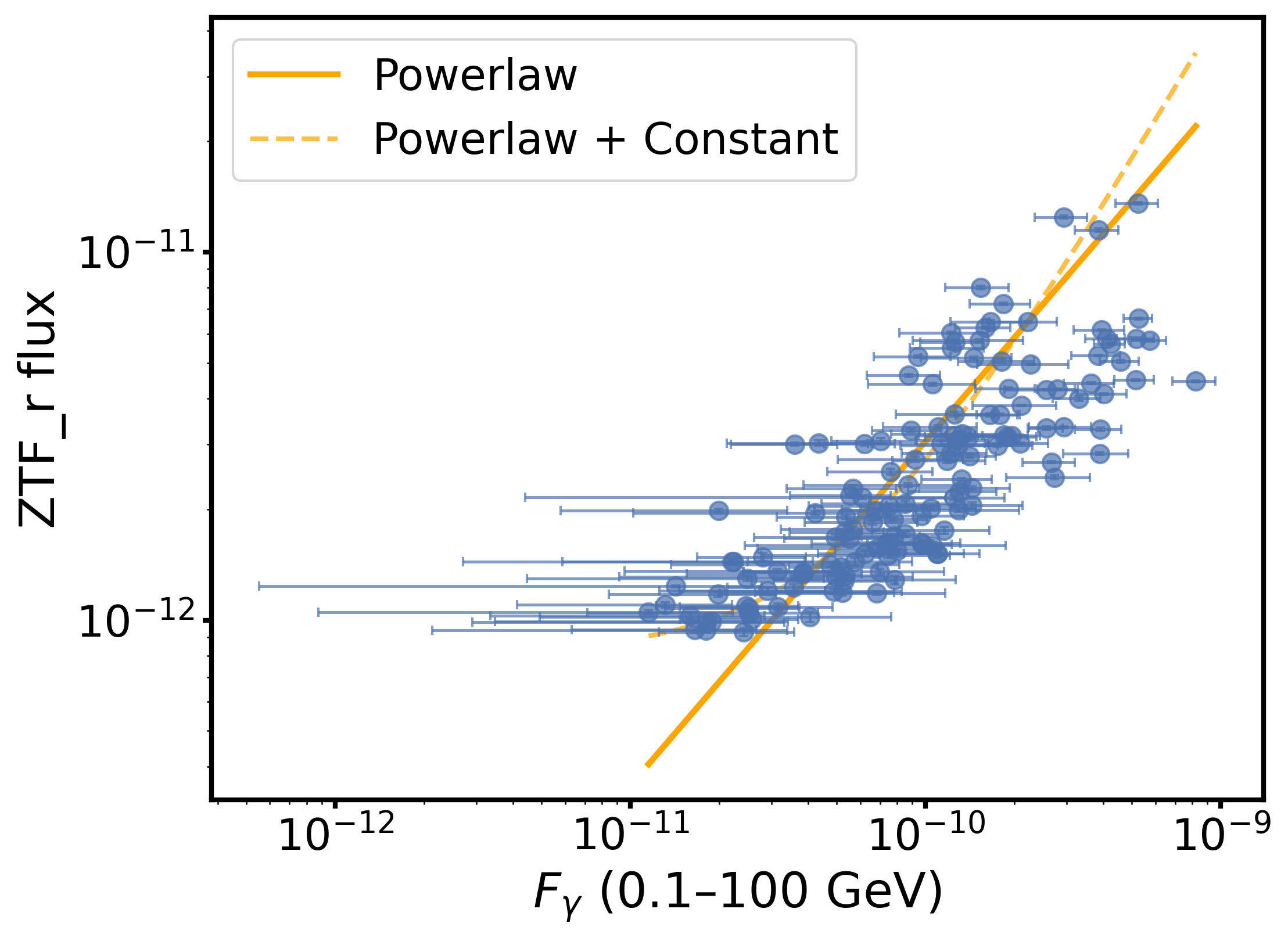}
    \includegraphics[width=0.32\textwidth]{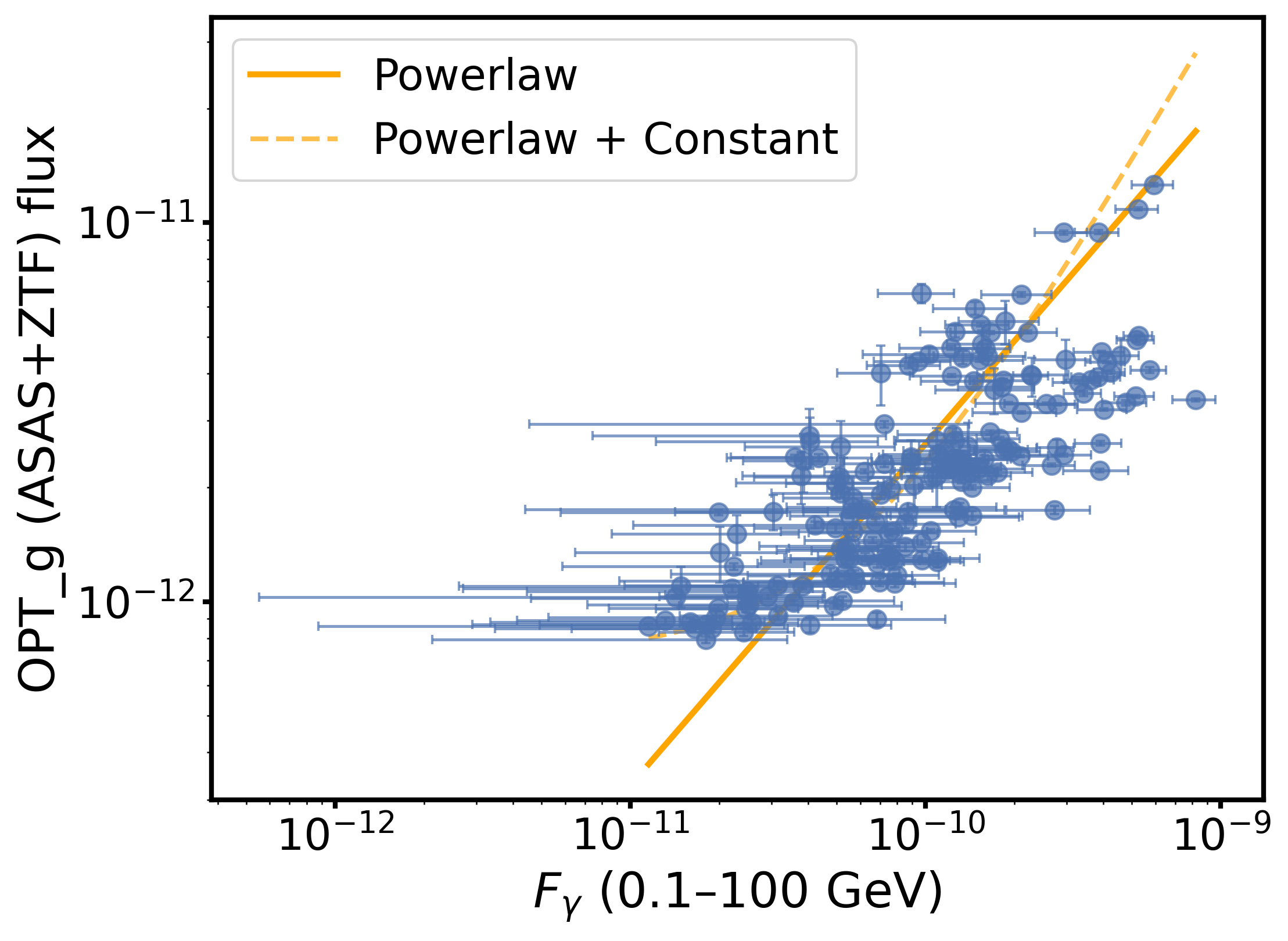}
    \includegraphics[width=0.32\textwidth]{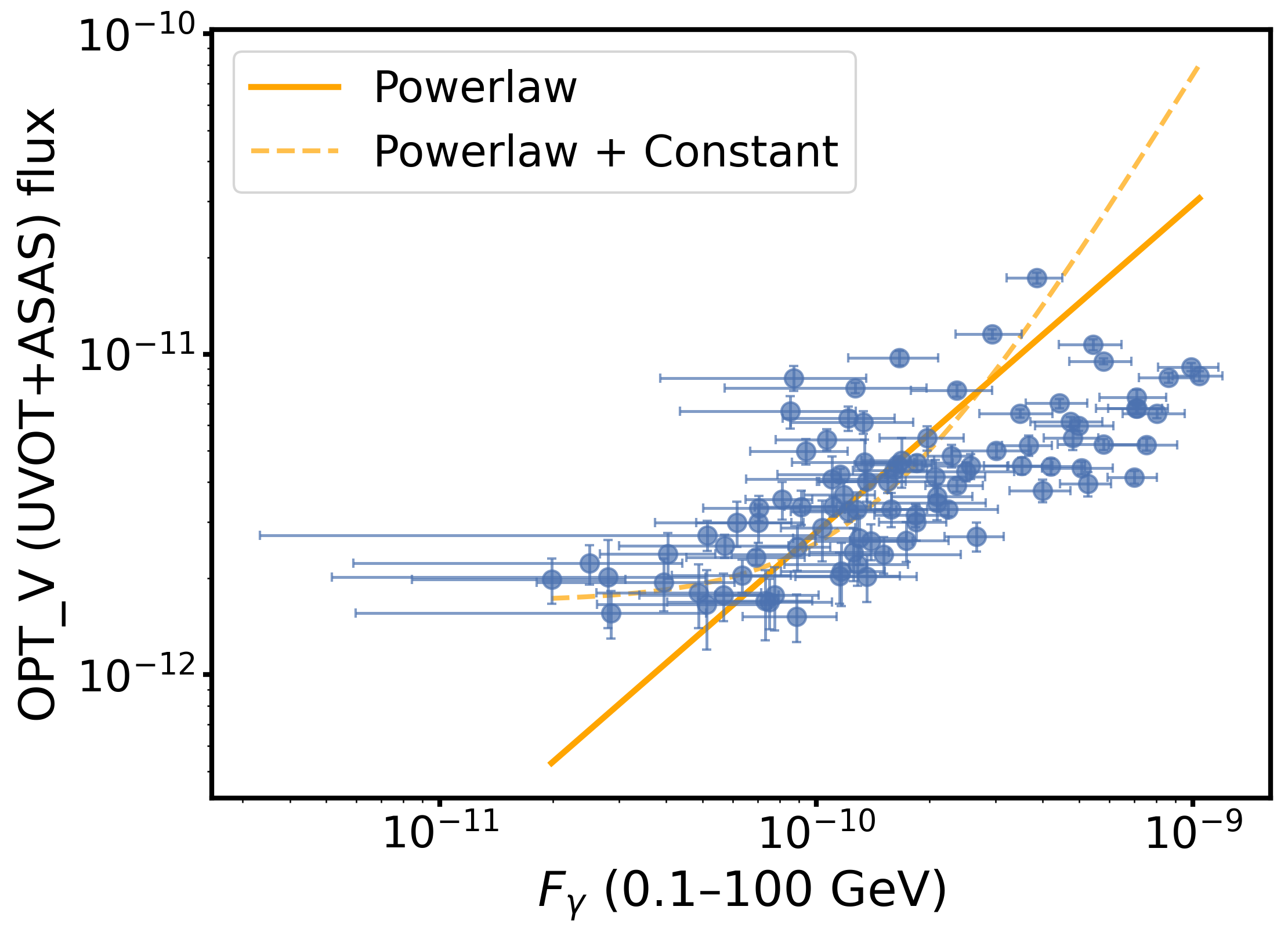}
    \includegraphics[width=0.32\textwidth]{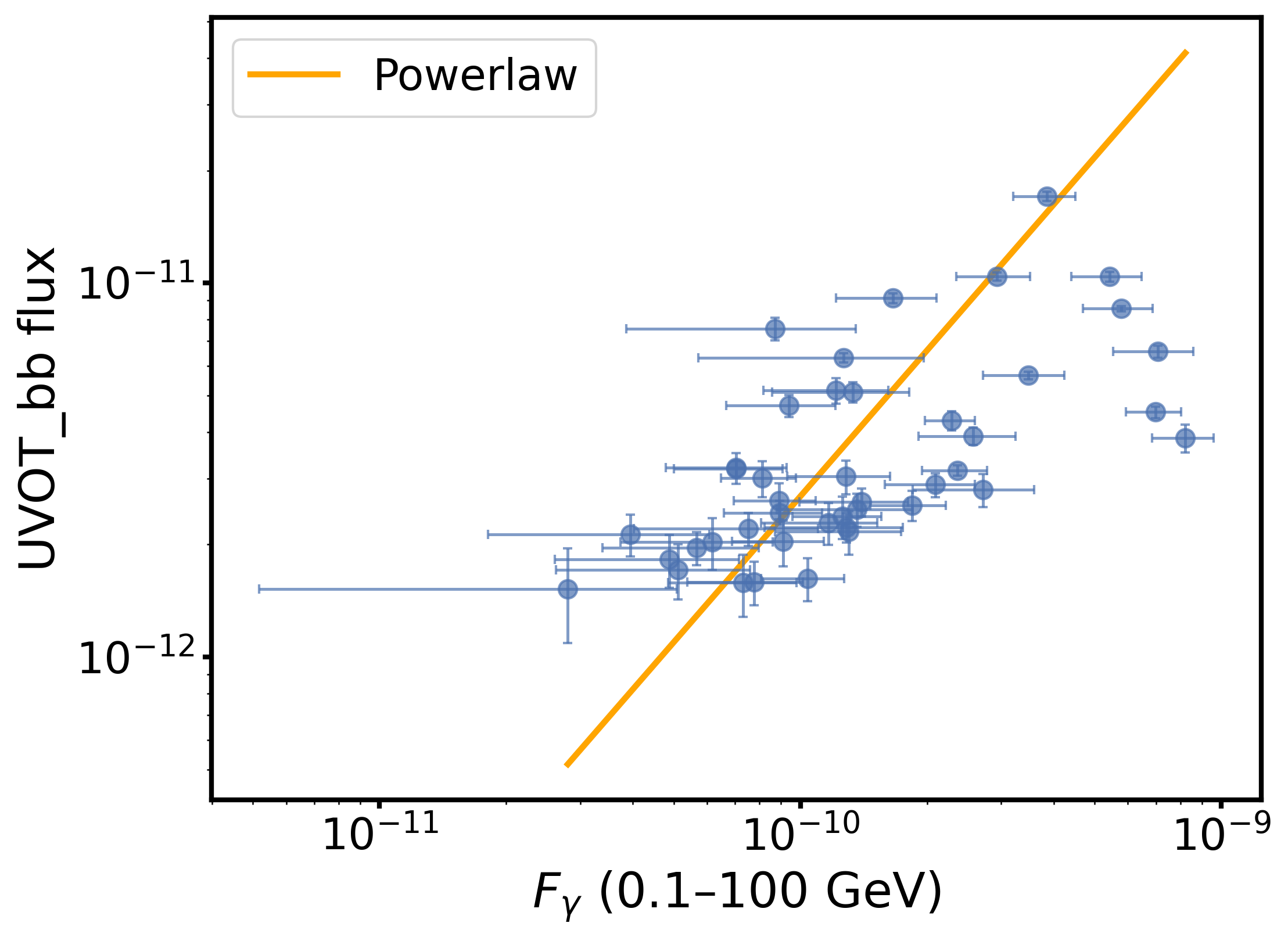}
    \includegraphics[width=0.32\textwidth]{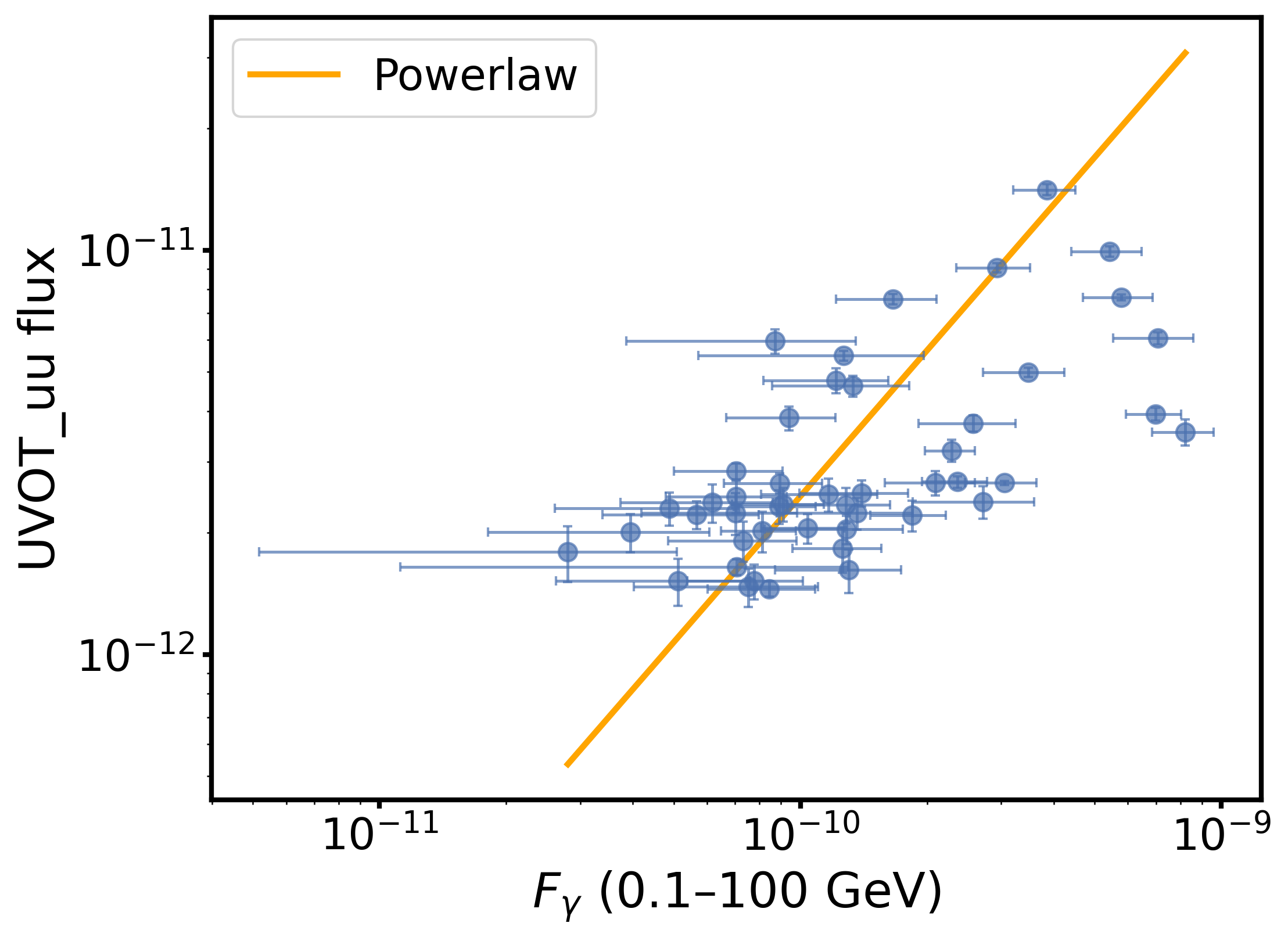}
    \includegraphics[width=0.32\textwidth]{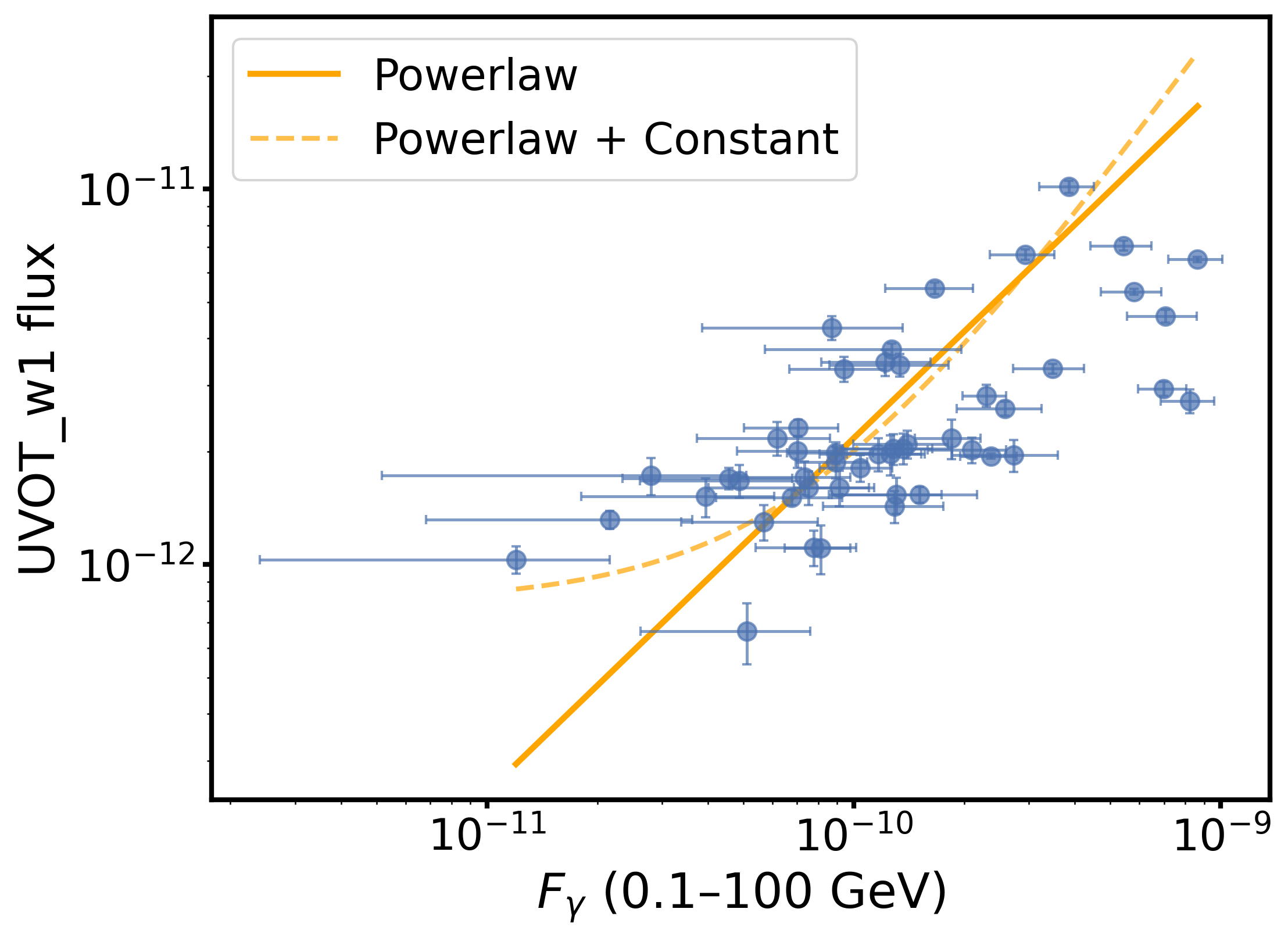}
    \includegraphics[width=0.32\textwidth]{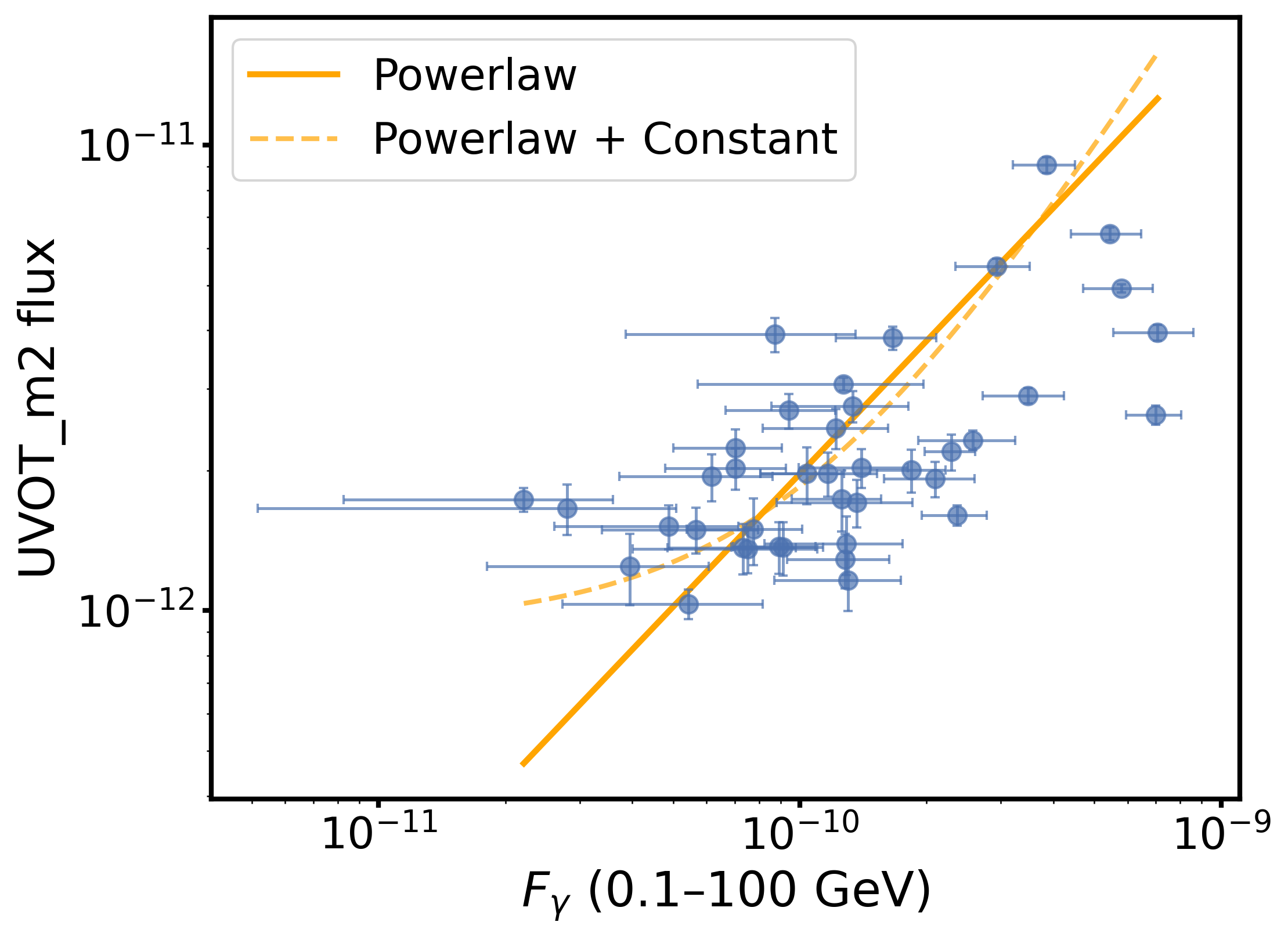}
    \includegraphics[width=0.32\textwidth]{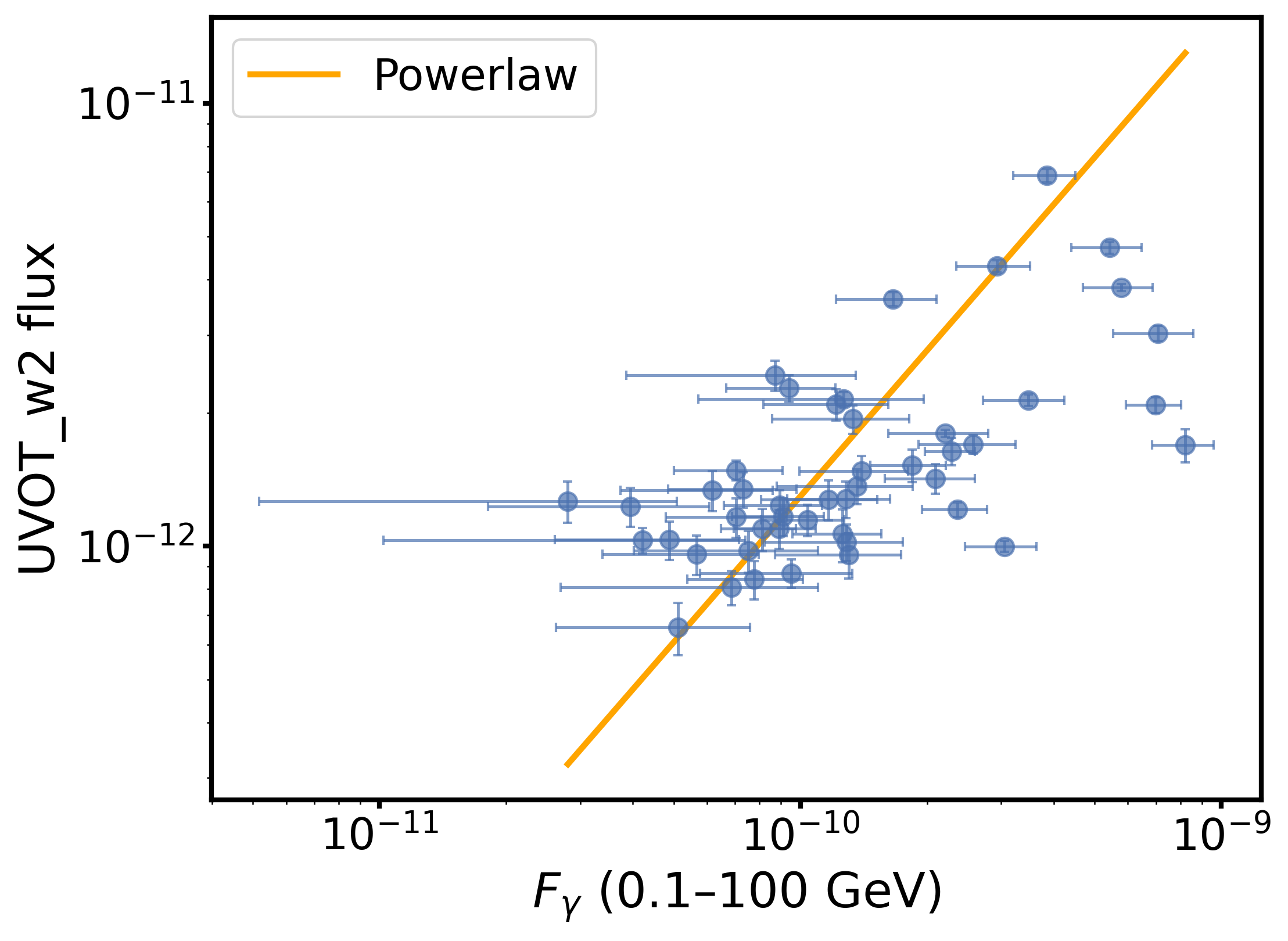}
     \includegraphics[width=0.32\textwidth]{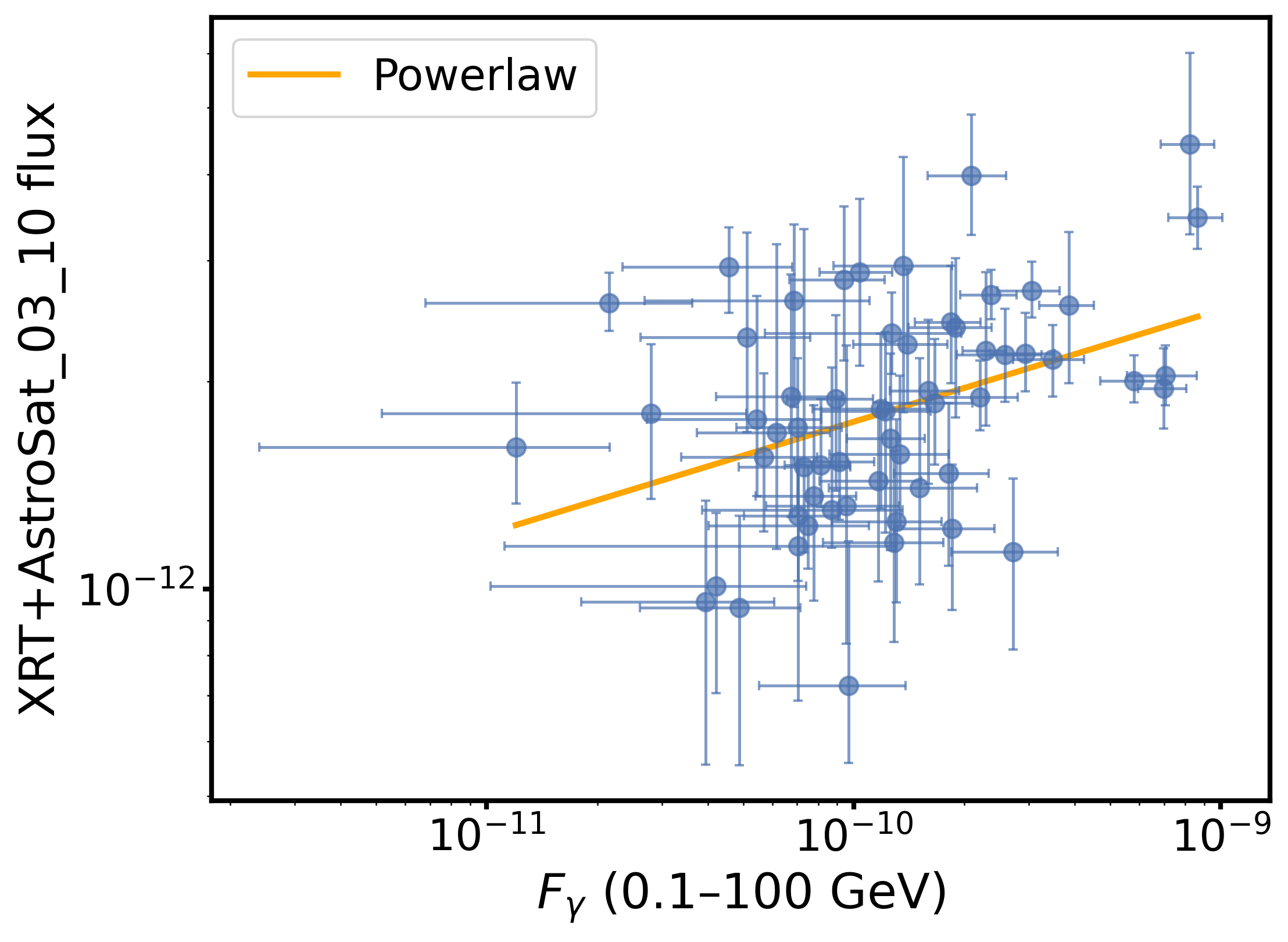}
    \caption{MWL correlation plots with respect to the integral \gray\ photon flux between 100 MeV and 100 GeV. From top to bottom and left to right, the quantities on the y-axis are: NEOWISE flux at $4.6\mu$m; NEOWISE flux at $3.4\ \mu$m; ZTF i filter flux; ZTF r filter flux; ASAS-SN and ZTF g filter flux; ASAS-SN and UVOT V filter flux; UVOT b filter flux; UVOT u filter flux; UVOT w1 filter flux; UVOT M2 filter flux; UVOT W2 filter flux; XRT and \textit{AstroSat}-SXT integral photon flux between 0.3 and 10 keV.}
    \label{fig:corr}
\vspace{0.5cm}
\resizebox{\textwidth}{!}{%
\begin{tabular}{lccccccccccc}
\hline
Band & $r_{\rm P}$ & $p_{\rm P}$ & $r_{\rm S}$ & $p_{\rm S}$ & $a_{\rm PL}$ & $b_{\rm PL}$ & $a_{\rm add}$ & $b_{\rm add}$ & $C_{\rm add}$ & $\Delta\chi^2$ & $p(\Delta\chi^2)$ \\
\hline
NEOWISE\_4.6u & 0.83 & $2.15\times10^{-4}$ & 0.83 & $2.51\times10^{-4}$ & $1.06\pm0.23$ & $-0.83\pm2.20$ & -- & -- & -- & -- & -- \\
NEOWISE\_3.4u & 0.84 & $1.47\times10^{-4}$ & 0.83 & $2.51\times10^{-4}$ & $0.99\pm0.21$ & $-1.44\pm2.01$ & -- & -- & -- & -- & -- \\
ZTF\_i & 0.85 & $8.72\times10^{-26}$ & 0.81 & $1.48\times10^{-21}$ & $1.25\pm0.08$ & $1.03\pm0.84$ & -- & -- & -- & -- & -- \\
ZTF\_r & 0.65 & $1.85\times10^{-18}$ & 0.85 & $3.25\times10^{-40}$ & $0.89\pm0.07$ & $-2.75\pm0.64$ & $1.36\pm0.19$ & $(0.77\,\pm\,3.25)\times10^{2}$ & $(8.03\,\pm\,1.51)\times10^{-13}$ & 62.1 & $3.27\times10^{-15}$ \\
V (UVOT+ASAS) & 0.56 & $3.12\times10^{-9}$ & 0.71 & $8.68\times10^{-16}$ & $0.78\pm0.08$ & $-3.83\pm0.78$ & $1.91\pm0.45$ & $(0.12\,\times\pm\,1.21)\times10^{8}$ & $(1.69\,\pm\,0.30)\times10^{-13}$ & 98.5 & $3.21\times10^{-23}$ \\
g (ASAS+ZTF) & 0.64 & $1.12\times10^{-21}$ & 0.79 & $4.66\times10^{-39}$ & $0.90\pm0.06$ & $-2.78\pm0.63$ & $1.34\pm0.17$ & $(0.40\,\pm\,1.53)\times10^{2}$ & $(7.11\,\pm\,1.17)\times10^{-13}$ & 97.1 & $6.50\times10^{-23}$ \\
UVOT\_bb & 0.48 & $1.17\times10^{-3}$ & 0.68 & $5.17\times10^{-7}$ & $1.10\pm0.19$ & $-0.78\pm1.84$ & -- & -- & -- & -- & -- \\
UVOT\_uu & 0.51 & $2.60\times10^{-4}$ & 0.65 & $7.24\times10^{-7}$ & $1.07\pm0.18$ & $-1.08\pm1.75$ & -- & -- & -- & -- & -- \\
UVOT\_w1 & 0.58 & $1.09\times10^{-5}$ & 0.69 & $4.45\times10^{-8}$ & $0.89\pm0.13$ & $-2.91\pm1.25$ & $1.35\pm0.31$ & $(0.38\,\pm\,2.62)\times10^{2}$ & $(7.91\,\pm\,2.75)\times10^{-13}$ & 41.0 & $1.53\times10^{-10}$ \\
UVOT\_m2 & 0.60 & $4.23\times10^{-5}$ & 0.61 & $3.46\times10^{-5}$ & $0.99\pm0.17$ & $-2.03\pm1.64$ & $1.42\pm0.48$ & $(0.15\,\pm\,1.59)\times10^{3}$ & $(9.28\,\pm\,4.36)\times10^{-13}$ & 20.2 & $7.01\times10^{-6}$ \\
UVOT\_w2 & 0.52 & $1.69\times10^{-4}$ & 0.61 & $5.49\times10^{-6}$ & $1.02\pm0.18$ & $-1.86\pm1.71$ & -- & -- & -- & -- & -- \\
XRT+SXT & 0.48 & $1.14\times10^{-4}$ & 0.39 & $2.53\times10^{-3}$ & $0.21\pm0.05$ & $-9.64\pm0.46$ & -- & -- & -- & -- & -- \\
\hline
\end{tabular}}
\captionof{table}{Results from the MWL correlations with respect to the integral \gray\ photon flux between 100 MeV and 100 GeV. We provide the Pearson coefficient $r_{\rm P}$ with its probability $p_{\rm P}$, the Spearman coefficient $r_{\rm S}$ with its probability $p_{\rm S}$, the best-fit values for the power-law fit, and the best-fit values for the fit of a power-law with a constant (the latter only when it is favored by $p<10^{-3}$). \label{table:corr}}
\end{figure*}

    We then fit the correlations with a power-law function:
    \begin{equation}
        \nu F_{\nu, band}\ \rm{or}\  F_{X}[0.3-10 \rm{keV}] = b_{PL} \times (F_{\gamma}[0.1-100 \rm{GeV}])^{a_{PL}}
    \end{equation}
    The results of the fit are shown in Figure \ref{fig:corr} and Table \ref{table:corr}. In some of the plots (i.e. r, g), there is evidence for a deviation from this simple power-law. We test for this behavior by adding a constant to the power-law: 
    \begin{equation}
    \nu F_{\nu, band}\ \rm{or}\  F_{X}[0.3-10 \rm{keV}] = C_{add} + b_{add}  \times (F_{\gamma}[0.1-100 \rm{GeV}])^{a_{add}}
    \end{equation}
    This additional constant can represent a quasi-steady component (e.g. host galaxy, the accretion disk, or emission from the extended jet) on top of the variable jet contribution. In Figure \ref{fig:corr} we show this alternative fit only when it improves significantly over the simpler powerlaw fit. \\

\begin{figure*}
    \centering
    \includegraphics[width=0.32\textwidth]{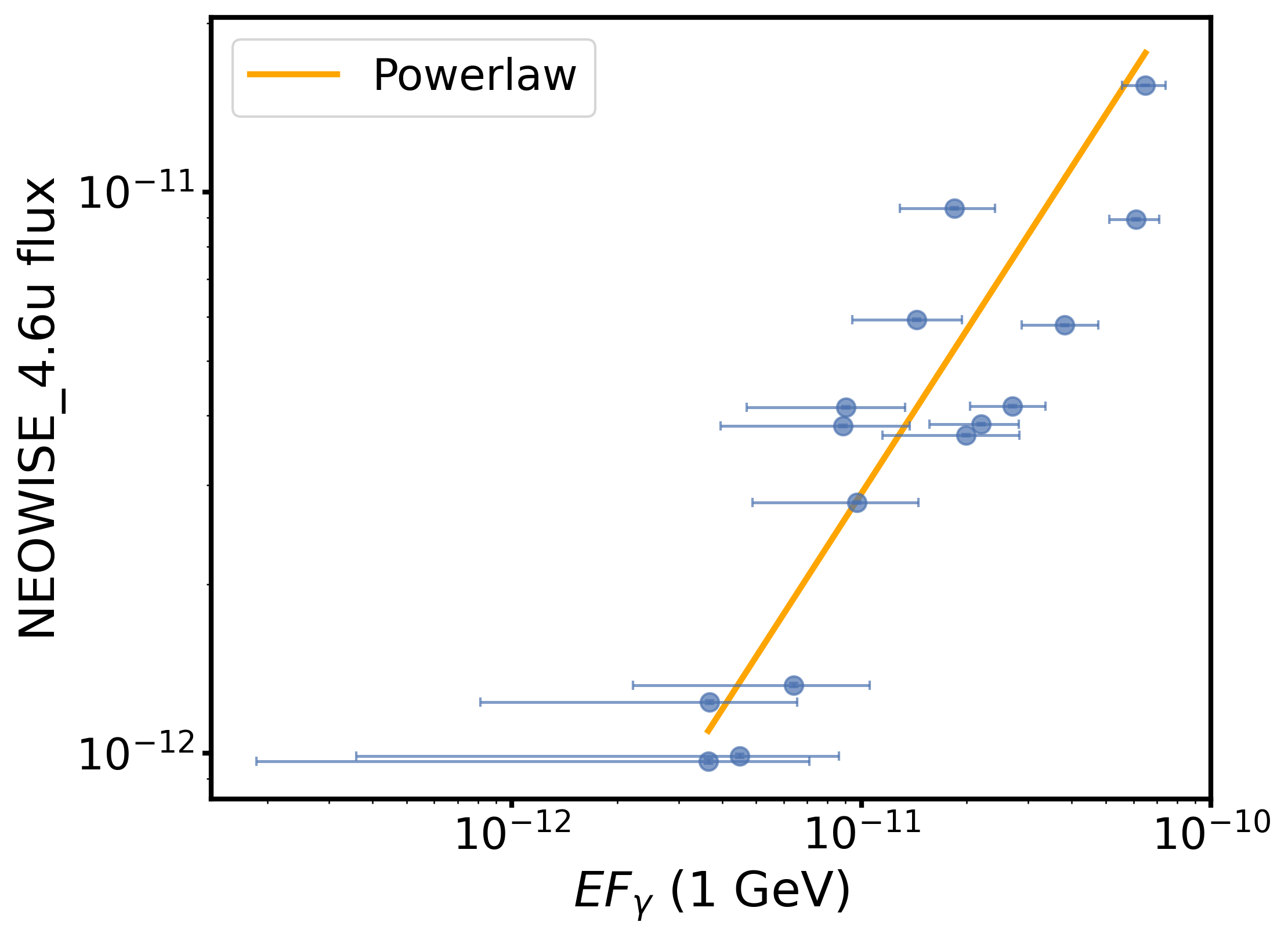}
    \includegraphics[width=0.32\textwidth]{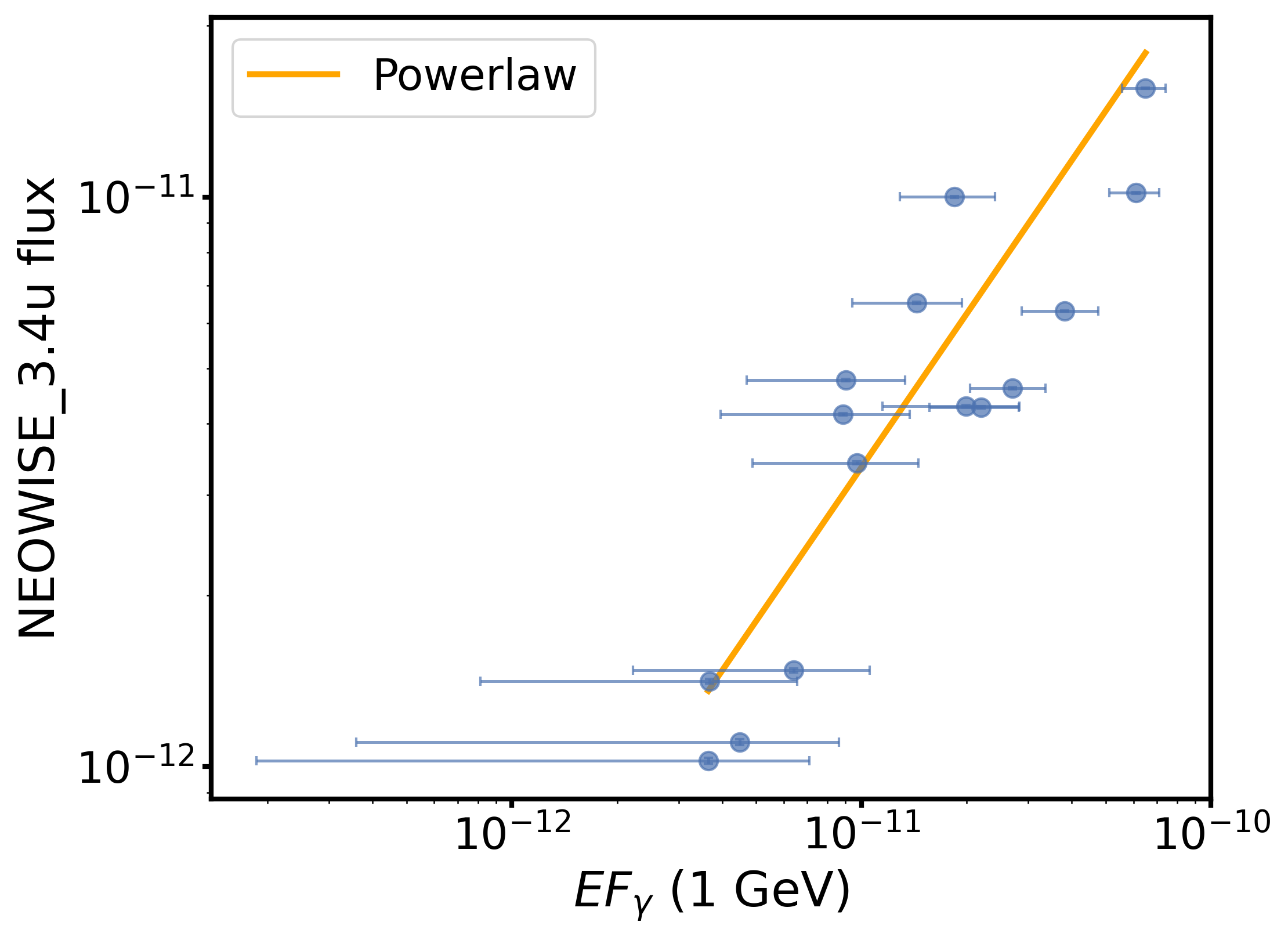}
    \includegraphics[width=0.32\textwidth]{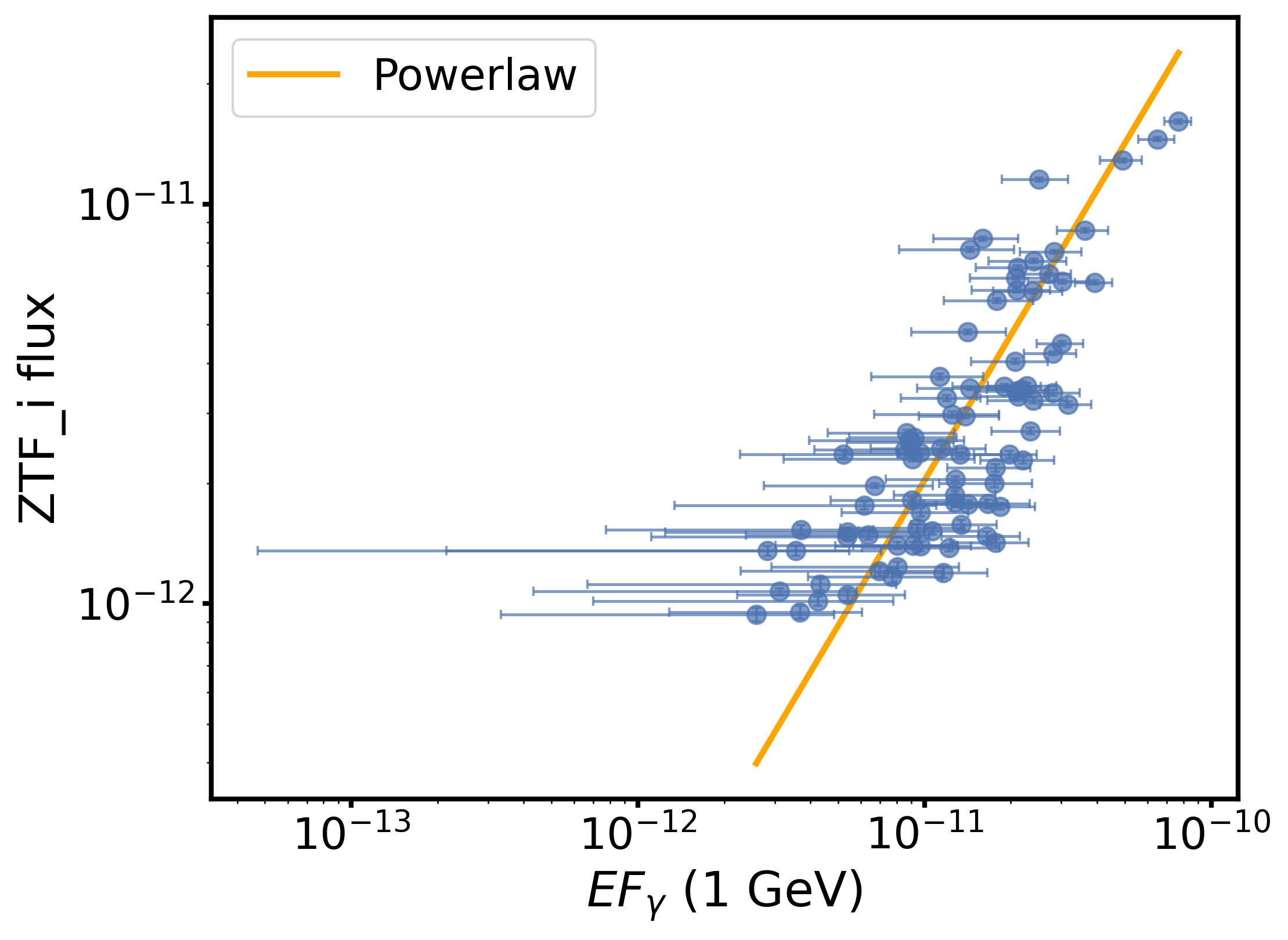}
    \includegraphics[width=0.32\textwidth]{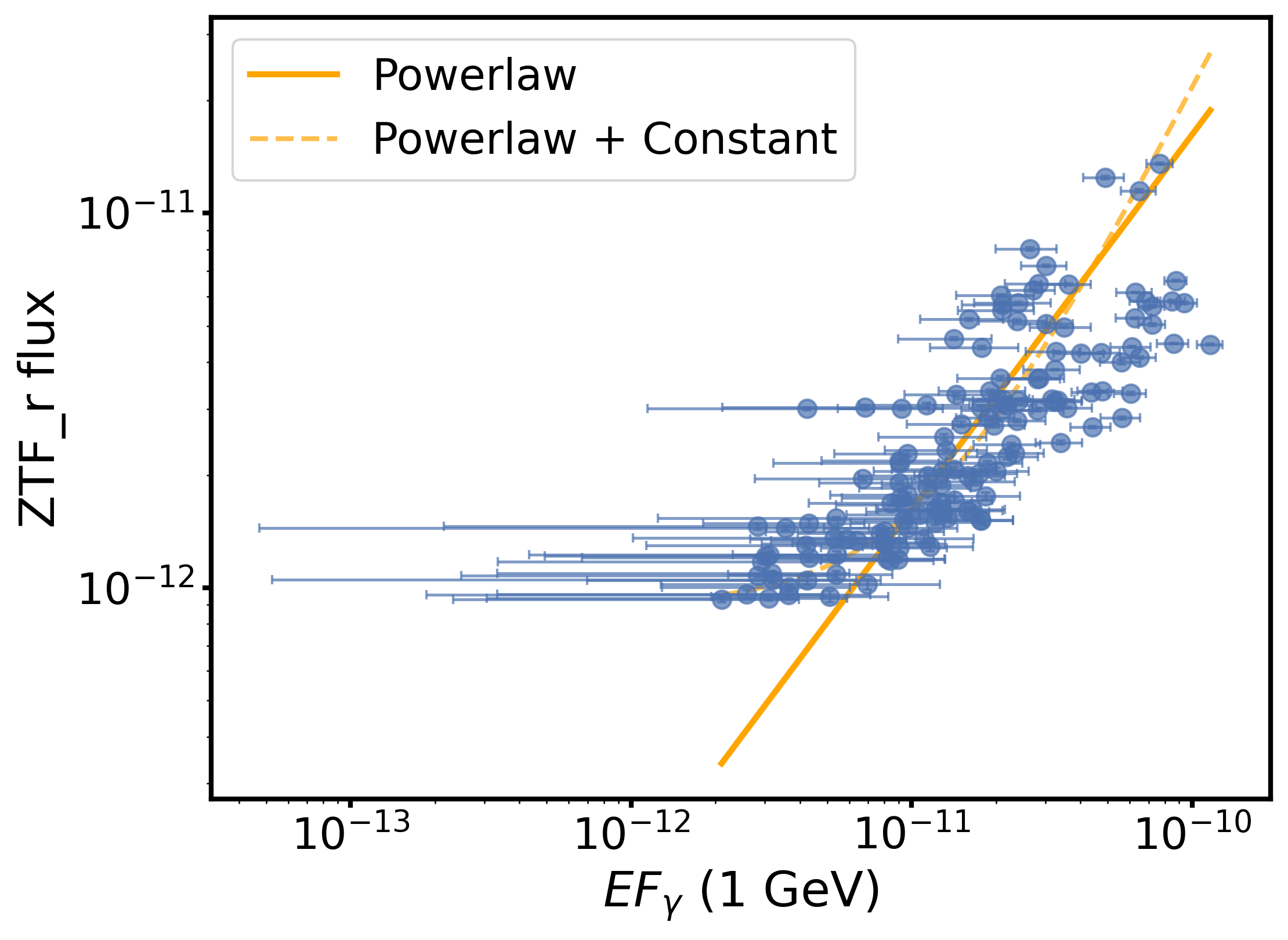}
    \includegraphics[width=0.32\textwidth]{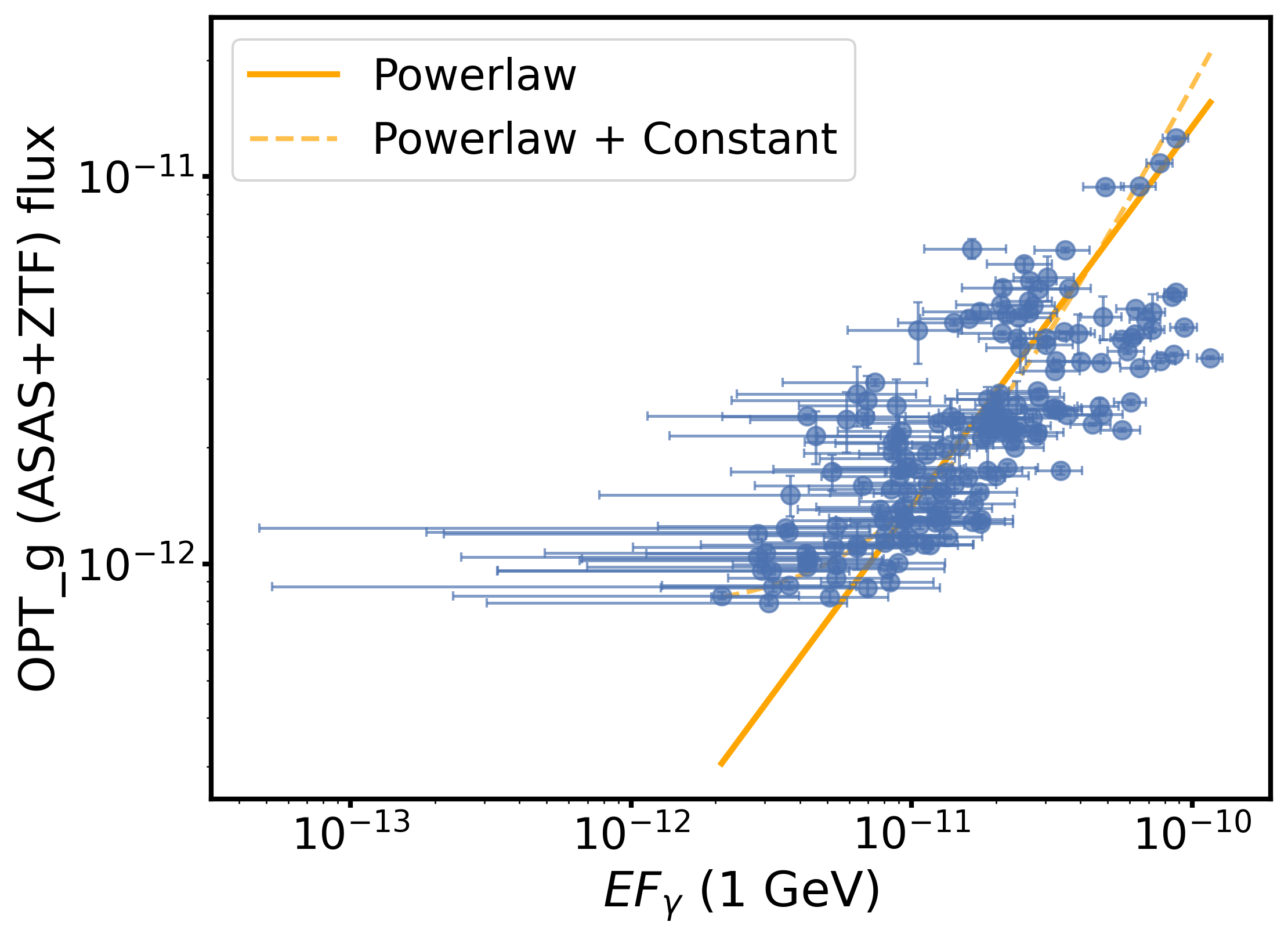}
    \includegraphics[width=0.32\textwidth]{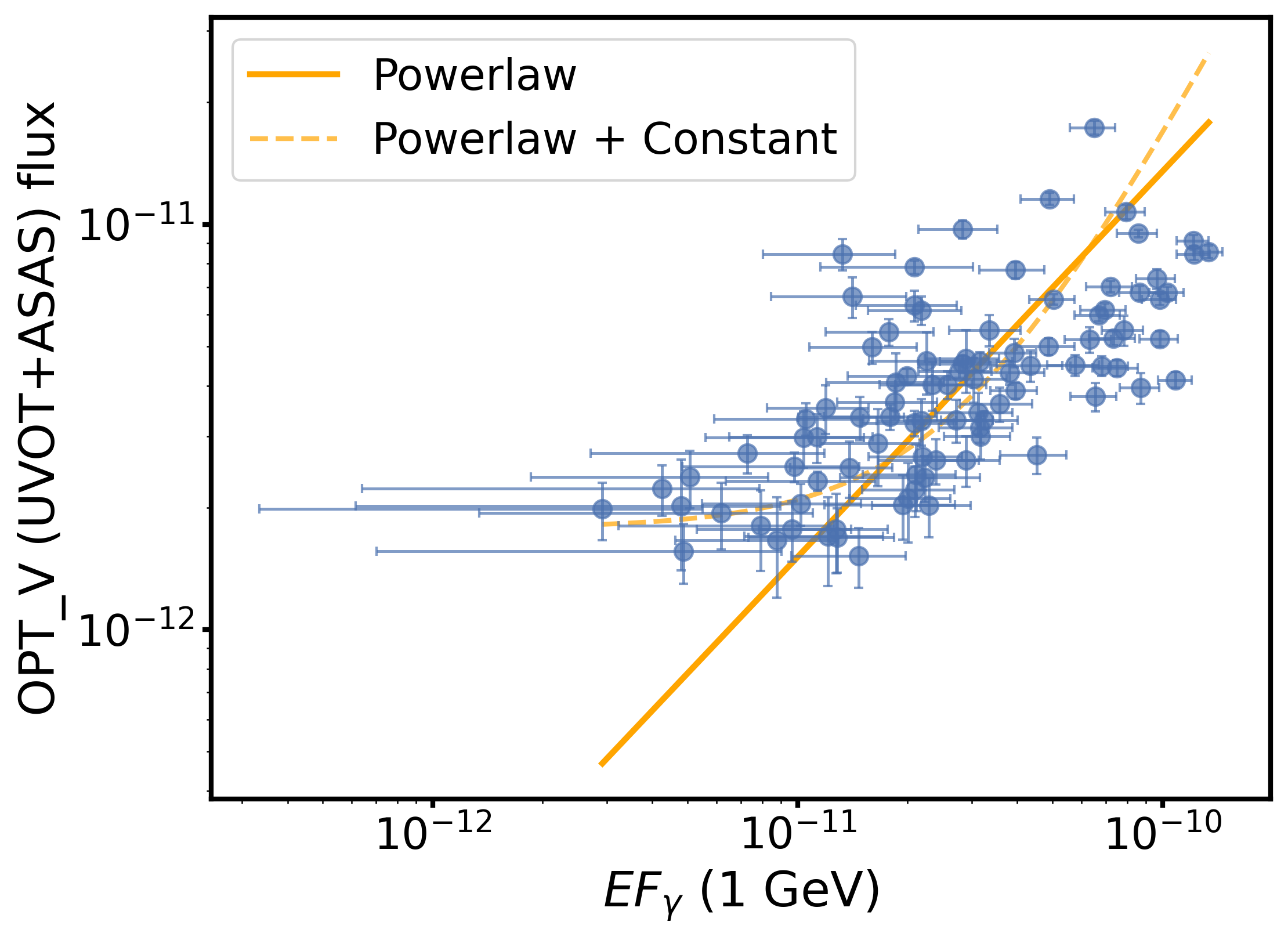}
    \includegraphics[width=0.32\textwidth]{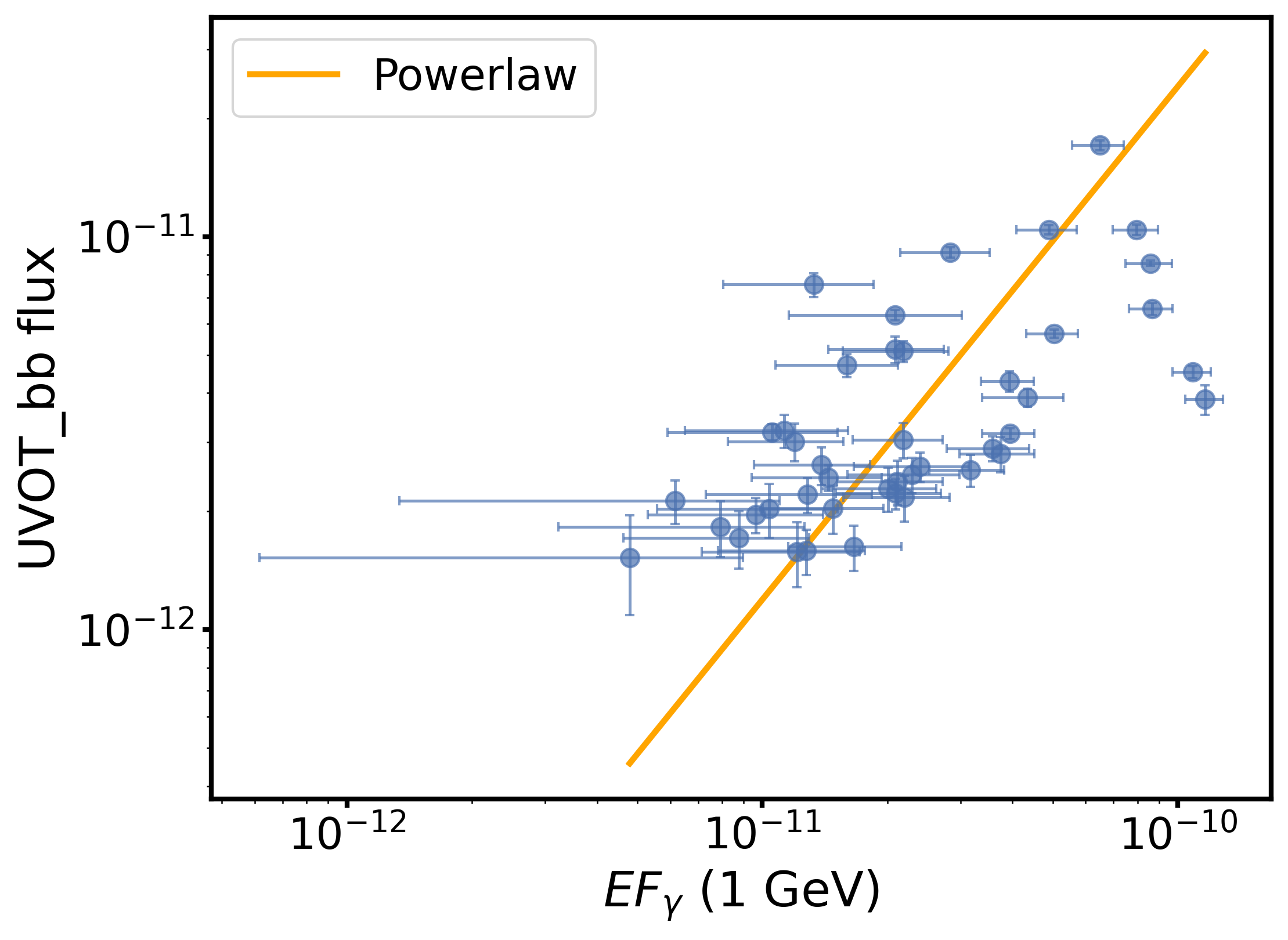}
    \includegraphics[width=0.32\textwidth]{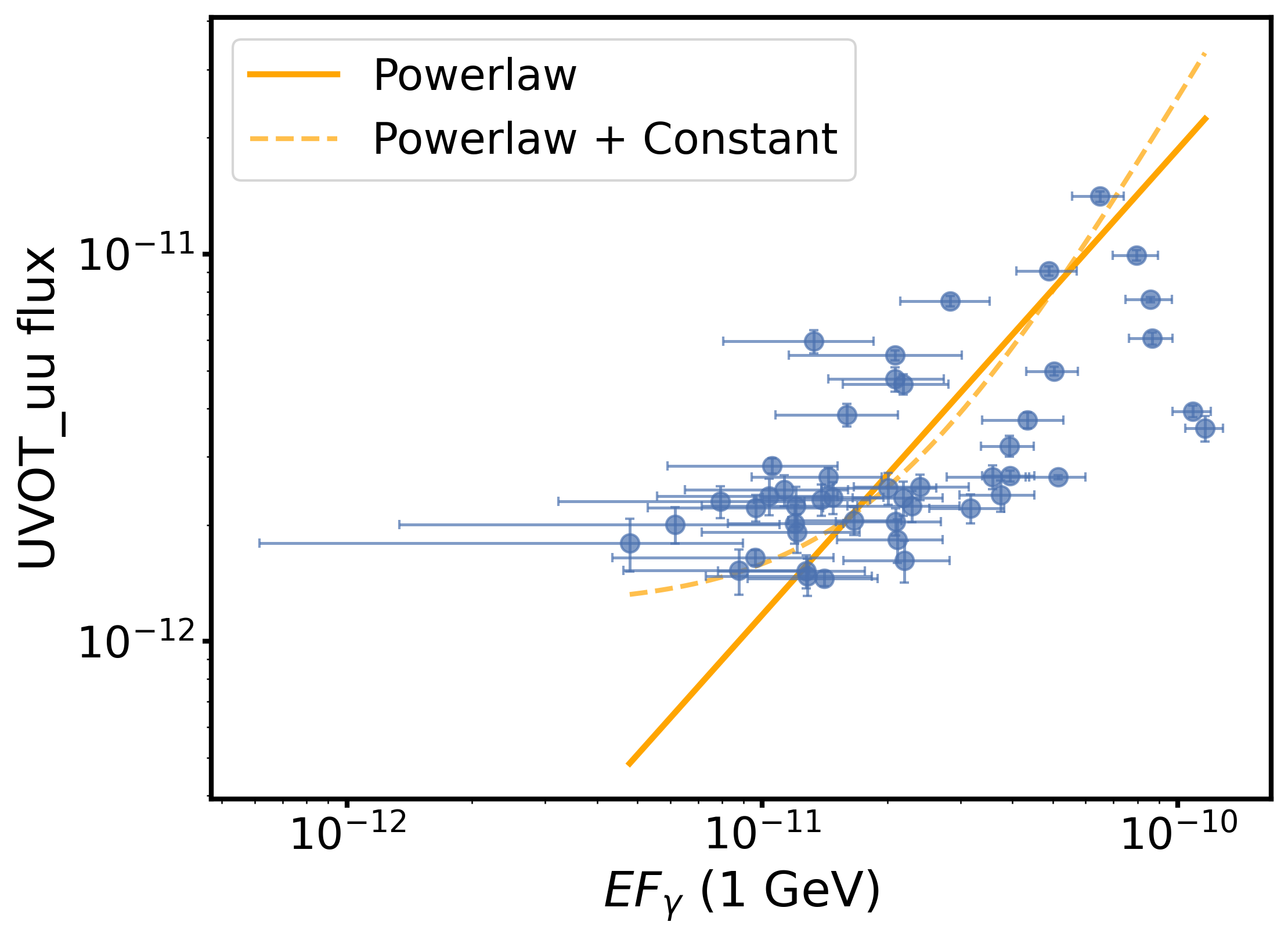}
    \includegraphics[width=0.32\textwidth]{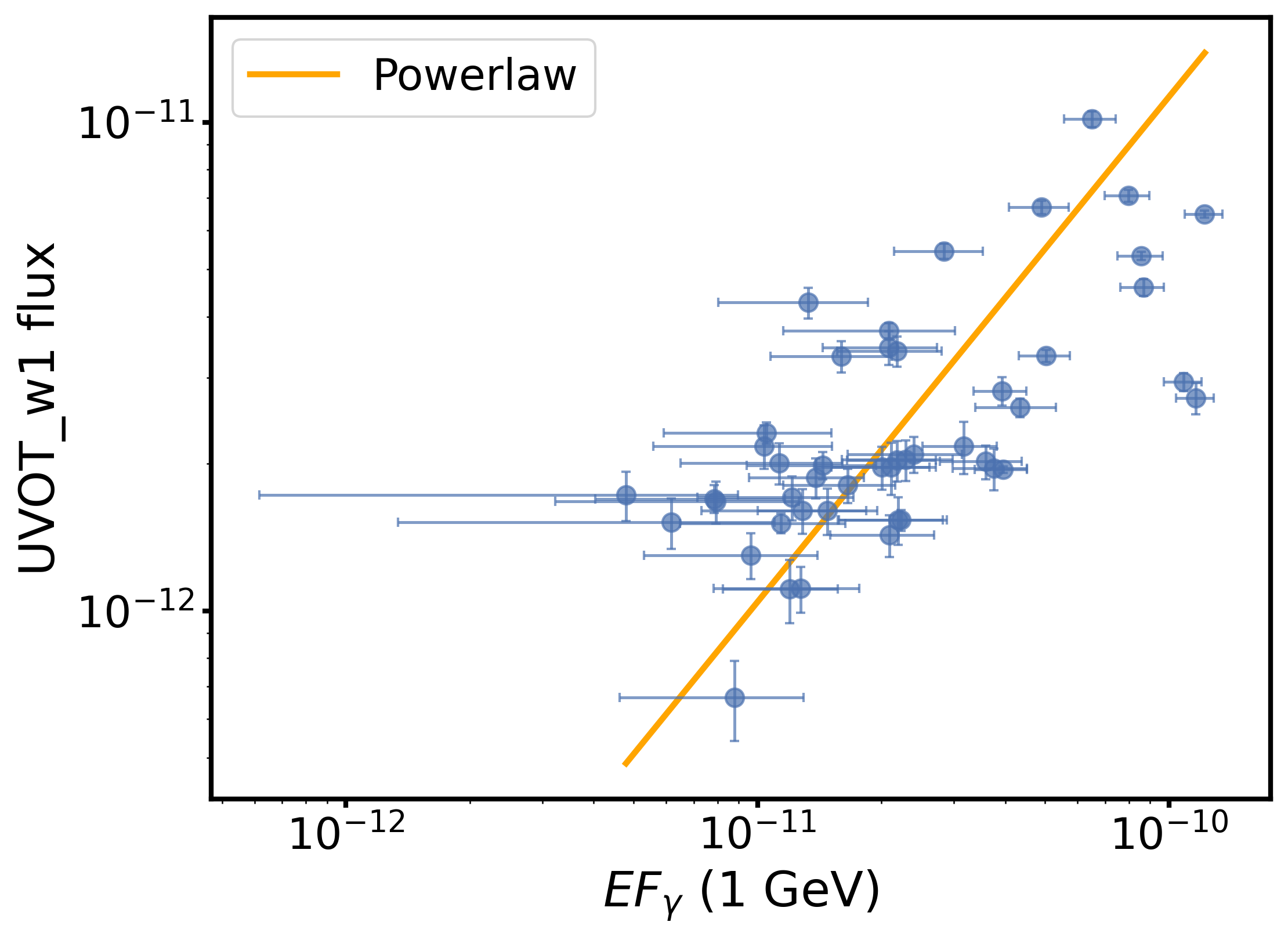}
    \includegraphics[width=0.32\textwidth]{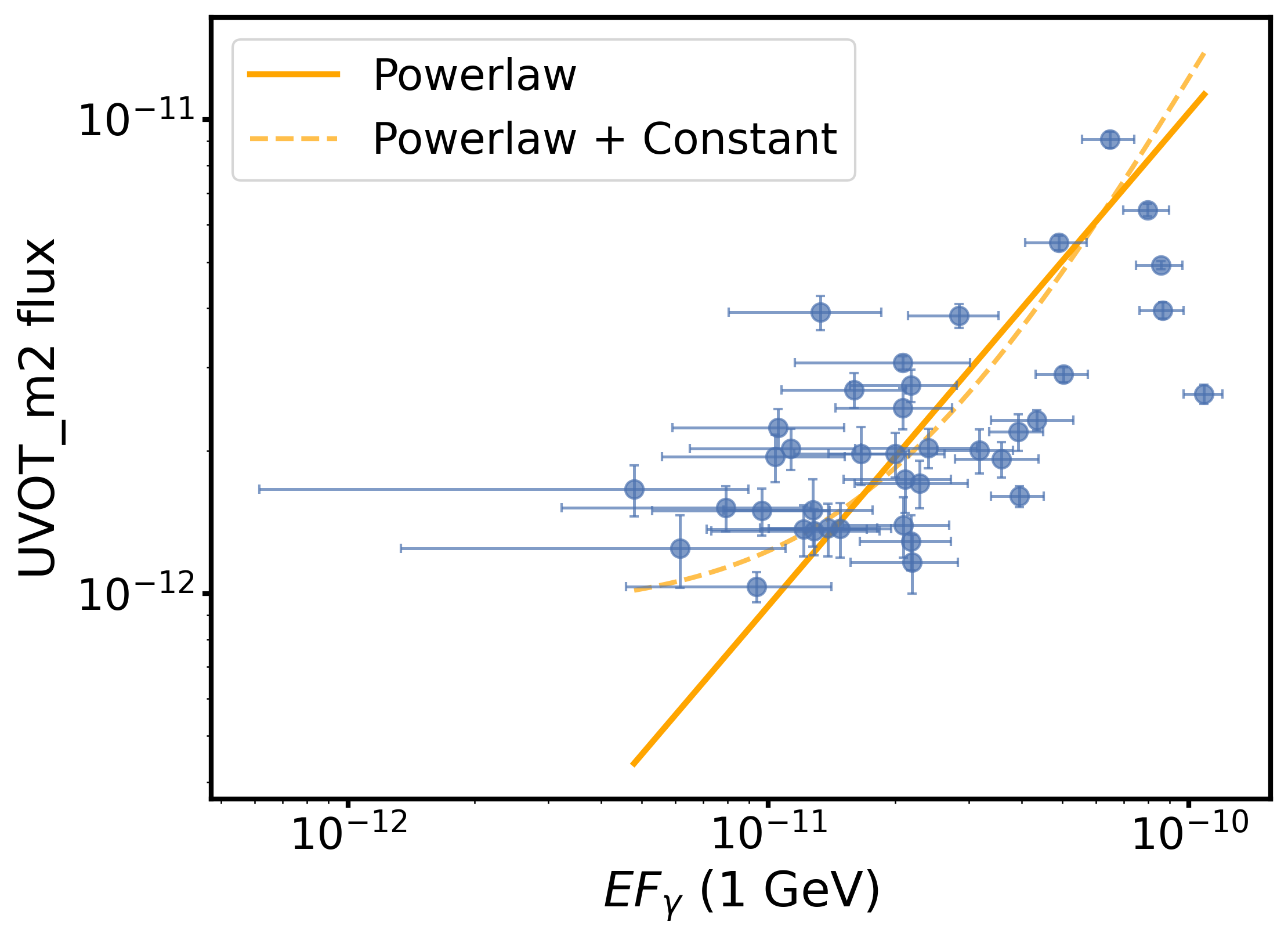}
    \includegraphics[width=0.32\textwidth]{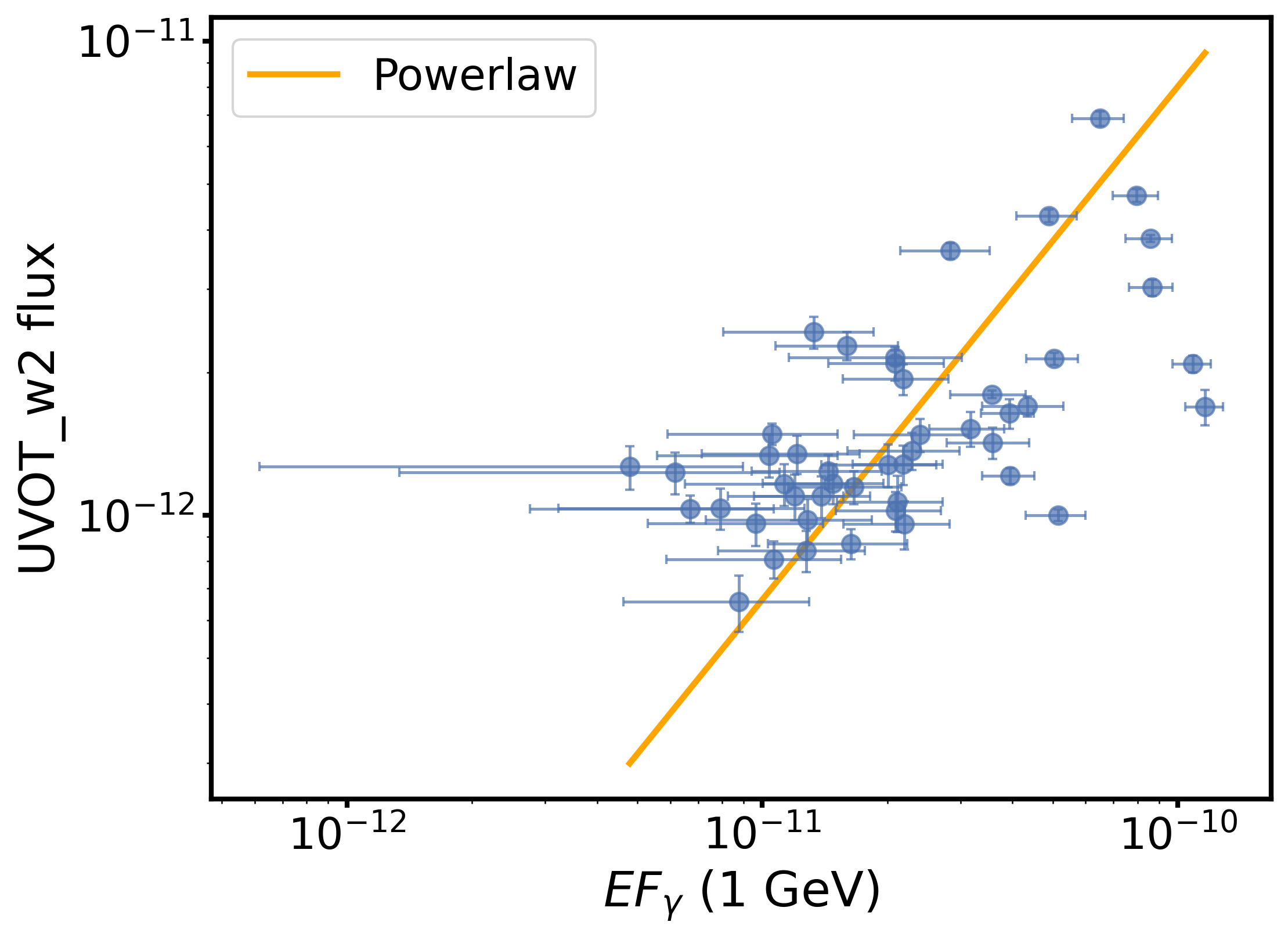}
     \includegraphics[width=0.32\textwidth]{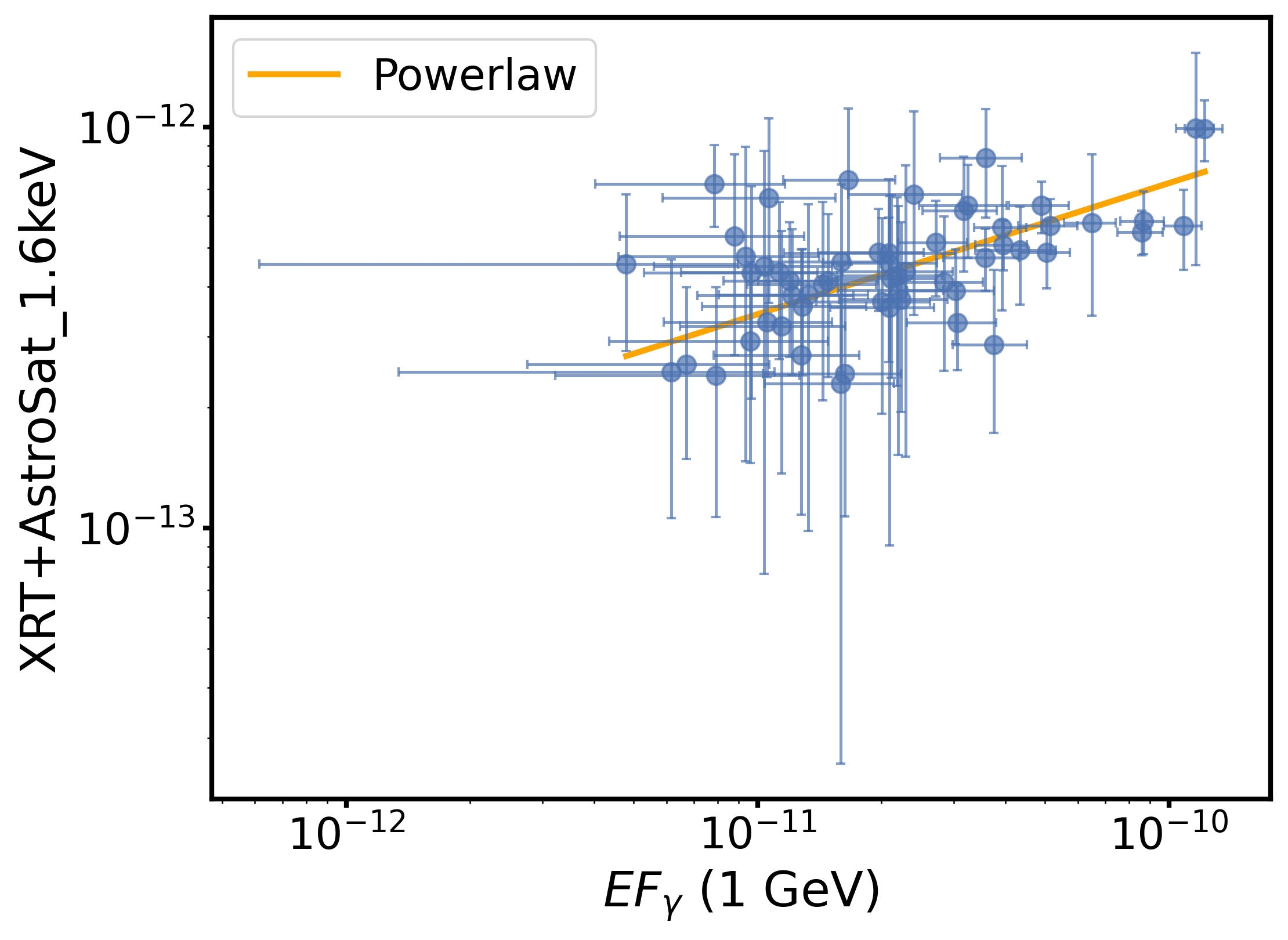}
    \caption{MWL correlation plots with respect to the \gray\ spectral energy distribution evaluated at 1 GeV. From top to bottom and left to right, the quantities on the y-axis are: NEOWISE flux at $4.6\mu$m; NEOWISE flux at $3.4\ \mu$m; ZTF i filter flux; ZTF r filter flux; ASAS-SN and ZTF g filter flux; ASAS-SN and UVOT V filter flux; UVOT b filter flux; UVOT u filter flux; UVOT w1 filter flux; UVOT M2 filter flux; UVOT W2 filter flux; XRT and \textit{AstroSat}-XRT spectral energy distribution evaluated at 1.6 keV.}
    \label{fig:corr_nfn}
\vspace{1cm}
\resizebox{\textwidth}{!}{%
\begin{tabular}{lccccccccccc}
\hline
Band & $r_{\rm P}$ & $p_{\rm P}$ & $r_{\rm S}$ & $p_{\rm S}$ & $a_{\rm PL}$ & $b_{\rm PL}$ & $a_{\rm add}$ & $b_{\rm add}$ & $C_{\rm add}$ & $\Delta\chi^2$ & $p(\Delta\chi^2)$ \\
\hline
NEOWISE\_4.6u & 0.844 & $7.54\times10^{-5}$ & 0.832 & $1.19\times10^{-4}$ & $1.04\pm0.18$ & $-0.28\pm1.90$ & -- & -- & -- & -- & -- \\
NEOWISE\_3.4u & 0.854 & $5.06\times10^{-5}$ & 0.843 & $7.97\times10^{-5}$ & $0.97\pm0.16$ & $-0.98\pm1.70$ & -- & -- & -- & -- & -- \\
ZTF\_i & 0.848 & $2.45\times10^{-24}$ & 0.792 & $2.71\times10^{-19}$ & $1.33\pm0.09$ & $2.84\pm0.91$ & -- & -- & -- & -- & -- \\
ZTF\_r & 0.669 & $2.22\times10^{-19}$ & 0.861 & $4.26\times10^{-42}$ & $1.01\pm0.08$ & $-0.86\pm0.88$ & $1.46\pm0.17$ & $8.71\times10^{3}\,\pm\,3.55\times10^{4}$ & $8.83\times10^{-13}\,\pm\,1.11\times10^{-13}$ & 47.11 & $6.71\times10^{-12}$ \\
OPT\_V (UVOT+ASAS) & 0.584 & $6.30\times10^{-10}$ & 0.706 & $1.84\times10^{-15}$ & $0.90\pm0.10$ & $-2.04\pm1.02$ & $1.68\pm0.35$ & $1.01\times10^{6}\,\pm\,8.38\times10^{6}$ & $1.78\times10^{-12}\,\pm\,3.22\times10^{-13}$ & 78.77 & $6.99\times10^{-19}$ \\
OPT\_g (ASAS+ZTF) & 0.633 & $7.70\times10^{-21}$ & 0.776 & $2.76\times10^{-36}$ & $1.08\pm0.09$ & $-0.25\pm0.94$ & $1.38\pm0.16$ & $1.11\times10^{3}\,\pm\,4.40\times10^{3}$ & $7.42\times10^{-13}\,\pm\,1.27\times10^{-13}$ & 74.38 & $6.45\times10^{-18}$ \\
UVOT\_bb & 0.513 & $4.36\times10^{-4}$ & 0.672 & $8.11\times10^{-7}$ & $1.22\pm0.24$ & $1.31\pm2.44$ & -- & -- & -- & -- & -- \\
UVOT\_uu & 0.536 & $1.02\times10^{-4}$ & 0.643 & $1.11\times10^{-6}$ & $1.22\pm0.23$ & $1.27\pm2.36$ & $1.83\pm0.66$ & $4.80\times10^{7}\,\pm\,7.51\times10^{8}$ & $1.23\times10^{-12}\,\pm\,5.74\times10^{-13}$ & 16.66 & $4.48\times10^{-5}$ \\
UVOT\_w1 & 0.597 & $9.56\times10^{-6}$ & 0.655 & $5.75\times10^{-7}$ & $1.01\pm0.16$ & $-1.10\pm1.64$ & -- & -- & -- & -- & -- \\
UVOT\_m2 & 0.622 & $2.41\times10^{-5}$ & 0.602 & $5.03\times10^{-5}$ & $1.15\pm0.21$ & $0.37\pm2.21$ & $1.56\pm0.53$ & $4.26\times10^{4}\,\pm\,5.37\times10^{5}$ & $9.15\times10^{-13}\,\pm\,4.40\times10^{-13}$ & 20.26 & $6.77\times10^{-6}$ \\
UVOT\_w2 & 0.547 & $6.91\times10^{-5}$ & 0.595 & $1.03\times10^{-5}$ & $1.14\pm0.21$ & $0.11\pm2.16$ & -- & -- & -- & -- & -- \\
XRT+XRT\_1.6keV & 0.619 & $2.84\times10^{-7}$ & 0.503 & $6.69\times10^{-5}$ & $0.32\pm0.05$ & $-8.94\pm0.52$ & -- & -- & -- & -- & -- \\
\hline
\end{tabular}}
\captionof{table}{Results from the MWL correlations with respect to the \gray\ spectral energy distribution evaluated at 1 GeV. We provide the Pearson coefficient $r_{\rm P}$ with its probability $p_{\rm P}$, the Spearman coefficient $r_{\rm S}$ with its probability $p_{\rm S}$, the best-fit values for the power-law fit, and the best-fit values for the fit of a power-law with a constant (the latter only when it is favored by $p<10^{-3}$).\label{table:corr_nfn}}

\end{figure*}

    Because integral quantities can mix flux variability with spectral variability that is present in the data, as presented in the previous Section, we repeated this method using pivot-energy proxies, namely $\nu F_{\nu, 1 \rm{GeV}}$ for \fermi,  and $\nu F_{\nu, 1.6 \rm{keV}}$ for \textit{Swift}-XRT and \textit{AstroSat}. The choice of 1 GeV and 1.6 keV as pivot-energies is made considering that these values are close to the typical decorrelation energy, where the flux estimate is least sensitive on the spectra hypothesis. The optical/UV bands are unchanged. The results of this second analysis are shown in Figure \ref{fig:corr_nfn} and Table \ref{table:corr_nfn}.
    
    Besides the fact that this choice reduces the dependence on the LAT bandpass integration and on possible photon-index fluctuations, this analysis produces results that are easier to interpret from a theoretical point of view. It is in fact well known that for synchrotron and synchrotron-self-Compton (SSC) radiation, the SED at a given frequency $\nu_0$, varies as the Doppler factor to the power of $3+\alpha_{\nu_0}$such that:
    \begin{equation}
        \nu F_{\nu} \propto \delta^{3+\alpha_{\nu_0}},
    \end{equation}
     where $\alpha$ is the spectral index defined by 
$F_\nu \propto \nu^{-\alpha}$). If the radiation is produced via external-inverse-Compton (EIC) on a target photon field that is boosted into the jet, the dependency becomes:
     \begin{equation}
     \nu F_{\nu} \propto \delta^{4+ 2\alpha_{\nu_0}}.
     \end{equation}

In the simplest one-zone interpretation, or more generally if the emitting regions responsible for the different bands share the same viewing-angle, variations of the Doppler factor are expected to produce correlated MWL variability. In this case, each band responds to the same underlying change in $\delta$, although with a different Doppler exponent depending on the radiative process and spectral slope. Assuming that the \gray\ emission is due to EIC, and all other bands are due to synchrotron or SSC, the correlation coefficient is thus predicted to be equal to $(3+\alpha_{band}) / (4+2\alpha_{LAT})$, or $(2+\Gamma_{band}) / (2+2\Gamma_{LAT})$ when expressed using the photon index $\Gamma$. Using range values of $\Gamma_{LAT} = (2,3)$, $\Gamma_{X} = (1,2)$, and $\Gamma_{optical} = (2.5, 3.2)$, we thus expect a correlation between X-ray and LAT with an index of ($0.4,0.7$) and between optical filters and LAT with an index of about ($0.6,0.9$).\\

\begin{figure*}[!t]
    \centering
    \includegraphics[width=0.33\textwidth]{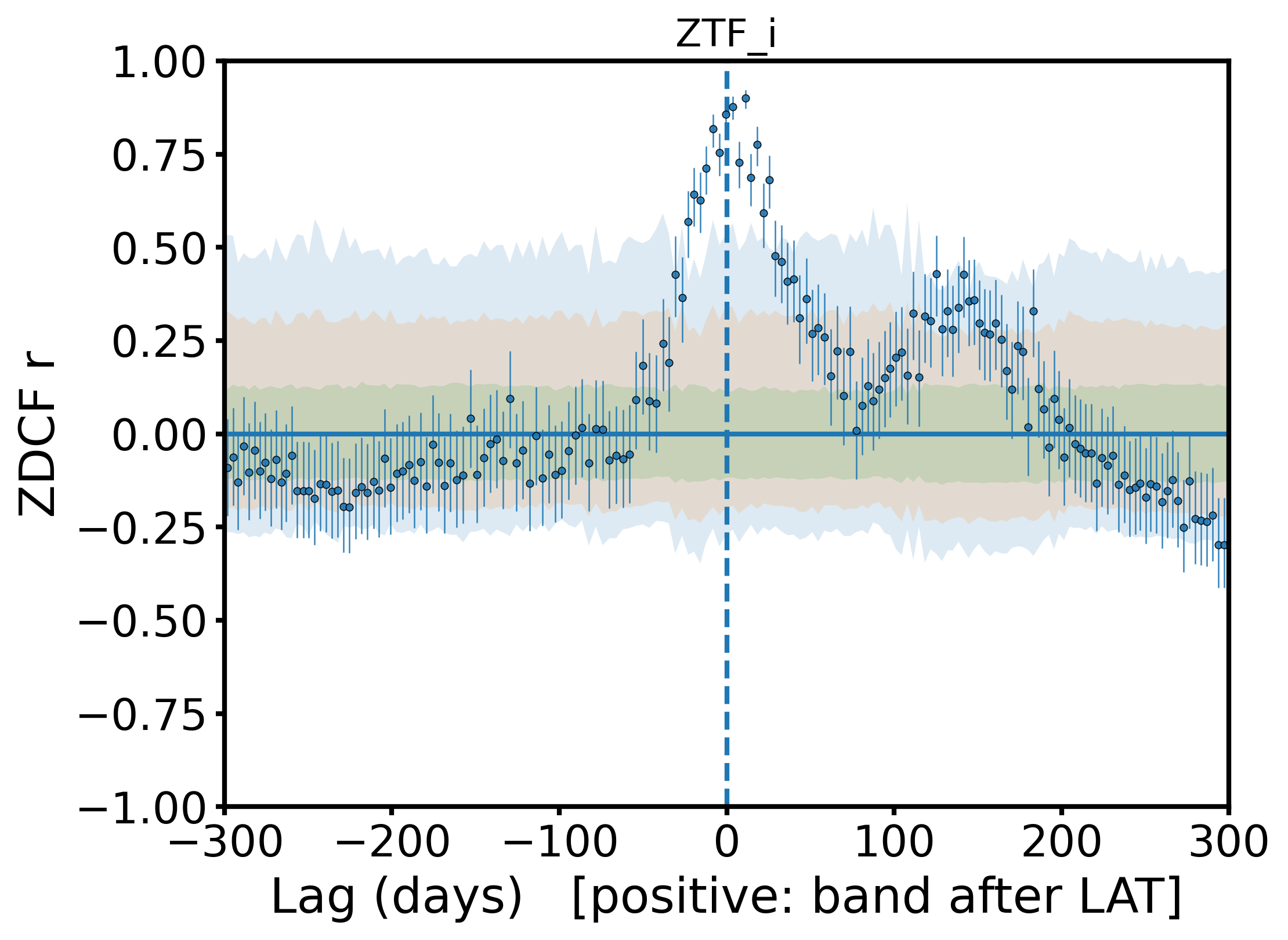}
    \includegraphics[width=0.33\textwidth]{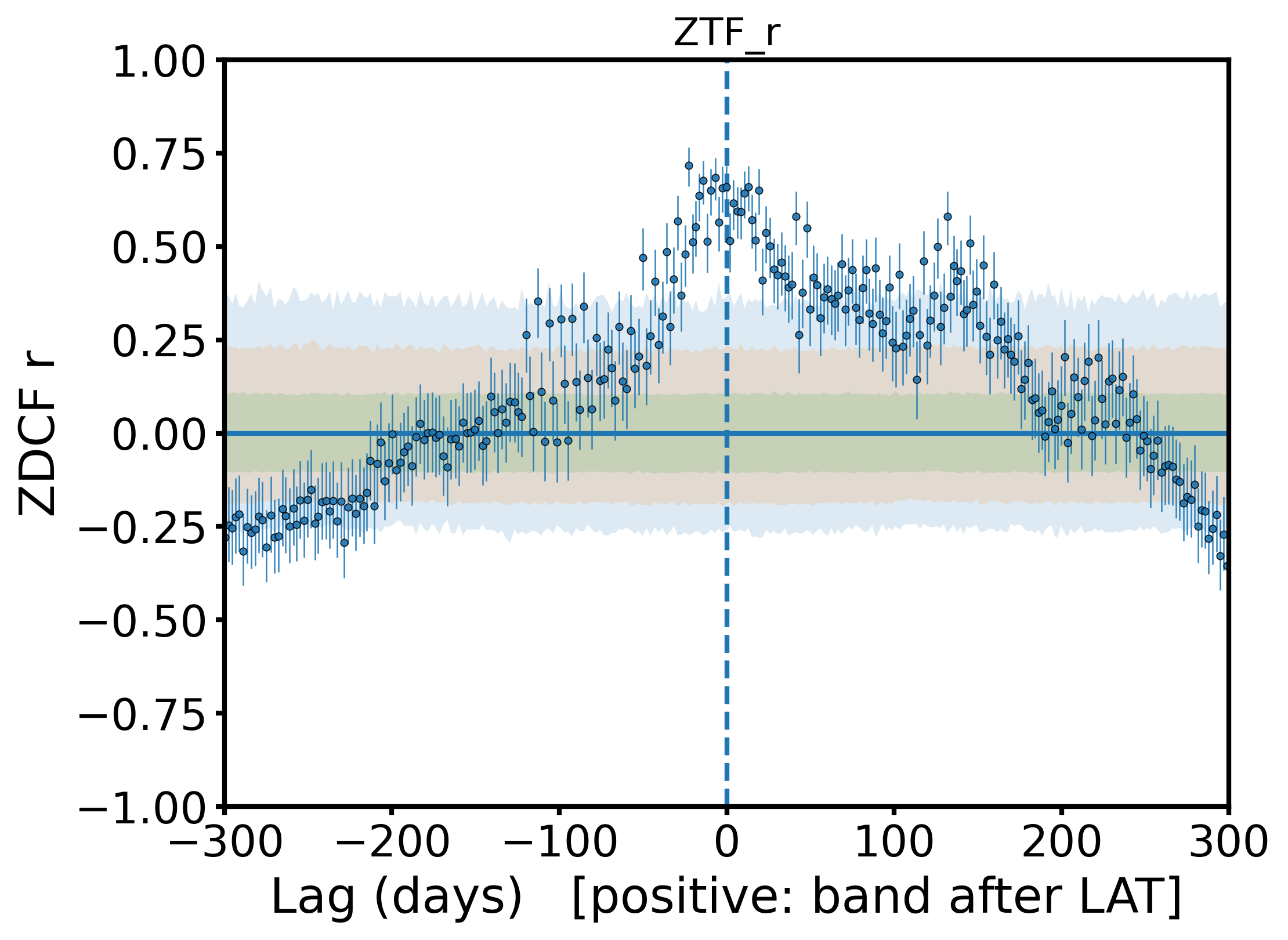}
    \includegraphics[width=0.33\textwidth]{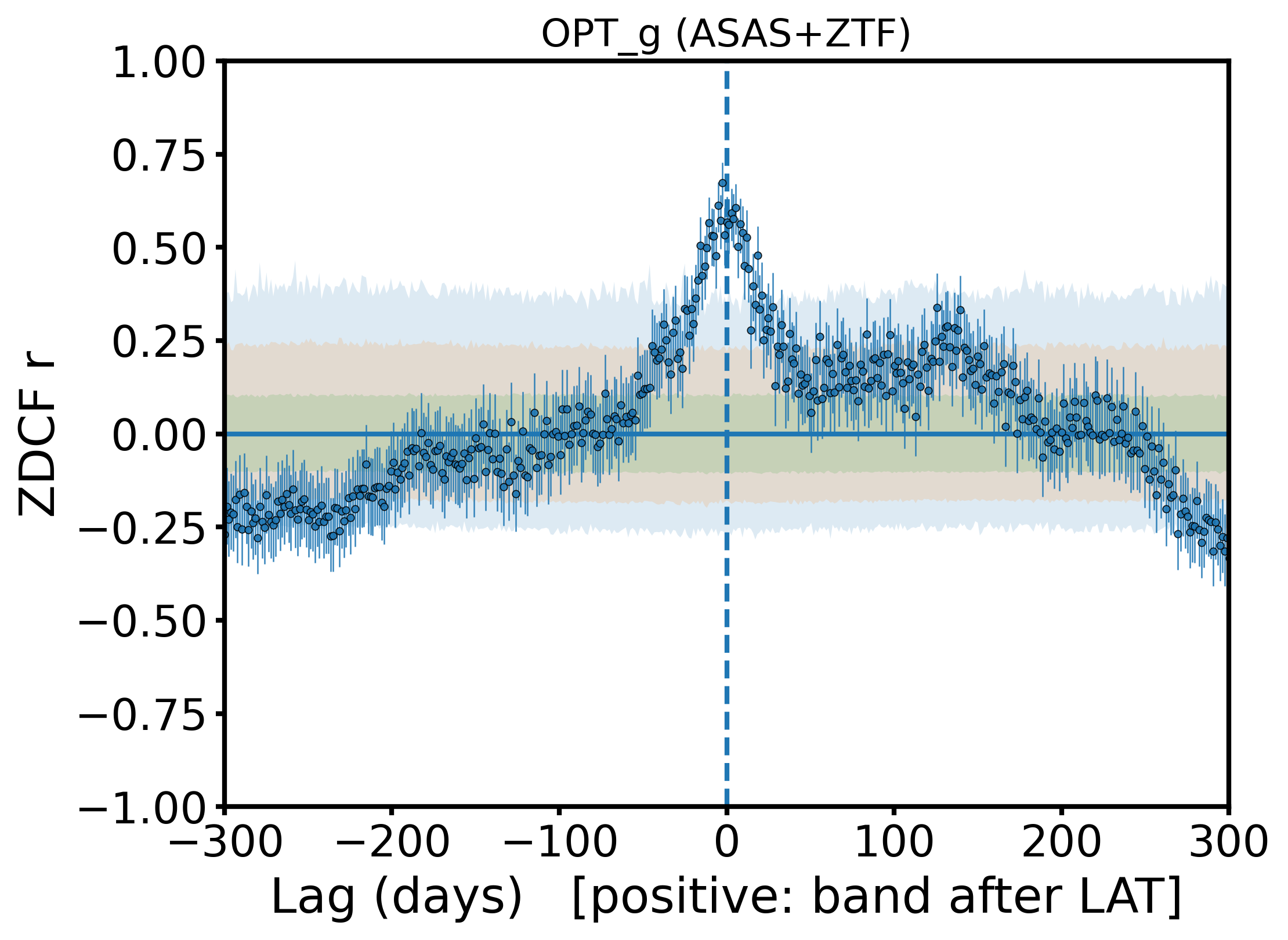}
    \includegraphics[width=0.33\textwidth]{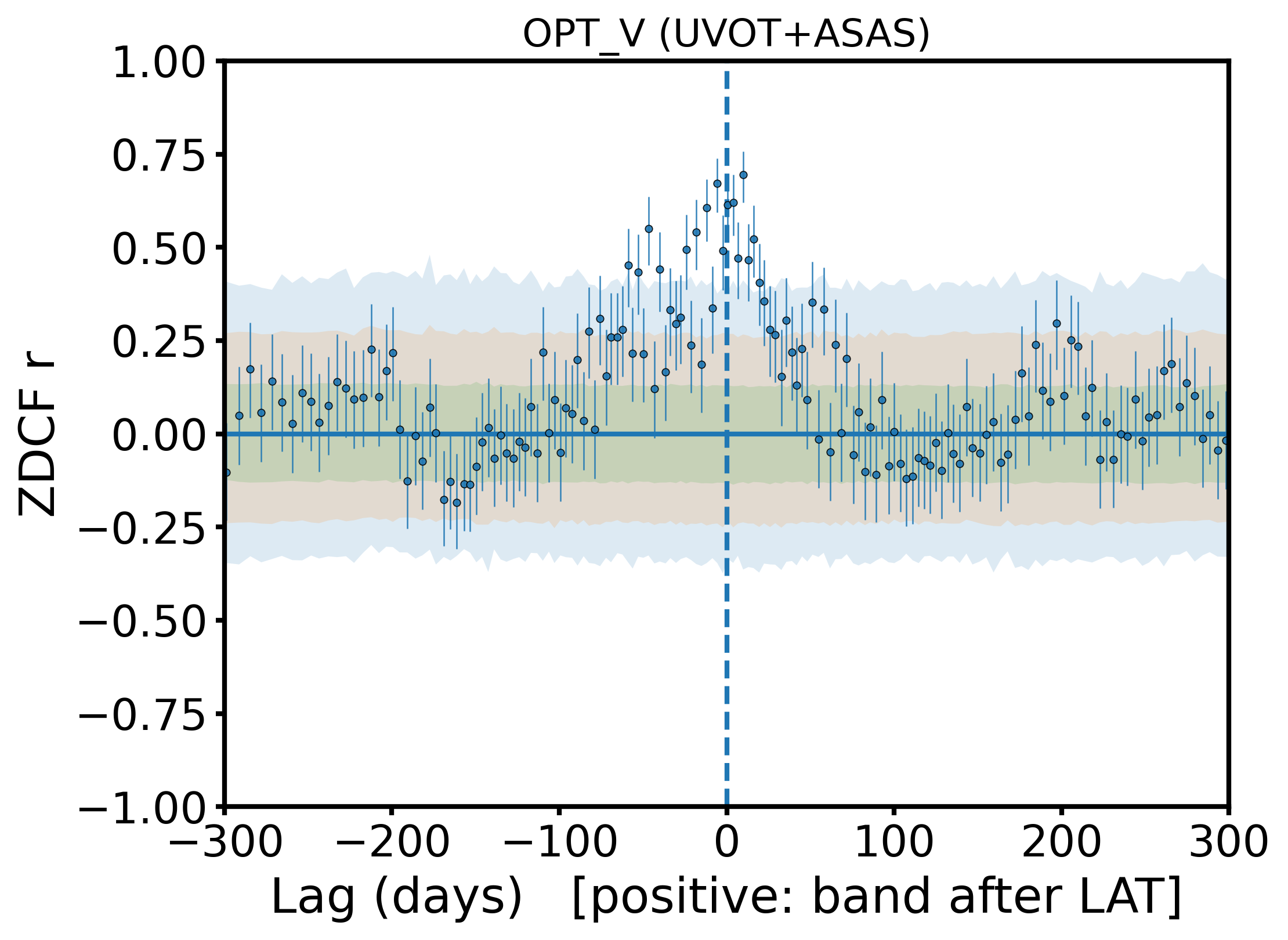}
    \includegraphics[width=0.33\textwidth]{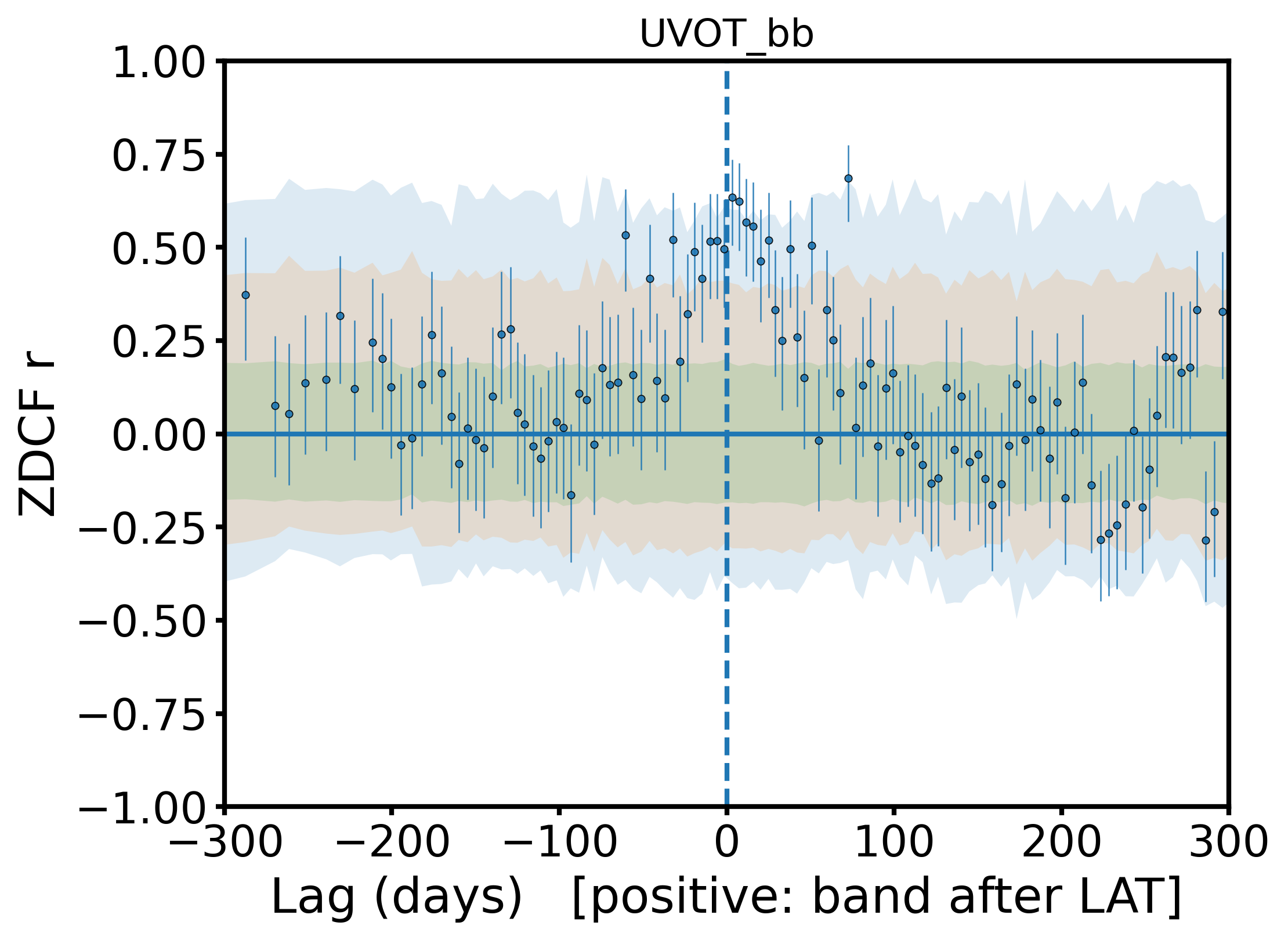}
    \includegraphics[width=0.33\textwidth]{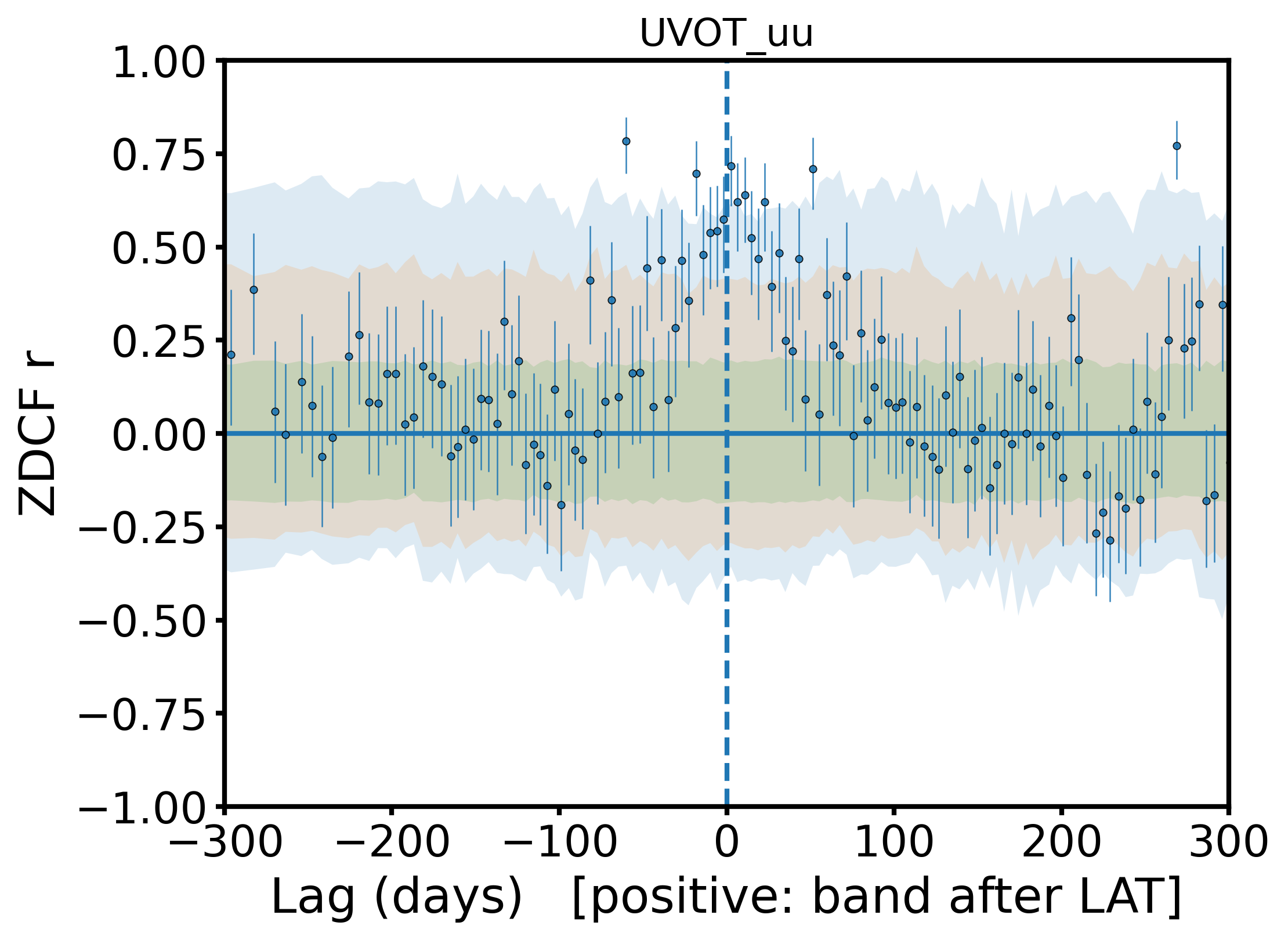}
    \includegraphics[width=0.33\textwidth]{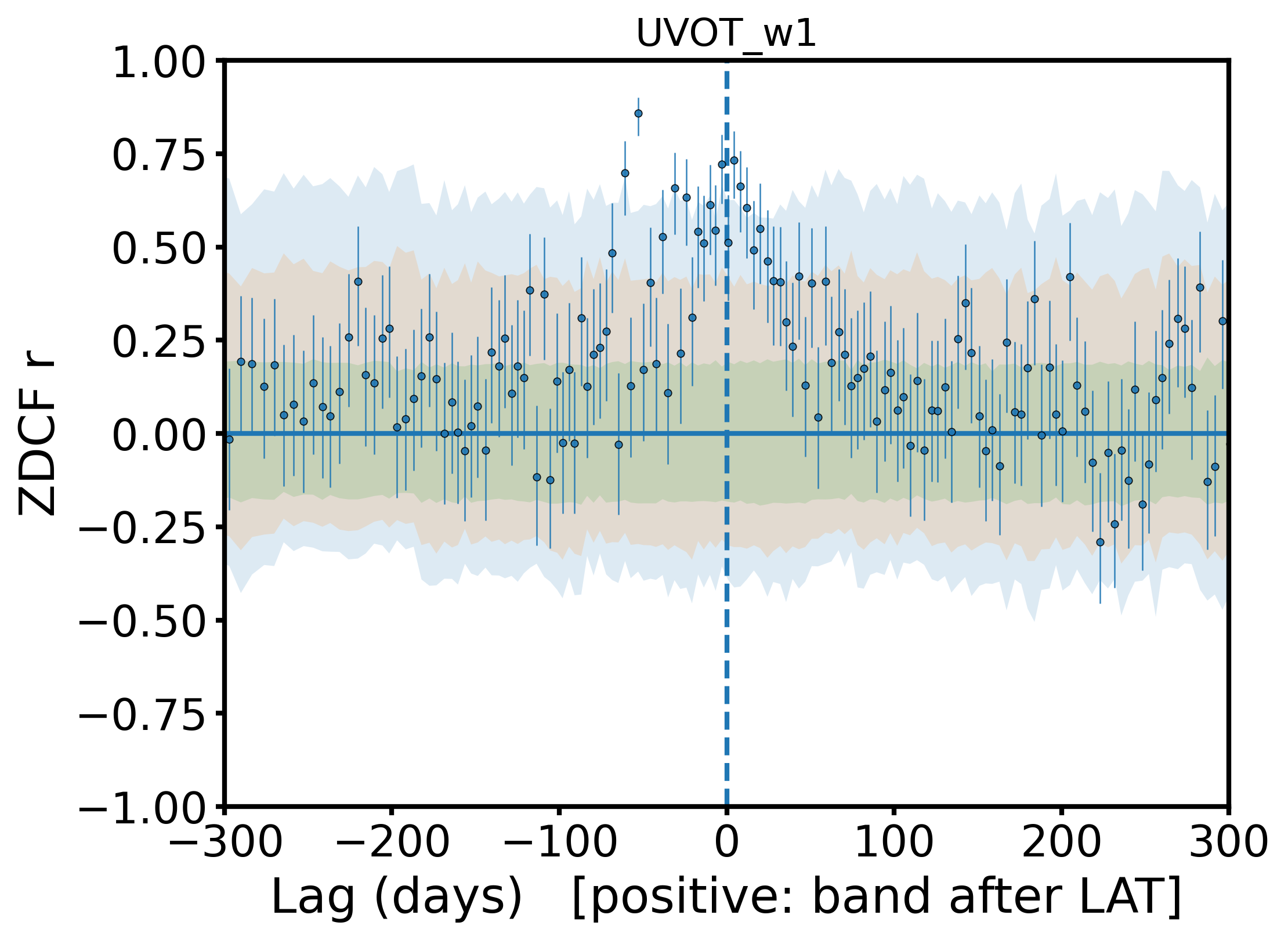}
    \includegraphics[width=0.33\textwidth]{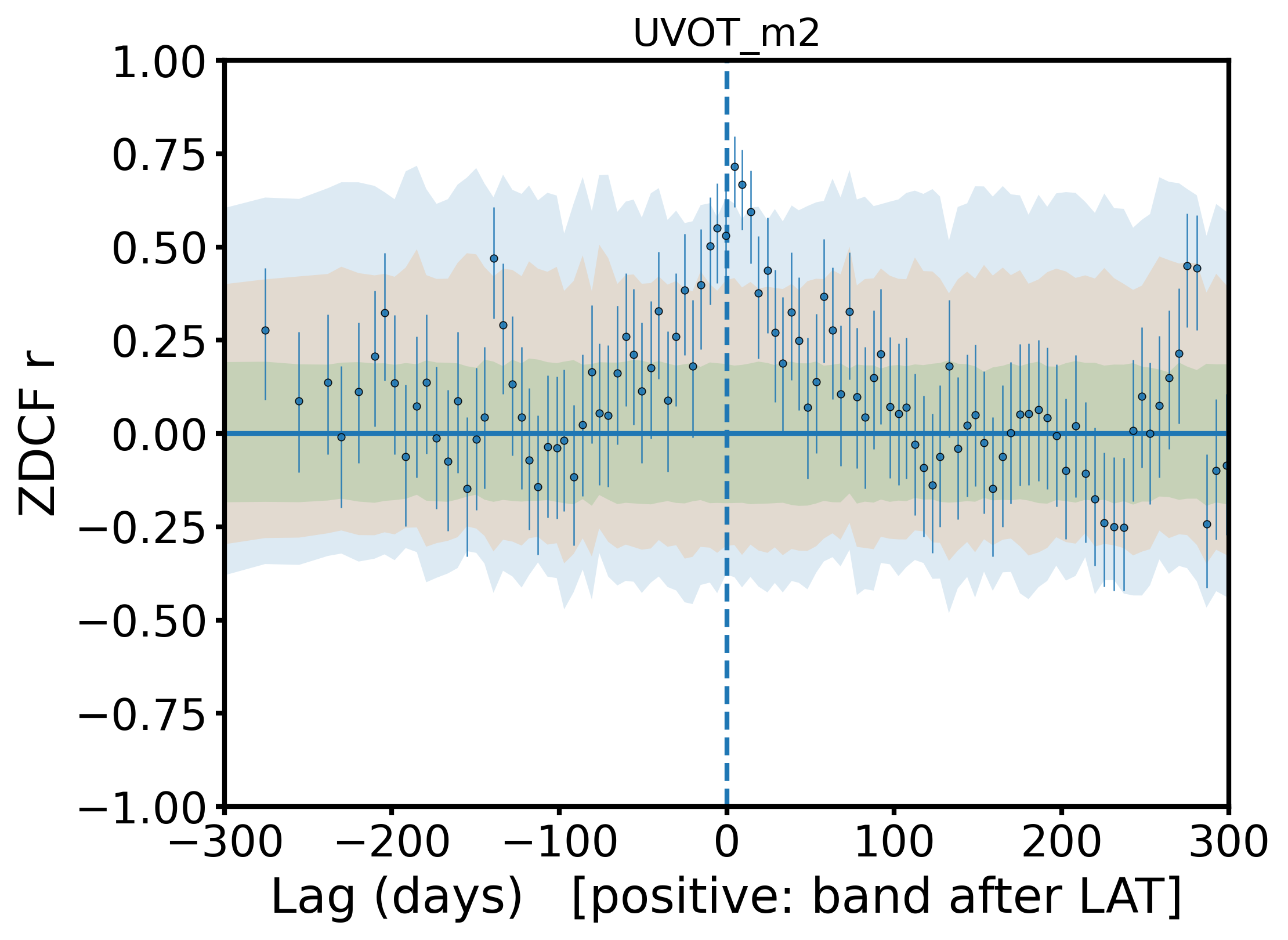}
    \includegraphics[width=0.33\textwidth]{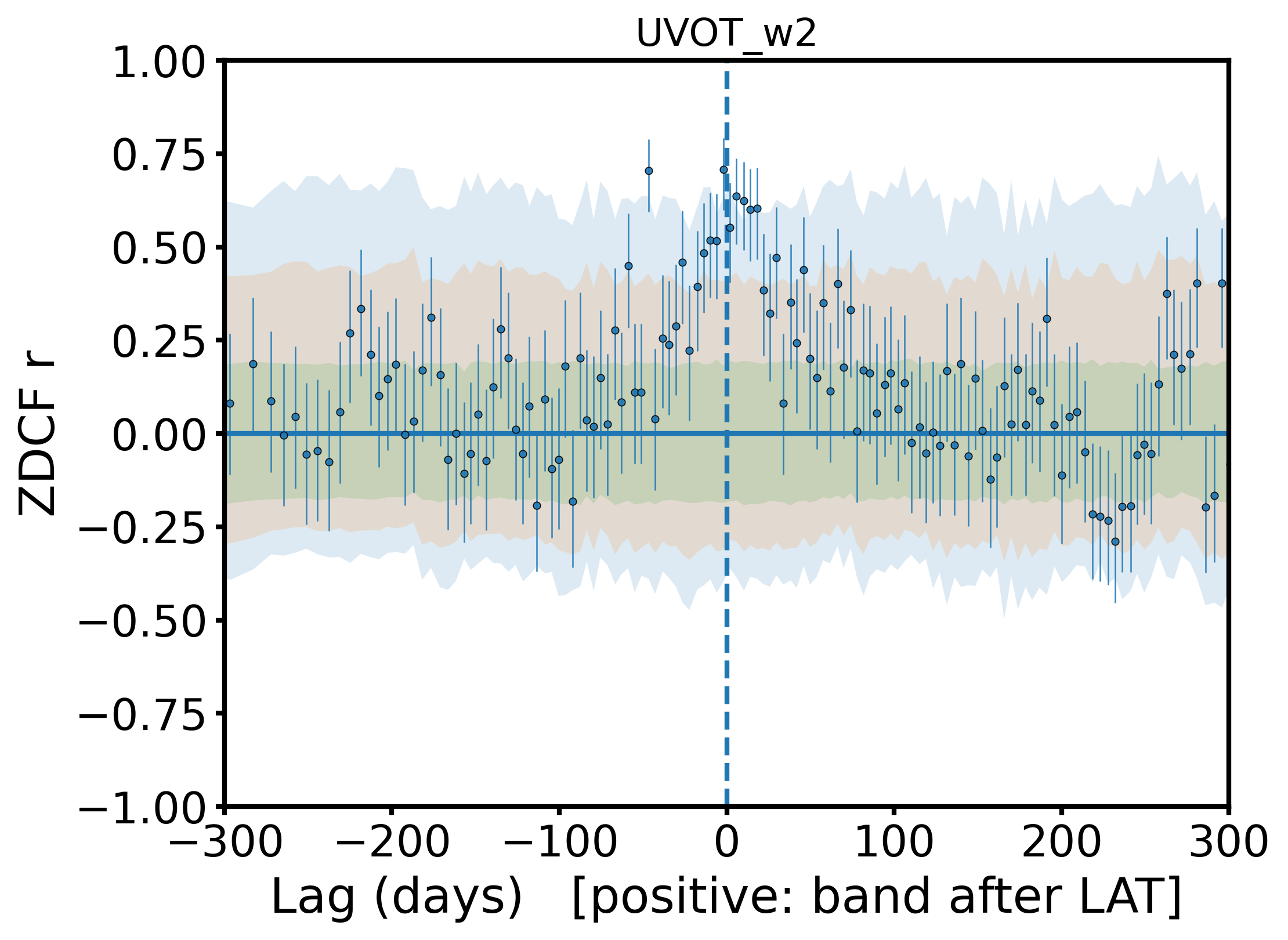}
    \includegraphics[width=0.33\textwidth]{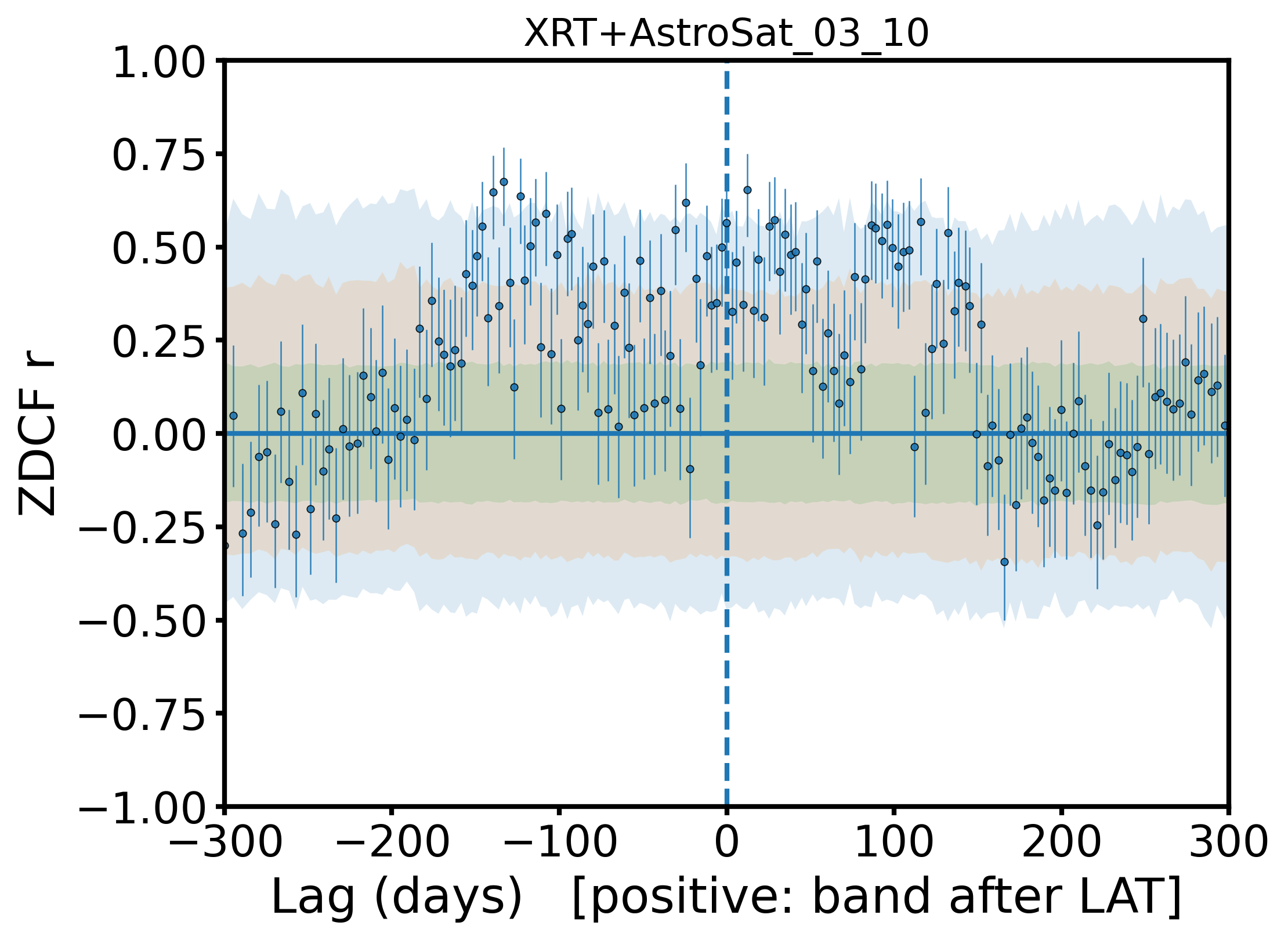}
    \caption{z-transformed discrete correlation function between the MWL bands and the \gray\ integral photon flux. From top to bottom and left to right, the MWL band used is: ZTF i filter flux; ZTF r filter flux; ASAS-SN and ZTF g filter flux; ASAS-SN and UVOT V filter flux; UVOT b filter flux; UVOT u filter flux; UVOT w1 filter flux; UVOT M2 filter flux; UVOT W2 filter flux; XRT and \textit{AstroSat}-XRT integral photon flux between 0.3 and 10 keV.}
    \label{fig:zdcf}
\end{figure*}

\subsection{Time-lags}
The correlation study presented above is limited by the fact that it assumes that there is no time-lag between the MWL bands we study. To further investigate this option, we performed a cross-correlation analysis of the \fermi\ light curve with each MWL light curve. Given the strongly uneven sampling of the data and the heterogeneous cadence across instruments, standard cross-correlation techniques based on regularly sampled time series are not applicable. We therefore adopted the z-transformed discrete correlation function \citep[ZDCF,][]{ZDCF}, which is specifically designed to handle sparsely and irregularly sampled light curves. 
For each MWL band, we computed the ZDCF between the LAT integral flux light curve and the corresponding band-integrated flux. Positive time lags are defined such that the MWL band follows the \gray\ emission. The maximum lag range was chosen to be $\pm$300 days, ensuring sufficient coverage to probe both short and long delays while preserving adequate statistics in each lag bin. Due to the different sampling of the MWL light curves, the minimum number of data pairs in each ZDCF bin is set to 90 for the $r$ and the $g$ filter, to 60 for the $i$ and the $v$ filter, and to 30 for all other data. NEOWISE measurements are not considered for this analysis, due to their limited number of data points. To provide an empirical reference for the expected level of spurious correlations, we used a permutation test. For each band, we generated 5000 surrogate datasets by randomly reshuffling in time the flux values of the MWL light curve. This procedure preserves the sampling pattern and the observed flux distribution, but destroys any physical temporal association with the $\gamma$-ray light curve. The resulting $1$-,$2$-, and $3$-$\sigma$ envelopes should therefore be interpreted as empirical null envelopes, not as formal confidence intervals that would require simulation of artificial light curves with the same statistical properties as the data, in particular with the same power spectral density \citep{Timmer95, Emmanoulopoulos13}.

Figure \ref{fig:zdcf} shows the resulting ZDCF functions for all considered bands, together with the empirical null envelopes derived from the reshuffled realizations. In all optical/UV bands, the ZDCF displays a broad positive peak centered around zero lag, exceeding the $3\sigma$ curve, consistent with the strong flux–flux correlations discussed in the previous section. In contrast, X-ray show a flatter ZDCF profiles with larger uncertainties, which is more difficult to interpret. Overall, this analysis suggests that the MWL variability of S5 1044+71 is largely contemporaneous within the temporal resolution of the available data, and does not provide evidence for significant systematic delays between the \gray\ and lower-energy bands.

\begin{figure}[!t]
    \centering
    \includegraphics[width=0.48\textwidth]{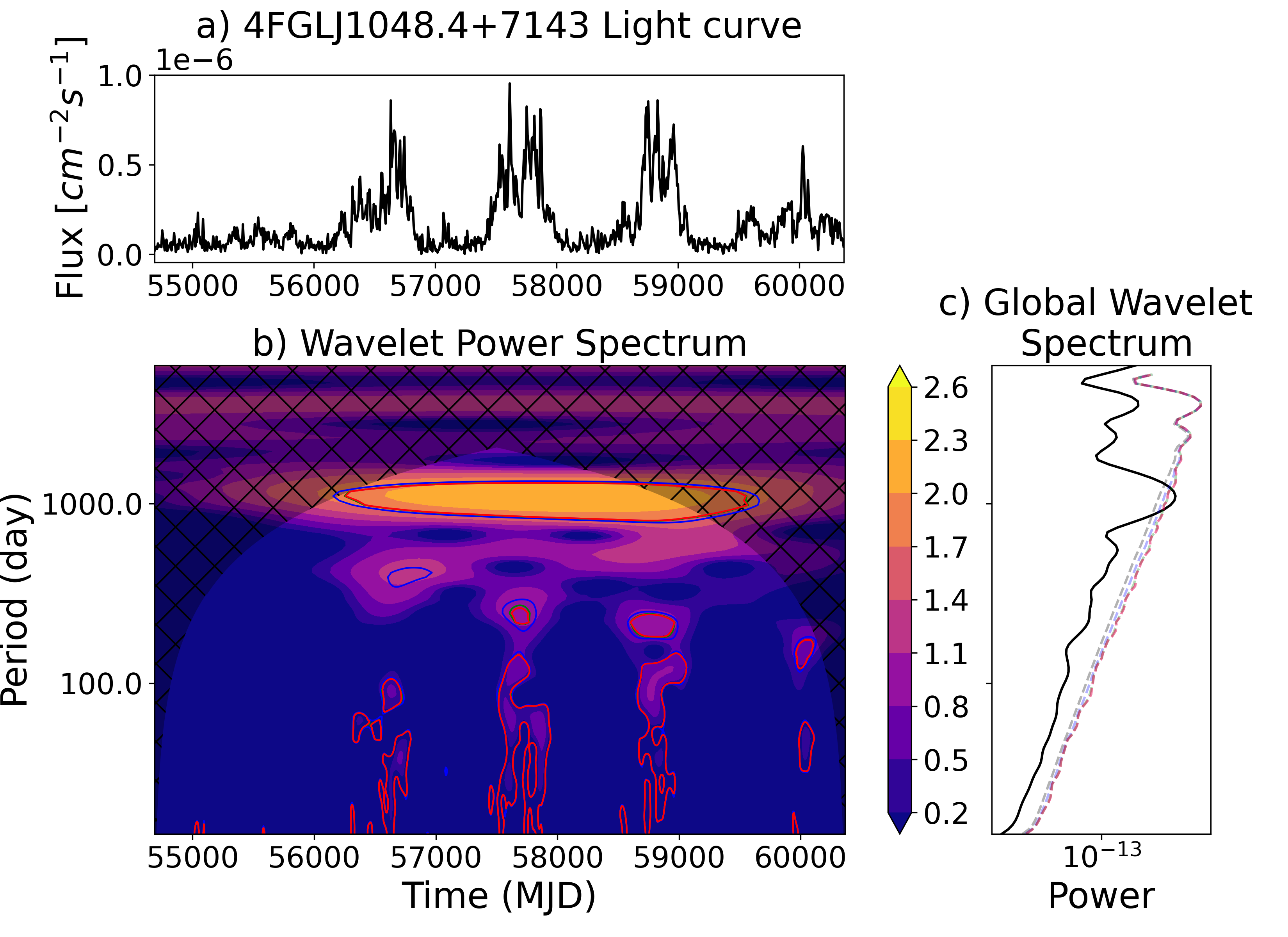}
    \caption{CWT for the monthly binned light curve of \sfive. a) id the \fermi\ light curve, b) is the wavelet power spectrum and c) the global wavelet spectrum.}
    \label{fig:CWT}
\end{figure}

    \subsection{Periodicity searches}

    For the QPO search in the \gray\ light curve, we follow the same approach as the one described in \cite{Ren23QPO}, using the new \fermi\ data. The Python package \texttt{PyCWT} developed by \cite{Torrence1998} is used to apply the Continuous Wavelet Transform (CWT) on the light curve. This technique can measure not only the potential constant QPOs, but also the transient and varying periodicities. We use the Morlet wavelet (see equation 1 in \cite{Torrence1998}) as the mother wavelet, and the source's light curve is convolved with the translation and dilation of the wavelet function to produce the wavelet power spectrum as presented in the bottom-left panel of Fig.~\ref{fig:CWT}). The shaded area in the wavelet spectrum is the cone of influence (COI) where the boarder effects become important. The Global wavelet spectrum, presented in the right panel of Fig.~\ref{fig:CWT}, is the time-averaged wavelet spectrum and corresponds to the true power spectrum of the light curve. 
    
    To estimate the significance of the wavelet spectrum and its post-trial significance, we follow the same method as described in Section 4.2 of \cite{Ren23QPO}. We simulated 10$^4$ fake light curves using the algorithm from \citet{Emmanoulopoulos13}, with the Python implementation provided by \citet{Connolly15}. The later produces artificial time series with the same statistical and variability properties as the original light curve of \sfive, meaning that the simulated light curves have the same probability distribution function (PDF) and power spectral density (PSD). The confidence levels are then obtained from the histogram of the global wavelet spectrum of all artificial light curves. The post-trial probability is calculated as follows:
    \begin{equation}
    	P_G = 1-(1-P_L^{\ a})^n, 
    \end{equation}
    with $P_L$ being the pre-trial probability. $a$ and $n$ are specific for each mother wavelet, and in this case, following \cite{Auchere2016}'s parameterization, we have: $a = 0.810 ~(N_{\rm{out}}~ \delta j)^{0.011}$ and $n = 0.491 ~(N_{\rm{out}}~ \delta j)^{0.926}$ for the local wavelet spectrum; and $a = 0.805 + 0.45 \times 2^{-S_{\rm{out}} \delta j }$ and $n = 1.136 ~(S_{\rm{out}} ~\delta j)^{1.2}$ for the global wavelet spectrum. The parameters  $N_{\rm{out}} = 787$ and $S_{\rm{out}}=16$ are respectively the time-period bins and the period bins of the wavelet spectrum outside the COI, where border effects become significant. 
    
    Figure~\ref{fig:CWT} shows the source's light curve in panel a), its CWT map in panel b), with the significance contour lines already in post-trial, and the global wavelet spectrum in panel c). One can notice that the period at around 1100~days exceeds the $5-\sigma$ significance contour in both the CWT maps and the global spectrum. After fitting a Gaussian function to the global wavelet spectrum curve, we obtain a period of 1114$\pm$213~days, where the uncertainty is calculated as the half of the Full Width at Half Maximum.

\section{Modeling of $\gamma$-ray light curve}
\label{sec:four}
    
    \begin{figure*}
        \centering
        \includegraphics[width=0.981\linewidth]{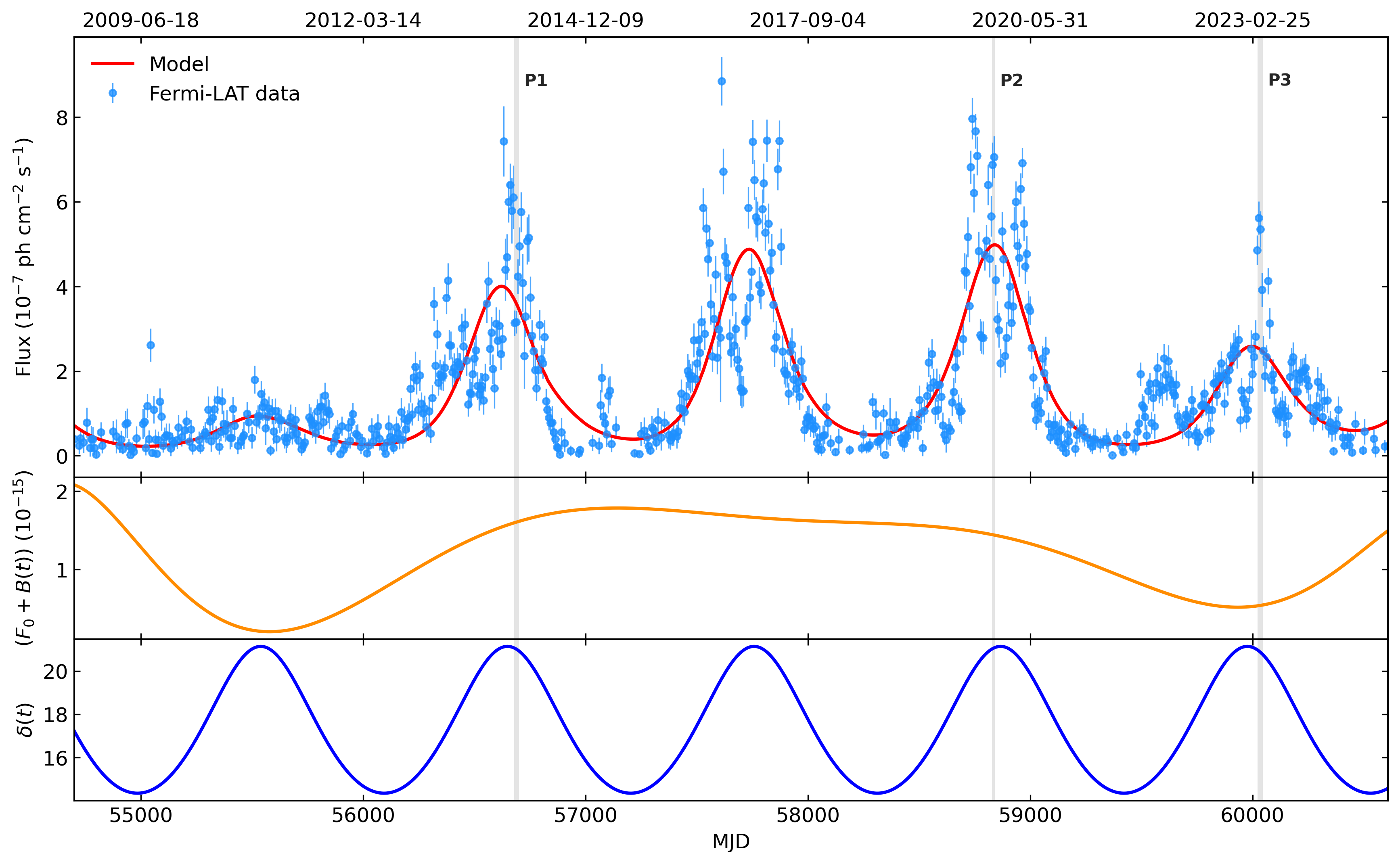}
        \caption{\textit{Top panel:} \fermi\ \gray\ light curve of \sfive\ compared with the best-fit jet-precession model (solid line). The model accounts for long-term quasi-periodic modulation through periodic variations of the Doppler factor, while a low-frequency polynomial baseline captures slower trends not reproduced by pure geometrical effects. \textit{Middle panel:} Evolution of the normalization term $F_0+B(t)$, representing the phenomenological low-frequency baseline that accounts for long-term variability not reproduced by Doppler modulation alone.} \textit{Bottom panel:} Temporal evolution of the Doppler factor $\delta(t)$ inferred from the precession geometry.
        \label{fig:prec_model}
    \end{figure*}
    
    The long-term \gray\ flux variability of \sfive\ shown in panels~a and~b) of Figure~\ref{fig:lc} and in Figure~\ref{fig:CWT} exhibits pronounced repeating out-bursting periods on year-like timescales. There are several mechanisms that can explain such long-timescale QPOs in blazars, including accretion-disk instabilities, jet–disk coupling, binary supermassive black hole systems, and geometrical effects associated with helical or precessing jets. Geometric models are attractive because they can produce large-amplitude, quasi-periodic flux modulations without requiring strong intrinsic changes in the system. In a helical-jet scenario \citep[e.g.,][]{1993ApJ...411...89C, 1999A&A...347...30V,2004A&A...419..913O, 2017Natur.552..374R}, the emitting region follows a curved trajectory, periodically approaching and moving away from the line of sight observer, leading to alternating flux enhancements and dimming. In a jet-precession scenario \citep[e.g.,][]{2000A&A...360...57R, 2003MNRAS.341..405S, 2004MNRAS.349.1218C, Britzen23} the entire jet axis undergoes slow precession, causing the angle between the jet velocity and the line of sight to vary periodically.     
    In this work, we focus on the jet precession model to reproduce the repeating outbursts observed in the \gray\ light curve of \sfive. Within this model, the observed flux variability is driven by periodic changes in the viewing angle $\theta(t)$ between the jet axis and the line of sight. This change induces time-dependent Doppler boosting. For a jet moving with bulk Lorentz factor $\Gamma_{\rm bulk}$, the Doppler factor is given by
    \begin{equation}
        \delta(t)=\frac{1}{\Gamma_{\rm bulk}\left(1-\beta\mu(t)\right)},
    \qquad
    \beta=\sqrt{1-1/\Gamma_{\rm bulk}^{2}},
    \end{equation}
    where $\mu(t)=\cos\theta(t)$. When the jet viewing angle is small, even small variations in $\theta(t)$ can result in significant changes in $\delta(t)$, and thus in the observed flux. The jet orientation is defined using the standard precession geometry, where the instantaneous viewing angle is computed as:
    \begin{equation}
        \mu(t)\equiv\cos\theta(t) = \cos i\:\cos\psi+\sin i\:\sin\psi\, \cos\left[\frac{2\pi (t-t_0)}{P}+\phi\right],
    \end{equation}
    where $i$ is the inclination of the precession axis relative to the line of sight, $\psi$ is the half-opening angle of the precession cone, $P$ is the precession period (for the observer, corresponding to an intrinsic period $P_{\rm rest} = P/(1+z)$ with $z=1.15$), $\phi$ is a phase offset and $t_0$ is a reference epoch. It is important to mention that the function is invariant by swapping  $i$ and $\psi$.\\

    In order to connect Doppler boosting with the observed \gray\ variability, we need to assume a flux scaling as a function of $\delta(t)$ as:
    \begin{equation}
      F(t)=\bigl[F_0+B(t)\bigr]\times\delta(t)^{p(t)},  
    \end{equation}
    where $F_0$ is a constant normalization and $B(t)$ is a low-frequency baseline described by a polynomial function, which takes into account the blazar variability not captured by pure geometrical modulation. We stress that \(B(t)\) is introduced as a phenomenological term at the light-curve level. It accounts for long-term variability that is not reproduced by the purely geometrical Doppler modulation, and may therefore include intrinsic changes in the emitting plasma. The model should not be interpreted as a unique decomposition of the observed flux into a purely geometrical component and a physically specified intrinsic component. Rather, the Doppler factor \(\delta(t)\) inferred from the fit provides the geometrical modulation required by the precession scenario, while \(B(t)\) absorbs the residual low-frequency variability. The functional form of the index $p(t)$ depends on the dominant emission mechanism, which in this case is assumed to be external inverse-Compton scattering of external photons. As discussed in the previous Section, in this case $\nu F_\nu$ is proportional to $\delta^{4+2\alpha}$, where $\alpha$ is the energy spectral index ($F_\nu\propto\nu^{-\alpha}$). This scaling also holds for the integrated flux within the LAT energy band, under the assumption that the spectrum does not deviate significantly from a power-law function. Using the standard relation $\alpha=\Gamma_\gamma-1$, this yields:
    \begin{equation}
         p(t)=4+2\alpha=2+2\Gamma_\gamma(t),
    \end{equation}
    where $\Gamma_\gamma(t)$ is the smoothed photon index in the \gray\ band, inferred from the light curve. 
    
    The model depends significantly on $\Gamma_{\rm bulk}$, for which we tested eight different values from 5 to 22.5.  We fit the \gray\ light curve by sampling the posterior distribution of the nonlinear parameter set $\{i,\psi,P,\phi\}$ using nested sampling. 
    In order to successfully fit the first suppressed maximum in the LAT light curve, we increase the degree of the polynomial $B(t)$ up to the eight order. The flux normalization $F_0$ and the coefficients of $B(t)$ are determined simultaneously within the fit.
    The best-fit parameters are provided in table \ref{tab:precession_gamma_scan} for all scanned values of $\Gamma_{\rm bulk}$. In table \ref{tab:precession_polynomial_scan} we provide the results of the fit when varying the degree of the polynomial $B(t)$, showing that the fit improves up to the eight degree. In the upper panel of Figure \ref{fig:prec_model} we show the result for $\Gamma_{\rm bulk}=15$, a value typical for FSRQ modeling \citep[see e.g.][]{2015MNRAS.448.1060G}. The fit yields $i=(0.76\pm0.13)^\circ$, $\psi=(3.23\pm1.08)^\circ$, $P=(1109.2\pm5.7)~\mathrm{d}$ ($\simeq3.04~\mathrm{yr}$),
    and $\phi=(11.92\pm1.85)^\circ$. As discussed above, the best-fit values of $i$ and $\psi$ can be swapped, resulting in a larger angle to the line-of-sight with smaller precession cone. The inferred minimum angular separation between the line of sight and the precession cone, $\chi_{\min}=|i-\psi|=2.47^\circ$, is smaller than the characteristic relativistic beaming angle, $1/\Gamma_{\rm bulk}=3.82^\circ$, implying that the observer periodically enters the beaming cone. In the same way, the maximum angular separation is larger than the beaming angle, explaining the deep minima. The inferred precessing period $\simeq3.04~\mathrm{yr}$ is in the same range as inferred from the previous studies and as estimated from our CWT analysis. The model also allows to compute the Doppler boosting in different periods. The evolution of the Doppler boosting over the time is shown in lower panel of Figure \ref{fig:prec_model}. The corresponding maximum Doppler factor reaches $\delta_{\max}\simeq21.14$ while the lowest is $\delta_{\min}\simeq14.36$.

The light-curve fit was performed on the \fermi\ data because they provide the most complete dataset among the various instruments. A similar study can be done on the optical light-curves with some caveats: first of all none of the optical datasets covers more than two full cycles of the three-year periodicity due to the multiple filters used in the observations; and then there is no instantaneous measurement of the photon index $\Gamma_{optical}$ as we have for the \fermi\ data. As consistency check, we tried to apply the same precession framework to the ZTF optical light-curve, for the $r$ and $g$ filters separately (the ones that have more time coverage). Contrarily to the fit the $\gamma$-ray data, we adopted in this case a synchrotron beaming factor ($3 + \alpha$, with $\alpha=2$). The results of the fit for the $r$ ($g$) filter is:  $i=1.10^\circ\pm0.18^\circ$, $\psi=3.00^\circ\pm1.15^\circ$, $P=1270\pm53$ days, and $\phi=71^\circ\pm21^\circ$ ($i=1.62^\circ\pm0.32^\circ$, $\psi=2.70^\circ\pm1.03^\circ$, $P=1306\pm38$ days, and $\phi=80^\circ\pm14^\circ$). The results of the fit to ZTF data are broadly in agreement with the fit to \fermi\ data, although they have larger uncertainties. The period estimate is larger in optical data, but it is expected being estimated only from two full cycles and highly sensitive to the exact time of the optical maximum.

\begin{table*}
\centering
\caption{Best-fit parameters of the precessing-jet light-curve model (EIC, polynomial order 8) for different assumed bulk Lorentz factors $\Gamma$. Quoted uncertainties correspond to posterior standard deviations.}
\label{tab:precession_gamma_scan}
\begin{tabular}{cccccccc}
\hline\hline
$\Gamma$ & $i$ (deg) & $\psi$ (deg) & $|i-\psi|$ (deg) & $P$ (days) & $\phi$ (deg) & $\delta_{\rm max}$ & $\ln \mathcal{Z}$ \\
\hline
5.0  &
$3.30 \pm 0.30$ &
$4.32 \pm 0.48$ &
1.02 &
$1112.3 \pm 5.8$ &
$14.3 \pm 1.7$ &
9.82 &
$10050.23$ \\

7.5 &
$2.02 \pm 0.28$ &
$3.29 \pm 0.65$ &
1.27 &
$1110.4 \pm 5.8$ &
$13.0 \pm 1.6$ &
14.54 &
$10050.86$ \\

10.0 &
$1.37 \pm 0.25$ &
$2.99 \pm 0.86$ &
1.62 &
$1109.9 \pm 5.8$ &
$12.4 \pm 1.7$ &
18.49 &
$10049.98$ \\

12.5 &
$0.98 \pm 0.19$ &
$3.09 \pm 1.04$ &
2.11 &
$1109.4 \pm 5.8$ &
$12.2 \pm 1.7$ &
20.61 &
$10049.13$ \\

15.0 &
$0.76 \pm 0.13$ &
$3.23 \pm 1.08$ &
2.47 &
$1109.2 \pm 5.7$ &
$11.9 \pm 1.9$ &
21.14 &
$10048.74$ \\

17.5 &
$0.65 \pm 0.11$ &
$2.76 \pm 1.00$ &
2.10 &
$1108.6 \pm 5.9$ &
$11.6 \pm 2.0$ &
24.78 &
$10047.96$ \\

20.0 &
$0.55 \pm 0.07$ &
$3.38 \pm 1.08$ &
2.83 &
$1109.3 \pm 6.1$ &
$12.1 \pm 1.9$ &
20.27 &
$10048.18$ \\

22.5 &
$0.50 \pm 0.05$ &
$3.52 \pm 1.19$ &
3.02 &
$1109.4 \pm 5.9$ &
$12.4 \pm 2.1$ &
18.73 &
$10047.25$ \\
\hline
\end{tabular}
\end{table*}

\begin{table*}
\centering
\caption{Best-fit parameters of the precessing-jet light-curve model (EIC, Lorentz factor = 15) for different assumed degree $b$ of the polynomial function $B(t)$. Quoted uncertainties correspond to posterior standard deviations.}
\label{tab:precession_polynomial_scan}
\begin{tabular}{cccccccc}
\hline\hline
$b$ & $i$ (deg) & $\psi$ (deg) & $|i-\psi|$ (deg) & $P$ (days) & $\phi$ (deg) & $\delta_{\rm max}$ & $\ln \mathcal{Z}$ \\
\hline
1  &
$0.72 \pm 0.05$ &
$7.85 \pm 0.23$ &
7.13 &
$1131.9 \pm 8.3$ &
$24.5 \pm 2.9$ &
6.71 &
$9815.42$ \\

2 &
$0.71 \pm 0.04$ &
$5.31 \pm 0.88$ &
4.60 &
$1132.1 \pm 5.5$ &
$15.7 \pm 1.7$ &
12.29 &
$10018.51$ \\

3 &
$0.71 \pm 0.04$ &
$4.82 \pm 0.96$ &
4.11 &
$1128.5 \pm 5.6$ &
$12.4 \pm 1.7$ &
13.89 &
$10019.53$ \\

4 &
$0.71 \pm 0.08$ &
$4.16 \pm 1.17$ &
3.45 &
$1125.1 \pm 6.1$ &
$14.7 \pm 1.8$ &
16.54 &
$10028.24$ \\

5 &
$0.70 \pm 0.08$ &
$4.09 \pm 1.15$ &
3.39 &
$1125.4 \pm 6.2$ &
$14.2 \pm 2.0$ &
16.81 &
$10029.68$ \\

6 &
$0.78 \pm 0.14$ &
$3.11 \pm 1.01$ &
2.33 &
$1116.4 \pm 5.5$ &
$12.8 \pm 1.8$ &
21.86 &
$10044.01$ \\

7 &
$0.76 \pm 0.12$ &
$3.38 \pm 1.10$ &
2.62 &
$1115.9 \pm 5.5$ &
$12.8 \pm 1.8$ &
20.40 &
$10044.73$ \\

8 &
$0.76 \pm 0.13$ &
$3.23 \pm 1.08$ &
2.47 &
$1109.2 \pm 5.7$ &
$11.9 \pm 1.9$ &
21.14 &
$10048.74$ \\

9 &
$0.76 \pm 0.15$ &
$3.18 \pm 1.12$ &
2.42 &
$1108.3 \pm 5.8$ &
$12.7 \pm 1.9$ &
21.41 &
$10049.02$ \\

\hline
\end{tabular}
\end{table*}

    \section{SED Modeling}
    \label{sec:five}

   The aim of the SED modeling is to test whether the MWL emission observed during selected epochs can be reproduced with physically reasonable one-zone leptonic parameters when the Doppler factor is fixed to the value implied by the precession model. The Doppler factor is therefore used as a geometrical constraint, not as a parameter describing the intrinsic variability of the emitting plasma. The SED models should thus be regarded as a consistency test of the precessing-jet interpretation rather than as a unique decomposition of geometrical and intrinsic variability.

    The MWL light curves shown in Figure \ref{fig:lc} allow for the identification of several epochs with sufficiently contemporaneous data and for the construction of time-resolved SEDs suitable for theoretical modeling. Performing SED modeling across distinct periods is essential, as it enables to track changes in the emission components. To reliably constrain both the synchrotron and inverse-Compton emission components, SEDs are constructed only for epochs in which X-ray observations are available. However, as noted in Section~\ref{sec:anal}, the source is relatively faint in the X-ray band, and a single short \textit{Swift}-XRT exposure typically does not provide sufficient signal-to-noise ratio to produce an SED suitable for modeling. Consequently, XRT observations that are close in time and show no evidence of significant variability are merged and re-analyzed. This allows to obtain spectra with adequate photon statistics and to extract a sufficient number of data points for theoretical modeling. Similarly, to improve photon statistics in the HE \gray\ band, the adaptively binned light curve is segmented into intervals of approximately constant flux using the Bayesian Block algorithm \citep{2013ApJ...764..167S}. Within each block the flux is assumed to be statistically consistent with a constant level, allowing to perform spectral analysis over a longer integration time than one individual measurement and thereby obtain spectra that extend above the GeV band. The resulting Bayesian blocks are shown as red segments in panel a) of Figure~\ref{fig:lc}. The start and end times of these blocks are then used to search for contemporaneous observations in the other bands, enabling the construction of MWL SEDs based on overlapping time intervals. This procedure follows the SED/light-curve animation method described in \citet{2024AJ....168..289S}. 
    
    Based on this approach, the following intervals defined by the Bayesian-block boundaries are selected for SED modeling. For each period, all available \textit{Swift}-XRT pointings within the corresponding MJD range are merged into a single spectrum after verifying the absence of significant intra-interval variability, and \textit{Swift}-UVOT data are taken from the exposure providing full filter coverage. Specifically:
    
    \begin{itemize}
    
    \item \textit{Period (P1)}: MJD 56678.5–56699.8, an interval within the first \gray\ brightening episode. UVOT data are taken from the 26 January 2014 exposure.

    \item \textit{Period (P2)}: MJD 58827.5–58841.4, within the third \gray\ brightening and overlapping with an \textit{AstroSat} observation in 2020. UVOT data are taken from the 21 December 2019 exposure.
    
    \item \textit{Period (P3)}: MJD 60023.0–60045.28, a segment of the fourth \gray\ brightening, coincident with a \textit{NuSTAR} pointing and a second \textit{AstroSat} observation. UVOT data are taken from the 5 April 2023 exposure.
    
    \end{itemize}
    
    The SED corresponding to the second brightening phase is not modeled, as it is essentially identical to the one observed during the first brightening episode and therefore does not provide additional independent constraints. The three selected intervals correspond to active periods with sufficient MWL coverage for SED modeling. They are not intended to provide an exhaustive sampling of low, intermediate, and high brightness states.

    \subsection{One-zone leptonic external inverse Compton model}
    The double-peaked SED of FSRQs is commonly interpreted within the external inverse Compton (EIC) scenario, in which relativistic electrons in the jet up-scatter both the synchrotron photons produced within the jet by the same electron population, as well as the external radiation fields. 
    The dominant source of external seed photons depends on the location of the emitting region relative to the central black hole. Possible contributors include direct radiation from the accretion disk \citep{1992A&A...256L..27D, 1994ApJS...90..945D}, photons reprocessed in the broad-line region (BLR) \citep{1994ApJ...421..153S}, and infrared emission from the dusty torus \citep{2000ApJ...545..107B}.
    
    The EIC model adopted in this study is based on a convolutional neural network (CNN) surrogate approach that replaces computationally expensive numerical calculations, enabling efficient global inference of the model parameters. A detailed description of the model is provided in \citet{2024ApJ...971...70S}; here we summarize the key aspects.
    
    The emission region is assumed to be a spherical blob of radius $R$, located at a distance $R_{\rm diss}$ from the central black hole and filled with a uniform magnetic field. The blob moves relativistically along the jet with bulk Lorentz factor $\Gamma$; assuming a small viewing angle, the Doppler factor is $\delta \simeq \Gamma$. Assuming a conical jet with opening angle $\simeq 1/\Gamma$, the distance of the dissipation region from the central engine scales with the emission region size as $R_{\rm diss} \simeq \Gamma R$.

    Relativistic electrons are continuously injected into the emitting region according to a power-law distribution with an exponential cutoff,
        \begin{equation}
        Q_e(\gamma)=
        \begin{cases}
        Q_{e,0}\,\gamma^{-p}\exp\!\left(-\dfrac{\gamma}{\gamma_{\max}}\right),
        & \gamma \ge \gamma_{\min}, \\[4pt]
        0, & \text{otherwise},
        \end{cases}
        \end{equation}
        where $p$ is the electron spectral index, and $\gamma_{\min}$ and $\gamma_{\max}$ are the minimum and maximum Lorentz factors, respectively. The normalization $Q_{e,0}$ defines the total electron luminosity given by:
        \begin{equation}
            L_e = \pi R^2 \delta^2 m_e c^3 \int_1^{\infty} \gamma, Q_e(\gamma), d\gamma.
        \end{equation}

    External photon fields are modeled following standard prescriptions and include thermal emissions from the accretion disk, radiation reprocessed by the BLR, and infrared emissions from the dusty torus. The latter two components are assumed to scale with the accretion-disk luminosity ($L_d$). The spectra of the external photon fields are approximated as blackbody distributions, defined in the black hole frame and Lorentz-transformed into the comoving frame of the emitting region.
    Consequently, the model has nine free parameters: the emitting region radius $R$, Doppler factor $\delta$, magnetic field strength $B$, electron luminosity $L_e$, minimum and maximum electron Lorentz factors $\gamma_{\rm min}$ and $\gamma_{\rm max}$, the electron spectral index $p$, the accretion-disk luminosity $L_d$, and the black hole mass $M_{\rm BH}$, which determines the characteristic temperature of the accretion-disk emission. Considering a broad range for these parameters as presented in Table 1 of \citet{2024ApJ...971...70S}, we generate synthetic SEDs with the \texttt{SOPRANO} code \citep{2022MNRAS.509.2102G} to generate the training dataset for the CNN. After the training phase, the CNN accurately reproduces the radiative output of the EIC model while reducing the computational cost of generating individual SEDs by orders of magnitude. The trained CNN for the EIC model, together with analogous CNNs developed for synchrotron self-Compton (SSC; \citealt{2024ApJ...963...71B}) and lepto-hadronic models \citep{2025ApJ...990..222S}, are publicly available through the \texttt{MMDC} platform \citep{2024AJ....168..289S}.

    \subsection{Modeling results}
    The selected SEDs are modeled by coupling the CNN surrogate with the nested sampling algorithm \texttt{MultiNest} \citep{FHB09}, using 1500 live points and a tolerance parameter of 0.4 to ensure efficient sampling and convergence. Due to the lack of low-frequency data ($<10^{10}\,\mathrm{Hz}$), the minimum electron Lorentz factor is fixed at $\gamma_{\rm min}=100$ for all epochs. For each SED, the Doppler factor is inferred independently from jet precession modeling of the \gray\ light curve, which provides the temporal evolution of $\delta$ as well. The value of $\delta$ corresponding to the epoch of the \textit{Swift}-UVOT observation is then adopted in the SED modeling for each period; the resulting $\delta$ values are reported in Table \ref{tab:model_parameters}. Absorption by the extragalactic background light (EBL) is included in the model calculations using the model of \citet{2011MNRAS.410.2556D}. Other unknown parameters in the model are the black hole mass and the accretion disk luminosity. When a “blue bump” is observed in the optical/UV band, it can be attributed to direct thermal emission from the accretion disk and used to constrain these parameters \citep[see e.g.][]{2010MNRAS.405..387G}. However, no such feature is evident in the long-term SED of \sfive. We therefore constrained the accretion disk luminosity and its characteristic temperature (the black hole mass) by fitting a disk component to the full set of optical/UV observations of \sfive, requiring that the disk emission does not exceed the observed lower fluxes. This procedure yields $L_{\rm disk}=1.14\times10^{46}\mathrm{erg\:s^{-1}}$ and $M_{\rm BH}=2.6\times10^{9} M_\odot$. It should be noted that the inferred black hole mass is consistent with the value reported in \citet{2025RAA....25b5004W}, $M_{\rm BH}=3.49\times10^{9},M_\odot$, which was estimated from the observed QPO period. However, the inferred accretion disk luminosity is somewhat higher than the value estimated by \citet{2021ApJS..253...46P}, $L_{\rm disk}=1.38\times10^{45}\mathrm{erg\:s^{-1}}$, which was derived from the luminosity of the BLR. In our modeling, adopting $L_{\rm disk}=1.38\times10^{45}\,\mathrm{erg\,s^{-1}}$ together with $\delta \simeq 21$ (see Table \ref{tab:model_parameters}) requires extremely large electron powers, $L_{\rm e} \gtrsim 10^{48}\,\mathrm{erg\,s^{-1}}$, to reproduce the high $\gamma$-ray flux observed during the active periods. Such a scenario is not only problematic from the point of the required jet energetics, but also leads to an regime in which the SSC component dominates over the EIC component. Therefore, the derived disk luminosity used in this work should be considered as an upper limit, which we adopt as the maximum allowed contribution from the accretion disk. 

\begin{figure*}[!t]
            \centering
            \includegraphics[width=0.48\linewidth]{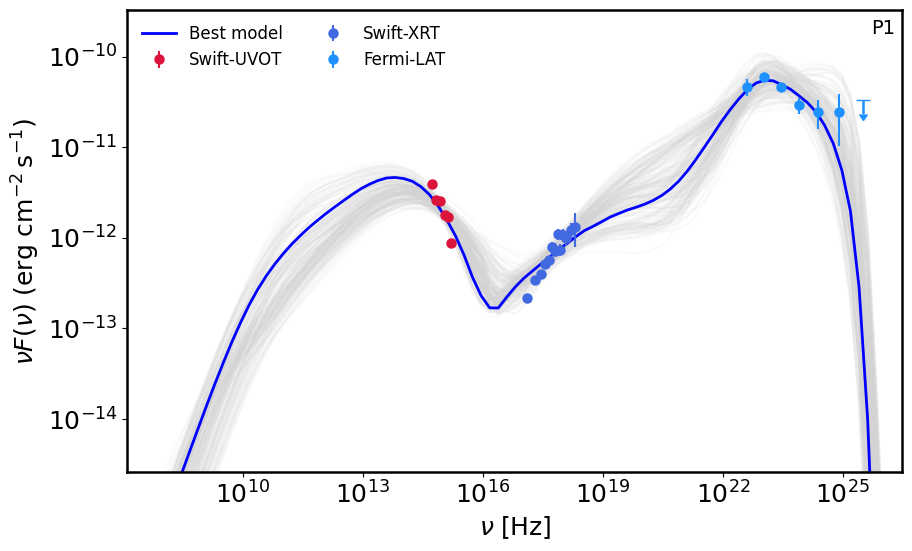}
            \includegraphics[width=0.48\linewidth]{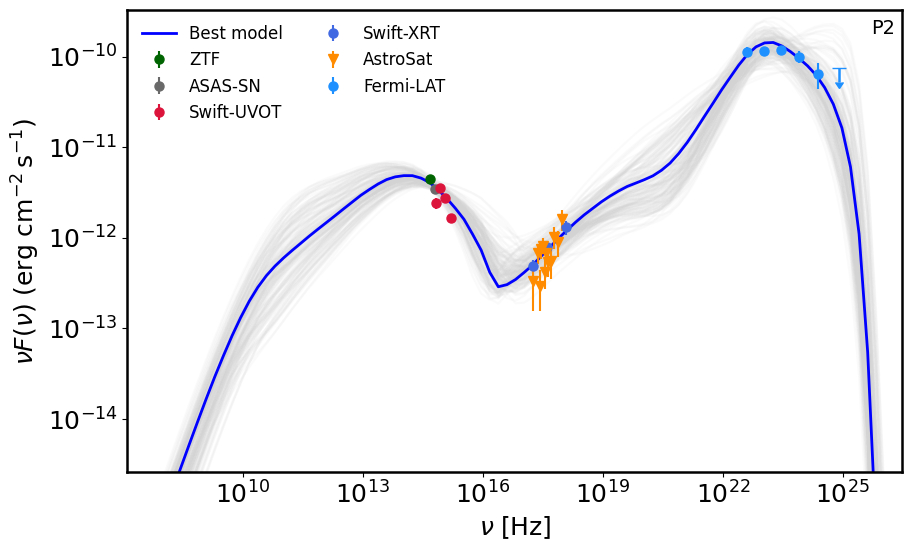}\\
            \includegraphics[width=0.48\linewidth]{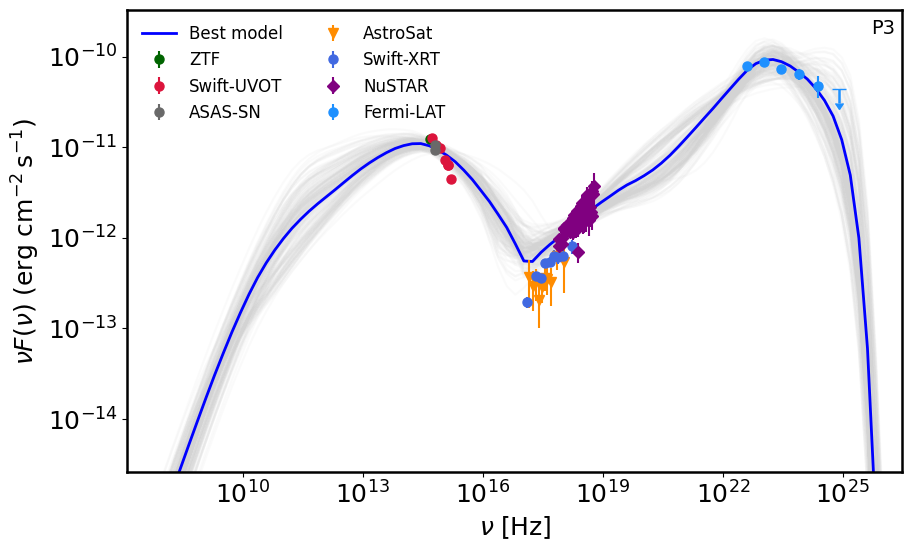}
            \caption{Broadband SEDs of \sfive\ during the three activity periods P1–P3. MWL data from ZTF, ASAS-SN, \textit{Swift}-UVOT, \textit{Swift}-XRT, \textit{AstroSat}, \textit{NuSTAR}, and \fermi\ are shown with different symbols and colors, as indicated in the legends. The blue solid line represents the best-fit leptonic emission model for each period, while the gray curves show the model uncertainty. In all cases, the model include attenuation due to EBL absorption, computed using the model of \citet{2011MNRAS.410.2556D}.}
            \label{fig:sed}
        \end{figure*}

    The modeling results of the SEDs for the three active periods (P1–P3) are shown in Figure~\ref{fig:sed}, and the corresponding best-fit parameters are listed in Table~\ref{tab:model_parameters}. The model corresponding to the best-fit parameters is shown as a blue solid line, while the associated model uncertainty is shown in gray. In all three active states, the emission up to the optical/UV band is dominated by synchrotron radiation from relativistic electrons. The X-ray emission is primarily produced via SSC scattering, whereas the HE $\gamma$-ray emission is explained by the EIC component. The MWL fitting is successful in optical/ UV and in $\gamma$-rays, while soft X-rays are over predicted, especially in P3. This is likely due to the fixed value of $\gamma_{\rm min}=100$ adopted in the fitting algorithm: higher values of this parameter will result in a modeling more in line with the hard X-ray spectrum. The three SEDs correspond to selected active periods with adequate MWL coverage, rather than to a set of states chosen to span the full observed brightness range of the source. Within this limited sample, the derived model parameters do not show large variations across the three periods once uncertainties in the fitted parameters are considered. The power-law index of the emitting electron population remains close to $\sim 2$ in all epochs. The cutoff Lorentz factor shows a modest increase from P1 to P3, with $\log \gamma_{\rm cut}$ changing from 4.02 to 4.56, indicating that electrons are accelerated to higher maximum energies during the later activity state. The magnetic field strength remains consistent within uncertainties, with $\log B \sim -1.1$ to $-1.4$, while the size of the emitting region is also stable, $\log R \simeq 17.6$. The electron power is comparable in P1 and P2, with $\log L_{\rm e}\approx 46.48$, and decreases moderately in P3 to $\log L_{\rm e}=46.33$. Finally, the magnetic power, computed as $L_{B}=\pi c R_b^2 \Gamma^2 U_{B}$, is lower than the electron power in all periods $\log L_{\rm B}\approx 44.4-45.1$, implying a particle-dominated emitting region during these outburst states.
   These parameters should not be interpreted as a unique description of the intrinsic source evolution. They rather show that, once the Doppler factor is fixed to the value inferred from the precession model, the observed SEDs can be reproduced with physical parameters typical of FSRQ jets.

        \begin{table}[ht!]
        \centering
        \caption{ Best-fit parameters of the leptonic emission model for the three activity emission periods (P1–P3) of \sfive. The Doppler factor $\delta$ is not a free parameter of the SED fit but is inferred from the jet precession modeling. The minimum electron Lorentz factor was fixed to $\gamma_{\rm min}=100$ in all fits. The accretion disk luminosity and black hole mass were also fixed to $L_{\rm disk}=1.14\times10^{46}\mathrm{erg\:s^{-1}}$ and $M_{\rm BH}=2.6\times10^{9}M_\odot$, respectively.
        }
        \label{tab:model_parameters}
        \begin{tabular}{l|ccc}
        \hline
        Parameter & P1 & P2 & P3 \\
        \hline
        $p$ & $2.02 \pm 0.16$ & $2.09 \pm 0.17$ & $2.03 \pm 0.10$ \\
        
        $\delta$ & $21.04$ & $21.07$ & $20.82$ \\
        
        $\log \gamma_{\rm cut}$ & $4.02 \pm 0.35$ & $4.33 \pm 0.33$ & $4.56 \pm 0.31$\\
        
        $\log B$ & $-1.29 \pm 0.60$  & $-1.44 \pm 0.47$ & $-1.10 \pm 0.59$ \\
        
        $\log R$ & $17.57 \pm 0.43$ & $17.55 \pm 0.27$ & $17.56 \pm 0.32$ \\
        
        $\log L_{\rm e}$ & $46.48 \pm 0.54$ & $46.48 \pm 0.53$ & $46.33 \pm 0.48$ \\
        $\log L_{\rm B}$ & $44.78$ & $44.43$ & $45.13$
        \end{tabular}
        \end{table}
        
\section{Discussions and Conclusions}
\label{sec:disc}

In this work we have presented a comprehensive MWL study of the FSRQ \sfive, based on contemporaneous observations covering the full electromagnetic spectrum. At high energies we combined \textit{Fermi}-LAT $\gamma$-ray monitoring with X-ray observations from \textit{Swift}-XRT, \textit{AstroSat}-SXT, and \textit{NuSTAR}, while the optical and near-infrared domain is sampled by ZTF, Pan-STARRS, ASAS-SN, and NEOWISE. Archival optical data were also considered to extend the temporal baseline. Although the historical data are too sparse to provide a constraint on the long-term light curve, they show clear variability in optical, with for example a bright flare during December 18, 1998. If we consider the best-fit periodicity of 1110 days, with maxima in the \textit{Fermi} era at around MJD 55550, 56660, 57770, 58880, and 59990, one of the previous maxima happened at MJD 51110, that is less than two months before that Palomar observations. Remarkably, none of the other historical observations are close to a predicted past maximum. 

The combined light curves reveal pronounced long-term variability, characterized by four major episodes of enhanced activity on year-long timescales. A time-frequency analysis based on the continuous wavelet transform (CWT), reveals a characteristic timescale of $\sim 1100$ days. This timescale is also recovered by the LAT light-curve fitting and is further supported by the recurrence of major $\gamma$-ray outbursts. The robustness of this timescale points to a genuine long-term modulation.

The long-term variability is accompanied by clear MWL correlations. Strong flux-flux correlations are observed between the $\gamma$-ray and optical/UV/IR bands, while correlations involving the X-ray emission are significantly weaker, although it is important to underline that X-ray observations are sparser and biased towards the LAT bright states. The ZDCF analysis indicates that the optical/UV variations are broadly contemporaneous with the $\gamma$-ray emission within uncertainties, and no significant time lag is detected. These results suggest that the dominant variability across the optical to $\gamma$-ray bands is driven by a common mechanism.
The absence of a significant time-lag in S5 1044+71 between the $\gamma$-ray light curve and the optical to X-ray light curves is consistent with the general behavior observed in many LAT blazars. Strong optical/$\gamma$-ray correlations with small or zero delays have been reported both in individual sources and in population studies \citep{Bonning09, Chatterjee12, Cohen14, Hovatta14, deJaeger23}.

The slopes of the observed flux-flux relations provide important physical constraints. In Sect.~3.2 we introduced the fact that, if long-term variability were dominated by changes in the Doppler factor, the correlation index can be predicted to be between $0.4$ and $0.7$ between X-ray and \gray, and to be between $0.6$ and $0.9$ between optical/UV and \gray. The values we observe are about $0.3$ and $1$, respectively, close but slightly outside these theoretical predictions. Part of the discrepancy is expected because the simple estimate assumes no spectral variability, which we do observe. In addition, and as mentioned above, the X-ray measurement are not coming from an unbiased monitoring, but are biased towards the LAT flaring states. The correlation study seem to globally support indeed a variability driven by changes in the Doppler factor. But, more interestingly, it can also be used to challenge alternative models. If the periodicity would be due to changes in the particle content in the jet, (i.e. variations of the electron normalization linked to accretion-rate fluctuations), a one-zone leptonic scenario would predict an approximately linear relation between synchrotron (optical) and external inverse-Compton (EIC; $\gamma$-ray) emission, and a nearly quadratic response of the SSC-dominated X-rays (again, ignoring here spectral variability as we did for the Doppler-factor variability discussion). This prediction is remarkably different compared to the Doppler factor change: X-rays, being SSC dominated, should have shown a much stronger variability than the other bands. The observed behavior is markedly different: while the optical-$\gamma$ correlation is close to linear and cannot be used to discriminate between variability processes, the X-$\gamma$ correlation is much flatter than expected in a particle-driven scenario. This discrepancy strongly disfavors models in which long-term variability is primarily driven by changes in the electron density and thus on the injection of particles in the emitting region.  

The spectral variability study further supports the picture of a Doppler-factor induced variability. In addition to the flux brightening, an increase of $\delta$ also induces a translation of the SED to higher energies. If the SED peaks are close to the observing bands, the spectral variability is naturally explained. In this scenario, the harder-when-brighter behaviour in LAT data is due to the EIC peak getting closer to the LAT energy band, flattening the broad-band spectrum. In the same way, the harder-when-brighter behaviour in X-ray data is due to the shift to higher energies of the curved low-energy part of the SSC component. The softer-when-brighter behaviour in optical spectra is on the other hand probably due to the emergence of a separate hard component (likely the accretion disk) during the low flux states from the jet. This effect is also supported by the fact that in several of the optical/\gray correlations there is preference for a constant function superposed to the powerlaw correlation.

To investigate the physical origin of the long-term modulation, we modeled the LAT light curve within the framework of a precessing relativistic jet. In this scenario, periodic changes in the angle between the jet axis and the line of sight lead to time-dependent Doppler boosting of the observed emission. Using Bayesian inference, we infer a precession period of $P \simeq 1109$~d ($\simeq 3.04$~yr), fully consistent with the characteristic timescale identified by the CWT analysis and with the recurrence timescale of the observed outbursts. The best-fit solution corresponds to a strongly aligned geometry, with an inclination $i \simeq 0.76^\circ$ and a precession cone half-opening angle $\psi \simeq 3.23^\circ$, assuming a bulk Lorentz factor of 15 (we note that for the model $i$ and $\psi$ can be inverted). The corresponding Doppler factor varies in the range $\delta \sim 14.4$ to $21.1$, which is sufficient to produce large-amplitude flux enhancements even if the comoving emissivity varies only modestly. Importantly, scanning the bulk Lorentz factor over a wide range yields consistent precession periods and similar likelihood values, indicating that the inferred period is data-driven and not an artifact of a particular choice of $\Gamma_{\rm bulk}$. We remind the reader that the fitted periodicity does not take into account cosmological redshift. The precessing period in the AGN frame is $P_{rest} \simeq 1110/2.15 \simeq 516$~days.

The precessing-jet model we present thus provides a natural explanation for all observational features of \sfive: the quasi-regular recurrence of major $\gamma$-ray outbursts, the strong optical-$\gamma$ correlations, and the comparatively weaker X-ray response. In this picture, Doppler-factor variations define the long-term envelope of the variability, while intrinsic changes in the electron distribution and additional emission components modulate the detailed spectral and temporal behavior.

We modeled the spectral energy distribution of \sfive\ during three active periods with sufficient MWL coverage using a one-zone leptonic EIC framework, implemented through a CNN-based surrogate coupled with \texttt{MultiNest} to efficiently explore the high-dimensional parameter space. The fits indicate an electron energy distribution with a power-law index $p \simeq 2$, consistent with standard diffusive shock acceleration. The cutoff Lorentz factor spans $\gamma_{\rm cut} \sim (1$--$3.6)\times 10^{4}$, with a modest increase toward the latest activity state, suggesting either more efficient acceleration or reduced radiative constraints during that epoch.
The inferred physical conditions of the emitting region remain remarkably stable: both the magnetic field strength and the size of the emission region are consistent within uncertainties across P1--P3. This implies that the outbursts are not driven by large-scale structural changes of the emission zone, but rather by variations in the high-energy tail and normalization of the electron population, combined with geometric Doppler modulation.

The SED modeling also provides constraints on the location of the emission region. In the adopted framework, the emitting-region size and distance along the jet are related through $R_{\rm diss} \simeq \Gamma R$. The inferred size, $\log R \simeq 17.6$, together with $\delta \simeq \Gamma \simeq 21$, places the dissipation region at $R_{\rm diss} \gtrsim 10^{18}$~cm from the central black hole. For the inferred accretion-disk luminosity $L_{\rm disk} = 1.14 \times 10^{46}\,\mathrm{erg\,s^{-1}}$, the expected BLR radius is $R_{\rm BLR} \simeq 3.4 \times 10^{17}$~cm, implying that the emission region is located well outside the BLR. In this regime, the dominant external photon field for inverse Compton scattering is provided by the dusty torus rather than the BLR. This configuration reduces internal $\gamma\gamma$ absorption and makes \sfive\ a plausible candidate for very-high-energy $\gamma$-ray emission during bright states. The redshift of the source ($z=1.15$) is larger than the current most-distant very-high-energy blazars \citep{0346, OP313}, but it is reachable by CTAO. The predictability of the flaring activity from \sfive\ can make it an important target to study propagation of very-high-energy $\gamma$-ray photons on cosmological distances. 

The detection of a stable, year-scale modulation naturally raises the question of its physical origin. Precession of the jet-launching region can arise from several mechanisms, including warped or tilted accretion disks \citep{Caproni04, Liska18}, or tidal torques in a black hole binary \citep{Larwood98}. Our geometric modeling constrains the effective precession parameters but does not, by itself, uniquely identify the underlying physical driver.

Recent works \citep{Kun22, Kun24} have proposed that the long-term modulation in \sfive\ may be linked to spin-orbit precession in a very compact, near-merger SMBH binary. In such scenarios, year-scale precession of the black hole spin becomes possible because relativistic spin-orbit torques grow rapidly in the late inspiral phase. While our results are compatible with such an interpretation, they do not require it. We do not observe any evidence for deviation from the $P\simeq1100$ days period, not in the CWT analysis (that is sensitive to QPO with shifting period), nor in the light-curve fit. In addition, a near-merger binary would plausibly perturb the broad-line region, potentially producing asymmetric or time-variable line profiles. Optical spectra of \sfive, including archival Steward Observatory data \footnote{see \url{https://james.as.arizona.edu/~psmith/Fermi/}}, appear consistent with those of a standard FSRQ and do not show obvious signatures of strong BLR disruption. This does not exclude a compact binary configuration, but it requires that the BLR is not affected by it. An alternative possibility is that the precession involves the inner accretion flow or jet-launching region rather than the black hole spin itself. In this case, a misalignment between the disk angular momentum and the black hole spin can lead to a precessing jet nozzle without invoking a near-coalescence binary. Such a scenario naturally explains the modest precession cone angle inferred from the light-curve modeling and is consistent with the absence of clear spectroscopic anomalies.

In summary, our MWL analysis of \sfive\ reveals a robust, long-term modulation with a characteristic timescale of $\sim 3$~yr, accompanied by strong optical-$\gamma$ correlations and weaker X-ray variability. Flux-flux correlations disfavor particle-content-driven scenarios and instead support a geometric Doppler-factor modulation as the dominant driver of the long-term variability. A precessing-jet model provides a self-consistent description of the long-term \fermi\ light curve, of the spectral variability observed in \grays, X-rays, and optical, of the flux-flux correlations, and of the absence of time-lags between bands. The SED modeling places the emission region beyond the BLR and supports an EIC-dominated high-energy component, making this object an interesting target for very-high-energy $\gamma$-ray observations.

Future progress will require continued long-term monitoring to test the phase stability of the modulation over additional cycles, as well as high-resolution VLBI observations to directly probe changes in jet orientation. Dedicated spectroscopic monitoring of the broad emission lines could provide an independent test of compact-binary scenarios. Based on our best-fit model of the \fermi\ light curve, the next maximum is expected during March 2026. Given the broadness of the activity period, high \gray\ state is expected for few months before and after\footnote{The new activity period in \fermi\ data is confirmed and visible at the Light-Curve Repository (\url{https://fermi.gsfc.nasa.gov/ssc/data/access/lat/LightCurveRepository/source.html?source_name=4FGL_J1048.4+7143}). The new maximum was reached in April 2026, while the manuscript was undergoing peer review in the journal. }.\\

\printcredits

\bibliographystyle{cas-model2-names}

\bibliography{S51044_JHEAP}



\end{document}